%% file: diff33.tex
\documentclass[structabstract]{00_CONFIG/cls/aa}
\newcommand*{\THRESmin}{20}
\newcommand*{\THRESave}{40}
\newcommand*{\THRESmax}{60}
\input{00_CONFIG/Layout/Colors}
\input{00_CONFIG/Layout/Graphics}
\input{00_CONFIG/Layout/Fonts}
\input{00_CONFIG/Layout/Abbreviations}
\input{00_CONFIG/Layout/References}
\input{00_CONFIG/Astro/Astro}
\input{97_IDL/SEDsexT\THRESave/TABLES/M33_Numbers}
\begin{document}
\title{Millimeter and Submillimeter Excess Emission in \MESSIER{33} \\ revealed by Planck and LABOCA.}
\subtitle{}
\author {%
         I. Hermelo\inst{1}       \and
         M. Rela\~no\inst{2,3}    \and
         U. Lisenfeld\inst{2,3}   \and
         S. Verley\inst{2,3}      \and
         C. Kramer\inst{1}        \and
         T. Ruiz-Lara\inst{2,3}   \and
         M. Boquien\inst{4}       \and
         E. M. Xilouris\inst{5}   \and
         M. Albrecht\inst{6}
        }
\institute{
           Instituto Radioastronom\'ia Milim\'etrica (IRAM), Av. Divina Pastora 7, N\'ucleo Central, 18012, Granada, Spain \\ \email{hermelo@iram.es}
           \and
           Departamento de F\'isica Te\'orica y del Cosmos, Universidad de Granada, 18071, Granada, Spain
           \and
           Instituto Universitario Carlos I de F\'isica Te\'orica y Computacional, Universidad de Granada, 18071 Granada, Spain
           \and
           Institute of Astronomy, University of Cambridge, Madingley Road, Cambridge CB3 0HA, UK
           \and
           Institute for Astronomy, Astrophysics, Space Applications \& Remote Sensing, National Observatory of Athens, P. Penteli, 15236, Athens, Greece
           \and
           Argelander-Institut f\"ur Astronomie, Universit\"at Bonn, Germany
           }

%
%
%
\abstract
{
%
%
Previous studies have shown the existence of an excess of emission
at submillimeter (submm) and millimeter (mm) wavelengths in the
spectral energy distribution (SED) of many low-metallicity galaxies.
The so-called ``submm excess'', whose origin remains unknown, challenges our
understanding of the dust properties in low-metallicity environments.
\vspace{0pt}}{
%
%
The goal of the present study is to model separately the emission from the
star forming (SF) component and the emission from the diffuse interstellar
medium (ISM) in the nearby spiral galaxy \MESSIER{33}, in order to check if
both components can be well fitted using radiation transfer models or
if there is an excess of submm emission associated to any or both of them.
\vspace{0pt}}{
%
%
We decomposed the observed SED of \MESSIER{33} into its
SF and diffuse components.
Mid-infrared (MIR) and far-infrared (FIR) fluxes were extracted from
\SPITZER{} and \HERSCHEL{} data.
At submm and mm wavelengths, we used  ground-based observations
from \APEX{} to measure the emission from the SF component
and data from the \PLANCK{} space telescope to estimate the diffuse emission.
Both components were separately fitted using radiation transfer models
based on standard dust properties (i.e. emissivity index \Eindex{=}{2})
and a realistic geometry.
The large amount of previous studies helped us to estimate
the thermal radio emission and to constrain an important part
of the input parameters of the models.
Both modeled SEDs were combined to build the global SED of \MESSIER{33}.
In addition, the radiation field necessary to power the dust emission in our
modeling was compared with observations from \GALEX{}, \SLOAN{}, and \SPITZER{}.
\vspace{0pt}}{%
%
%
Our modeling is able to reproduce the observations at MIR and FIR wavelengths,
but we found a strong excess of emission at submm and mm wavelengths,
where the model expectations severely underestimate the
\CAMERAmu{LABOCA}{} and \PLANCK{} fluxes.
We also found that the ultraviolet (UV) radiation escaping the galaxy is
\percent{\ESCaveYOUNG}
higher than the model predictions.
From the total mass of dust derived from our modeling and
the mass of atomic and molecular gas measured with
the \VLA{} and the IRAM \meters{30} telescope,
we determined a gas-to-dust mass ratio
\Gdust{\sim}{100},
significantly lower than the value expected from the
sub-solar metallicity of \MESSIER{33}.
\vspace{0pt}}{%
%
%
We discussed different hypotheses to explain the
discrepancies found in our study
(i.e.,
excess of emission at submm and mm wavelengths,
deficit of UV attenuation,
and abnormally low value of \Gdust{}{}),
concluding that different dust properties in \MESSIER{33} is
the most plausible explanation.
}
\maketitle
%
\input{01_INTRODUCTION/Introduction}
%
\input{02_DATA/Data}
%
\input{03_PHOTOMETRY/PhotometrySEX}
%
\input{04_MODELS/Models}
%
\input{05_RESULTS/Results}
%
\input{06_DISCUSSION/Discussion}
%
\input{07_SUMMARY/Summary}
%
%
\begin{acknowledgements}
We thank the second referee for her/his constructive report and helpful suggestions.
%
%
We thank F.-X. D\'esert for his invaluable help with \PLANCK{} data.
This work was partially supported by a Junta de Andaluc\'ia Grant FQM108,
a Spanish MEC Grant AYA-2011-24728 and AYA2014-53506-P,
Juan de la Cierva fellowship Program and
the European Reintegration Grant ERG HER-SFR.
TRL thanks the support of the Spanish Ministerio de Educaci\'on,
Cultura y Deporte by means of the FPU fellowship.
MA acknowledges support by the German Research Foundation (DFG) in the
framework of the priority program 1573, ``Physics of the Interstellar Medium'',
through grant number AL 1467/2-1.
This research made use of the NASA/IPAC Extragalactic Database (NED),
which is operated by the Jet Propulsion Laboratory, California Institute of Technology,
under contract with the National Aeronautics and Space Administration.
We also acknowledge the use of the HyperLeda database (\url{http://leda.univ-lyon1.fr}).
Funding for the creation and distribution of the SDSS Archive has been provided
by the Alfred P. Sloan Foundation, the Participating Institutions,
the National Aeronautics and Space Administration,
the National Science Foundation, the US Department of Energy,
the Japanese Monbukagakusho, and the Max Planck Society.
The SDSS Web site is \url{http://www.sdss.org/}.
The SDSS is managed by the Astrophysical
Research Consortium (ARC) for the Participating Institutions.
The Participating Institutions are The University of Chicago, Fermilab,
the Institute for Advanced Study, the Japan Participation Group,
The Johns Hopkins University, the Korean Scientist Group,
Los Alamos National Laboratory, the Max-Planck-Institute for Astronomy (MPIA),
the Max-Planck-Institute for Astrophysics (MPA), New Mexico State University,
University of Pittsburgh, University of Portsmouth, Princeton University,
the United States Naval Observatory, and the University of Washington.
This research made use of Montage, funded by the
National Aeronautics and Space Administration's Earth Science Technology Office,
Computational Technnologies Project, under Cooperative Agreement Number NCC5-626
between NASA and the California Institute of Technology.
The code is maintained by the NASA/IPAC Infrared Science Archive.
This research made use of python (\url{http://www.python.org}),
of Matplotlib \citep[][]{Hunter:2007}, a suite of open-source python modules
that provides a framework for creating scientific plots,
and Astropy, a community-developed core Python package for Astronomy
\citep[][]{2013A&A...558A..33A}.
This research made use of APLpy, an open-source plotting package for
Python hosted at \url{http://aplpy.github.com}.
\end{acknowledgements}
\bibliographystyle{00_CONFIG/bst/aa}
\bibliography{00_CONFIG/Bibliography/Bibliography.bib}
\appendix
%
\input{08_CC/CC}
%
\input{09_CMB/CMB}
%
\input{10_THRESHOLD/THRESHOLD}
%
%
\end{document}

%% file: 00_CONFIG/Layout/Colors.tex
\usepackage[usenames,dvipsnames]{xcolor}

%% file: 00_CONFIG/Layout/Graphics.tex
\usepackage[]{graphicx}
\usepackage[compatibility=false,style=base]{caption}
\usepackage[]{subcaption}
\usepackage{marvosym} 
\newcommand{\ImagePag} [4]  {\begin{figure*}[!htbp]
                             \centering
                             \includegraphics[width=#1\textwidth,keepaspectratio=true]{#2}
                             \caption{\label{#3}#4}
                             \end{figure*}
                             }
\newcommand{\ImageCol} [4]  {\begin{figure}[!htbp]%
                             \includegraphics[width=#1\textwidth,keepaspectratio=true]{#2}%
                             \caption{\label{#3} #4}
                             \end{figure}
                             }
\newcommand{\ImageBands} [9] {\begin{figure*}
                              \centering
                              #9
                                \begin{subfigure}[b]{0.38\textwidth}
                                  \includegraphics[width=\textwidth]{#1}
                                \end{subfigure}~
                                \begin{subfigure}[b]{0.38\textwidth}
                                  \includegraphics[width=\textwidth]{#2}
                                \end{subfigure}
                                \begin{subfigure}[b]{0.38\textwidth}
                                  \includegraphics[width=\textwidth]{#3}
                                \end{subfigure}~
                                \begin{subfigure}[b]{0.38\textwidth}
                                  \includegraphics[width=\textwidth]{#4}
                                \end{subfigure}
                                \begin{subfigure}[b]{0.38\textwidth}
                                  \includegraphics[width=\textwidth]{#5}
                                \end{subfigure}~
                                \begin{subfigure}[b]{0.38\textwidth}
                                  \includegraphics[width=\textwidth]{#6}
                                \end{subfigure}
                              \caption{\label{#7} #8}
                              \end{figure*}
                             }
\hbadness=\maxdimen
\vbadness=\maxdimen

%% file: 00_CONFIG/Layout/Fonts.tex
\usepackage[utf8]{inputenc}
\usepackage{txfonts}    
\usepackage{amsfonts}   
\newcommand{\Cenc}{T1}
\newcommand{\Cfont}{phv}

        \newcommand{\texthb}[1]{{\huge\usefont{\Cenc}{\Cfont}{b}{n}#1}}
        \newcommand{\textlr}[1]{{\Large\usefont{\Cenc}{\Cfont}{m}{n}#1}}
        
        \newcommand{\textlc}[1]{{\Large\usefont{\Cenc}{\Cfont}{m}{sc}#1}}

        \newcommand{\textnr}[1]{{\normalsize\usefont{\Cenc}{\Cfont}{m}{n}#1}}
        \newcommand{\textni}[1]{{\normalsize\usefont{\Cenc}{\Cfont}{m}{it}#1}}
        
        \newcommand{\textnb}[1]{{\normalsize\usefont{\Cenc}{\Cfont}{b}{n}#1}}

        \newcommand{\textmr}[1]{{\footnotesize\usefont{\Cenc}{\Cfont}{m}{n}#1}}

        \newcommand{\textbc}[1]{{\Large\usefont{\Cenc}{\Cfont}{b}{sc}#1}}
\usepackage{listings}

%% file: 00_CONFIG/Layout/Abbreviations.tex
\newcommand{\Tab}   [1]  {\ifx&#1&%
                            {Tab.}%
                          \else
                            {Tab.~\ref{#1}}%
                          \fi}
\newcommand{\Fig}   [1]  {\ifx&#1&%
                            {Fig.}%
                          \else
                            {Fig.~\ref{#1}}%
                          \fi}
\newcommand{\Sec}   [1]  {\ifx&#1&%
                            {Sect.}%
                          \else
                            {Sect.~\ref{#1}}%
                          \fi}
\newcommand{\Eq}    [1]  {\ifx&#1&%
                            {Eq.}%
                          \else
                            {Eq.~\ref{#1}}%
                          \fi}
\newcommand{\App}   [1]  {\ifx&#1&%
                            {Appendix}%
                          \else
                            {Appendix~\ref{#1}}%
                          \fi}

%% file: 00_CONFIG/Layout/References.tex
\usepackage{natbib}
    \bibpunct{(}{)}{;}{a}{}{,}
\usepackage[]{hyperref}
    \hypersetup{draft=false,
                hidelinks,
                colorlinks=true,
                linkcolor=Black, 
                citecolor=Black, 
                filecolor=Black, 
                urlcolor=RoyalBlue}

%% file: 00_CONFIG/Astro/Astro.tex
\newcommand{\Vlim}    [3]   {\ifmmode{\rm{#2\,^{\wedge\,#3}_{\vee\,#1}}}\else$\rm{#2\,^{\wedge\,#3}_{\vee\,#1}}$\fi}
\newcommand{\NGC}     [1]  {{NGC\,#1}}
\newcommand{\MESSIER} [1]  {{M#1}}

\newcommand{\GALEX}       [1]   {\ifx&#1&%
                                 {\textit{GALEX}}%
                                 \else
                                 {\textit{GALEX}\,#1}%
                                 \fi}
\newcommand{\SLOAN}       [1]   {\ifx&#1&%
                                 {\textit{Sloan}}%
                                 \else
                                 {\textit{Sloan\,#1}-band}%
                                 \fi}
\newcommand{\SPITZER}     [1]   {\ifx&#1&%
                                 {\textit{Spitzer}}%
                                 \else
                                 {\textit{Spitzer}\,#1\,$\mu$m}%
                                 \fi}
\newcommand{\HERSCHEL}    [1]   {\ifx&#1&%
                                 {\textit{Herschel}}%
                                 \else
                                 {\textit{Herschel}\,#1\,$\mu$m}%
                                 \fi}
\newcommand{\APEX}        [1]   {\ifx&#1&%
                                 {\textit{APEX}}%
                                 \else
                                 {\textit{APEX}\,#1\,$\mu$m}%
                                 \fi}
\newcommand{\PLANCK}      [1]   {\ifx&#1&%
                                 {\textit{Planck}}%
                                 \else
                                 {\textit{Planck}\,#1\,GHz}%
                                 \fi}
\newcommand{\NIKA}        [1]   {\ifx&#1&%
                                 {\textit{NIKA}}%
                                 \else
                                 {\textit{NIKA}\,#1\,mm}%
                                 \fi}
\newcommand{\EFFELSBERG}  [1]   {\ifx&#1&%
                                 {\textit{Effelsberg}}%
                                 \else
                                 {\textit{Effelsberg}\,#1\,GHz}%
                                 \fi}
\newcommand{\VLA}         [1]   {\ifx&#1&%
                                 {\textit{VLA}}%
                                 \else
                                 {\textit{VLA}\,#1\,GHz}%
                                 \fi}
\newcommand{\CAMERAmu}    [2]   {\ifx&#2&%
                                  {\textup{#1}}%
                                 \else
                                  {\textup{#1}\,#2\,$\mu$m}%
                                 \fi}
\newcommand{\CAMERAmm}    [2]   {\ifx&#2&%
                                  {\textup{#1}}%
                                 \else
                                  {\textup{#1}\,#2\,mm}%
                                 \fi}
\newcommand{\CAMERAcm}    [2]   {\ifx&#2&%
                                  {\textup{#1}}%
                                 \else
                                  {\textup{#1}\,#2\,cm}%
                                 \fi}
\newcommand{\CAMERAghz}   [2]   {\ifx&#2&%
                                  {\textup{#1}}%
                                 \else
                                  {\textup{#1}\,#2\,GHz}%
                                 \fi}
\newcommand{\dex}        [1]   {\ifx&#1&
                                   \ifmmode\rm{dex}\else$\rm{dex}$\fi
                                 \else
                                   \ifmmode\rm{#1\,dex}\else$\rm{#1\,dex}$\fi
                                 \fi}
\newcommand{\percent}    [1]   {\ifx&#1&
                                   \ifmmode\rm{\%}\else$\rm{\%}$\fi
                                 \else
                                   \ifmmode\rm{#1\%}\else$\rm{#1\%}$\fi
                                 \fi}
\newcommand{\km}         [1]   {\ifx&#1&
                                   \ifmmode\rm{km}\else$\rm{km}$\fi
                                 \else
                                   \ifmmode\rm{#1\,km}\else$\rm{#1\,km}$\fi
                                 \fi}
\newcommand{\meters}     [1]   {\ifx&#1&
                                   \ifmmode\rm{m}\else$\rm{m}$\fi
                                 \else
                                   \ifmmode\rm{#1\,m}\else$\rm{#1\,m}$\fi
                                 \fi}
\newcommand{\cm}         [1]   {\ifx&#1&
                                   \ifmmode\rm{cm}\else$\rm{cm}$\fi
                                 \else
                                   \ifmmode\rm{#1\,cm}\else$\rm{#1\,cm}$\fi
                                 \fi}
\newcommand{\mm}         [1]   {\ifx&#1&
                                   \ifmmode\rm{mm}\else$\rm{mm}$\fi
                                 \else
                                   \ifmmode\rm{#1\,mm}\else$\rm{#1\,mm}$\fi
                                 \fi}
\newcommand{\microns}    [1]   {\ifx&#1&
                                   \ifmmode{\rm{\mu m}}\else$\rm{\mu m}$\fi
                                 \else
                                   \ifmmode{#1\,\rm{\mu m}}\else$#1\,\rm{\mu m}$\fi
                                 \fi}
\newcommand{\nm}         [1]   {\ifx&#1&
                                   \ifmmode\rm{nm}\else$\rm{nm}$\fi
                                 \else
                                   \ifmmode\rm{#1\,nm}\else$\rm{#1\,nm}$\fi
                                 \fi}
\newcommand{\angstroms}  [1]   {\ifx&#1&
                                   \ifmmode{\mbox{\normalfont\AA}}\else$\mbox{\normalfont\AA}$\fi
                                 \else
                                   \ifmmode{#1\,\mbox{\normalfont\AA}}\else$#1\,\mbox{\normalfont\AA}$\fi
                                 \fi}
\newcommand{\GHz}        [1]   {\ifx&#1&
                                   \ifmmode\rm{GHz}\else$\rm{GHz}$\fi
                                 \else
                                   \ifmmode\rm{#1\,GHz}\else$\rm{#1\,GHz}$\fi
                                 \fi}
\newcommand{\pc}         [1]   {\ifmmode\rm{#1\,pc}\else$\rm{#1\,pc}$\fi}
\newcommand{\kpc}        [1]   {\ifmmode\rm{#1\,kpc}\else$\rm{#1\,kpc}$\fi}
\newcommand{\Mpc}        [1]   {\ifmmode\rm{#1\,Mpc}\else$\rm{#1\,Mpc}$\fi}
\newcommand{\seconds}    [1]    {\ifx&#1&
                                   \ifmmode{\rm{s}}\else${\rm{s}}$\fi
                                 \else
                                   \ifmmode{#1\,\rm{s}}\else${#1\,\rm{s}}$\fi
                                 \fi}
\newcommand{\degrees}    [1]   {\ifmmode\rm{#1^{\circ}}\else$\rm{#1^{\circ}}$\fi}
\newcommand{\arcseconds} [2]    {\ifx&#2&%
                                   \ifmmode\rm{#1\arcsec}\else$\rm{#1\arcsec}$\fi
                                \else
                                   \ifmmode\rm{#1\,\farcs#2}\else$\rm{#1\,\farcs#2}$\fi
                                \fi}
\newcommand{\arcminutes} [2]    {\ifx&#2&%
                                   \ifmmode\rm{#1\arcmin}\else$\rm{#1\arcmin}$\fi
                                \else
                                   \ifmmode\rm{#1\farcm#2}\else$\rm{#1\farcm#2}$\fi
                                \fi}
\newcommand{\Msun}       [2]   {\ifx&#1&%
                                   \ifx&#2&%
                                     \ifmmode{\rm{M_\odot}}\else$\rm{M_\odot}$\fi
                                   \else
                                     \ifmmode{10^#2\,\rm{M_\odot}}\else$10^#2\,\rm{M_\odot}$\fi
                                   \fi
                                \else
                                   \ifx&#2&%
                                     \ifmmode{\rm{#1}\,\rm{M_\odot}}\else$\rm{#1}\,\rm{M_\odot}$\fi
                                   \else
                                     \ifmmode{\rm{#1}{\times}10^#2\,\rm{M_\odot}}\else$\rm{#1}{\times}10^#2\,\rm{M_\odot}$\fi
                                   \fi
                                \fi}
\newcommand{\Lsun}       [2]   {\ifx&#1&%
                                   \ifx&#2&%
                                     \ifmmode{\rm{L_\odot}}\else$\rm{L_\odot}$\fi
                                   \else
                                     \ifmmode{10^#2\,\rm{L_\odot}}\else$10^#2\,\rm{L_\odot}$\fi
                                   \fi
                                \else
                                   \ifx&#2&%
                                     \ifmmode{\rm{#1}\,\rm{L_\odot}}\else$\rm{#1}\,\rm{L_\odot}$\fi
                                   \else
                                     \ifmmode{\rm{#1}{\times}10^#2\,\rm{L_\odot}}\else$\rm{#1}{\times}10^#2\,\rm{L_\odot}$\fi
                                   \fi
                                \fi}
\newcommand{\Zsun}       [1]   {\ifx&#1&%
                                   \ifmmode{\rm{Z_\odot}}\else$\rm{Z_\odot}$\fi
                                 \else
                                   \ifmmode{\rm{#1}\,\rm{Z_\odot}}\else$\rm{#1}\,\rm{Z_\odot}$\fi
                                 \fi}
\newcommand{\yr}         [1]   {\ifx&#1&%
                                   \ifmmode{\rm{yr}}\else$\rm{yr}$\fi
                                 \else
                                   \ifmmode{\rm{#1}\,\rm{yr}}\else$\rm{#1}\,\rm{yr}$\fi
                                 \fi}
\newcommand{\Myr}        [1]   {\ifx&#1&%
                                   \ifmmode{\rm{Myr}}\else$\rm{Myr}$\fi
                                 \else
                                   \ifmmode{\rm{#1}\,\rm{Myr}}\else$\rm{#1}\,\rm{Myr}$\fi
                                 \fi}
\newcommand{\Gyr}        [1]   {\ifx&#1&%
                                   \ifmmode{\rm{Gyr}}\else$\rm{Gyr}$\fi
                                 \else
                                   \ifmmode{\rm{#1}\,\rm{Gyr}}\else$\rm{#1}\,\rm{Gyr}$\fi
                                 \fi}
\newcommand{\Kelvin}     [1]   {\ifx&#1&%
                                   \ifmmode{\rm{K}}\else$\rm{K}$\fi
                                 \else
                                   \ifmmode{\rm{#1}\,\rm{K}}\else$\rm{#1}\,\rm{K}$\fi
                                 \fi}
\newcommand{\Temp}       [2]   {\ifx&#1&%
                                   \ifx&#2&%
                                     \ifmmode{T}\else$T$\fi
                                   \else
                                     \Kelvin{#2}%
                                   \fi
                                \else
                                   \ifx&#2&%
                                     \ifmmode{T{\rm #1}}\else$T{\rm #1}$\fi
                                   \else
                                     \ifmmode{T{\rm #1}\Kelvin{#2}}\else$T{\rm #1}\Kelvin{#2}$\fi
                                   \fi
                                \fi}
\newcommand{\density}    [1]   {\ifmmode\rm{#1\,cm^{-3}}\else$\rm{#1\,cm^{-3}}$\fi}
\newcommand{\NEFD}       [1]   {\ifmmode\rm{#1\,mJy\sqrt{s}}\else$\rm{#1\,mJy\sqrt{s}}$\fi}
\newcommand{\scanspeed}  [1]   {\ifmmode\rm{#1^{\prime\prime}/s}\else$#1^{\prime\prime}/s$\fi}
\newcommand{\pwv}        [1]   {\ifmmode\rm{#1\,mm}\else$\rm{#1\,mm}$\fi}
\newcommand{\Jy}         [1]   {\ifx&#1&
                                   \ifmmode{\rm{Jy}}\else$\rm{Jy}$\fi
                                 \else
                                   \ifmmode{#1\,\rm{Jy}}\else$#1\,\rm{Jy}$\fi
                                 \fi}
\newcommand{\mJy}        [1]   {\ifx&#1&
                                   \ifmmode{\rm{mJy}}\else$\rm{mJy}$\fi
                                 \else
                                   \ifmmode{#1\,\rm{mJy}}\else$#1\,\rm{mJy}$\fi
                                 \fi}
\newcommand{\magnitude}  [1]   {\ifmmode\rm{#1\,mag}\else$\rm{#1\,mag}$\fi}
\newcommand{\kms}        [1]   {\ifx&#1&
                                   \ifmmode\rm{km\,s^{-1}}\else$\rm{km\,s^{-1}}$\fi
                                 \else
                                   \ifmmode\rm{#1\,km\,s^{-1}}\else$\rm{#1\,km\,s^{-1}}$\fi
                                 \fi}
\newcommand{\JyK}        [1]   {\ifx&#1&
                                   \ifmmode\rm{Jy\,K^{-1}}\else$\rm{Jy\,K^{-1}}$\fi
                                 \else
                                   \ifmmode\rm{#1\,Jy\,K^{-1}}\else$\rm{#1\,Jy\,K^{-1}}$\fi
                                 \fi}
\newcommand{\PLAW}       [1]   {\ifx&#1&%
                                   \ifmmode{S_{\rm \nu}\propto\nu^{\,\alpha}}\else$S_{\rm \nu}\propto\nu^{\,\alpha}$\fi
                                \else
                                   \ifmmode{S_{\rm \nu}\propto\nu^{\,#1}}\else$S_{\rm \nu}\propto\nu^{\,#1}$\fi
                                \fi}
\newcommand{\BB}         [1]   {\ifx&#1&%
                                   \ifmmode{B_{\rm \nu}(T)}\else$B_{\rm \nu}(T)$\fi
                                \else
                                   \ifmmode{B_{\rm \nu}(#1\,K)}\else$B_{\rm \nu}(#1\,K)$\fi
                                \fi}
\newcommand{\MBB}        [2]   {\ifx&#1&%
                                   \ifx&#2&%
                                     \ifmmode{S_{\rm \nu}\propto B_{\rm \nu}(T)\,\nu^{\,\beta}}\else$S_{\rm \nu}\propto B_{\rm \nu}(T)\,\nu^{\,\beta}$\fi
                                   \else
                                     \ifmmode{S_{\rm \nu}\propto B_{\rm \nu}(T)\,\nu^{\,#2}}\else$S_{\rm \nu}\propto B_{\rm \nu}(T)\,\nu^{\,#2}$\fi
                                   \fi
                                \else
                                   \ifx&#2&%
                                     \ifmmode{S_{\rm \nu}\propto B_{\rm \nu}(#1\,K)\,\nu^{\,\beta}}\else$S_{\rm \nu}\propto B_{\rm \nu}(#1\,K)\,\nu^{\,\beta}$\fi
                                   \else
                                     \ifmmode{S_{\rm \nu}\propto B_{\rm \nu}(#1\,K)\,\nu^{\,#2}}\else$S_{\rm \nu}\propto B_{\rm \nu}(#1\,K)\,\nu^{\,#2}$\fi
                                   \fi
                                \fi}
\newcommand{\Sobs}       [3]   {\ifx&#2&%
                                   \ifx&#3&%
                                     \ifmmode{S_{\rm obs}^{\rm #1}}\else$S_{\rm obs}^{\rm #1}$\fi
                                   \else
                                     \ifmmode{\Jy{#3}}\else$\Jy{#3}$\fi
                                   \fi
                                \else
                                   \ifx&#3&%
                                     \ifmmode{S_{\rm obs}^{\rm #1}{\rm #2}}\else$S_{\rm obs}^{\rm #1}{\rm #2}$\fi
                                   \else
                                     \ifmmode{S_{\rm obs}^{\rm #1}{\rm #2}\Jy{#3}}\else$S_{\rm obs}^{\rm #1}{\rm #2}\Jy{#3}$\fi
                                   \fi
                                \fi}
\newcommand{\Snu}        [2]   {\ifx&#1&%
                                   \ifx&#2&%
                                     \ifmmode{S_{\rm \nu}}\else$S_{\rm \nu}$\fi
                                   \else
                                     \ifmmode{#2\,{\rm Jy}}\else$#2\,{\rm Jy}$\fi
                                   \fi
                                \else
                                   \ifx&#2&%
                                     \ifmmode{S_{\rm \nu}{\rm #1}}\else$S_{\rm \nu}{\rm #1}$\fi
                                   \else
                                     \ifmmode{S_{\rm \nu}{\rm #1}#2\,{\rm Jy}}\else$S_{\rm \nu}{\rm #1}#2\,{\rm Jy}$\fi
                                   \fi
                                \fi}
\newcommand{\SurBright}  [2]   {\ifx&#1&%
                                   \ifx&#2&%
                                     \ifmmode{S_{\rm{0}}}\else${S_{\rm{0}}}$\fi
                                   \else
                                     \ifmmode{#2\,\rm{erg/s/kpc^2}}\else${#2\,\rm{erg\,s^{-1}\,kpc^{-2}}}$\fi
                                   \fi
                                \else
                                   \ifx&#2&%
                                     \ifmmode{S_{\rm{0}}\rm{#1}}\else${S_{\rm{0}}\rm{#1}}$\fi
                                   \else
                                     \ifmmode{S_{\rm{0}}\rm{#1}#2\,\rm{erg\,s^{-1}\,kpc^{-2}}}\else${S_{\rm{0}}\rm{#1}#2\,\rm{erg\,s^{-1}\,kpc^{-22}}}$\fi
                                   \fi
                                \fi}
\newcommand{\ColDensity} [2]   {\ifx&#1&%
                                   \ifx&#2&%
                                     \ifmmode{N_{\rm HI}^{\rm PDR}}\else$N_{\rm HI}^{\rm PDR}$\fi
                                   \else
                                     \ifmmode{#2\,{\rm cm^{-2}}}\else$#2\,{\rm cm^{-2}}$\fi
                                   \fi
                                \else
                                   \ifx&#2&%
                                     \ifmmode{N_{\rm HI}^{\rm PDR}{\rm #1}}\else$N_{\rm HI}^{\rm PDR}{\rm #1}$\fi
                                   \else
                                     \ifmmode{N_{\rm HI}^{\rm PDR}{\rm #1}#2\,{\rm cm^{-2}}}\else$N_{\rm HI}^{\rm PDR}{\rm #1}#2\,{\rm cm^{-2}}$\fi
                                   \fi
                                \fi}
\newcommand{\AGE}        [2]   {\ifx&#1&%
                                   \ifx&#2&%
                                     \ifmmode\mathcal{T}\else$\mathcal{T}$\fi
                                   \else
                                     \Myr{#2}%
                                   \fi
                                \else
                                   \ifx&#2&%
                                     \ifmmode\mathcal{T}\rm{#1}\else$\mathcal{T}\rm{#1}$\fi
                                   \else
                                     \ifmmode\mathcal{T}\rm{#1}\Myr{#2}\else$\mathcal{T}\rm{#1}$\Myr{#2}\fi
                                   \fi
                                \fi}
\newcommand{\MET}        [2]   {\ifx&#1&%
                                   \ifx&#2&%
                                     \ifmmode\rm{Z}\else$\rm{Z}$\fi
                                   \else
                                     \Zsun{#2}%
                                   \fi
                                \else
                                   \ifx&#2&%
                                     \ifmmode\rm{Z}\rm{#1}\else$\rm{Z}\rm{#1}$\fi
                                   \else
                                     \ifmmode\rm{Z}\rm{#1}\Zsun{#2}\else$\rm{Z}\rm{#1}$\Zsun{#2}\fi
                                   \fi
                                \fi}
\newcommand{\PRS}        [2]   {\ifx&#1&%
                                   \ifx&#2&%
                                     \ifmmode{p_{\rm 0}}\else$p_{\rm 0}$\fi
                                   \else
                                     \ifmmode{#2}\else$#2$\fi
                                   \fi
                                \else
                                   \ifx&#2&%
                                     \ifmmode{p_{\rm 0}{\rm #1}}\else$p_{\rm 0}{\rm #1}$\fi
                                   \else
                                     \ifmmode{p_{\rm 0}{\rm #1}#2}\else$p_{\rm 0}{\rm #1}#2$\fi
                                   \fi
                                \fi}
\newcommand{\CMP}        [2]   {\ifx&#1&%
                                   \ifx&#2&%
                                     \ifmmode\mathcal{C}\else$\mathcal{C}$\fi
                                   \else
                                     #2%
                                   \fi
                                \else
                                   \ifx&#2&%
                                     \ifmmode\mathcal{C}\rm{#1}\else$\mathcal{C}\rm{#1}$\fi
                                   \else
                                     \ifmmode\mathcal{C}\rm{#1}#2\else$\mathcal{C}\rm{#1}$#2\fi
                                   \fi
                                \fi}
\newcommand{\COV}        [2]   {\ifx&#1&%
                                   \ifx&#2&%
                                     \ifmmode{f_{\rm cov}}\else$f_{\rm cov}$\fi
                                   \else
                                     \ifmmode{#2}\else$#2$\fi
                                   \fi
                                \else
                                   \ifx&#2&%
                                     \ifmmode{f_{\rm cov}\rm{#1}}\else$f_{\rm cov}\rm{#1}$\fi
                                   \else
                                     \ifmmode{f_{\rm cov}\rm{#1}#2}\else$f_{\rm cov}\rm{#1}#2$\fi
                                   \fi
                                \fi}
\newcommand{\ESC}        [2]   {\ifx&#1&%
                                   \ifx&#2&%
                                     \ifmmode{f_{\rm esc}}\else$f_{\rm esc}$\fi
                                   \else
                                     \ifmmode{#2}\else$#2$\fi
                                   \fi
                                \else
                                   \ifx&#2&%
                                     \ifmmode{f_{\rm esc}{\rm #1}}\else$f_{\rm esc}{\rm #1}$\fi
                                   \else
                                     \ifmmode{f_{\rm esc}{\rm #1}#2}\else$f_{\rm esc}{\rm #1}#2$\fi
                                   \fi
                                \fi}
\newcommand{\CLUMP}      [2]   {\ifx&#1&%
                                   \ifx&#2&%
                                     \ifmmode{F}\else$F$\fi
                                   \else
                                     \ifmmode{#2}\else$#2$\fi
                                   \fi
                                \else
                                   \ifx&#2&%
                                     \ifmmode{F{\rm #1}}\else$F{\rm #1}$\fi
                                   \else
                                     \ifmmode{F{\rm #1}#2}\else$F{\rm #1}#2$\fi
                                   \fi
                                \fi}
\newcommand{\TAUb}       [2]   {\ifx&#1&%
                                   \ifx&#2&%
                                     \ifmmode{\tau_{\rm B}^{\rm f}}\else$\tau_{\rm B}^{\rm f}$\fi
                                   \else
                                     \ifmmode\rm{#2}\else$\rm{#2}$\fi
                                   \fi
                                \else
                                   \ifx&#2&%
                                     \ifmmode{\tau_{\rm B}^{\rm f}{\rm #1}}\else$\tau_{\rm B}^{\rm f}{\rm #1}$\fi
                                   \else
                                     \ifmmode{\tau_{\rm B}^{\rm f}{\rm #1}#2}\else$\tau_{\rm B}^{\rm f}{\rm #1}#2$\fi
                                   \fi
                                \fi}
\newcommand{\SFR}        [2]   {\ifx&#1&%
                                   \ifx&#2&%
                                     \ifmmode{SFR}\else$SFR$\fi
                                   \else
                                     \ifmmode{#2\,{\rm M_{\odot}\,yr^{-1}}}\else$#2\,{\rm M_{\odot}\,yr^{-1}}$\fi
                                   \fi
                                \else
                                   \ifx&#2&%
                                     \ifmmode{SFR{\rm #1}}\else$SFR{\rm #1}$\fi
                                   \else
                                     \ifmmode{SFR{\rm #1}#2\,{\rm M_{\odot}\,yr^{-1}}}\else$SFR{\rm #1}#2\,{\rm M_{\odot}\,yr^{-1}}$\fi
                                   \fi
                                \fi}
\newcommand{\SFRp}       [2]   {\ifx&#1&%
                                   \ifx&#2&%
                                     \ifmmode{SFR^{\,\prime}}\else$SFR^{\,\prime}$\fi
                                   \else
                                     \ifmmode{#2\,{\rm M_{\odot}\,yr^{-1}}}\else$#2\,{\rm M_{\odot}\,yr^{-1}}$\fi
                                   \fi
                                \else
                                   \ifx&#2&%
                                     \ifmmode{SFR^{\,\prime}\rm{#1}}\else$SFR^{\,\prime}\rm{#1}$\fi
                                   \else
                                     \ifmmode{SFR^{\,\prime}\rm{#1}#2\,{\rm M_{\odot}\,yr^{-1}}}\else$SFR^{\,\prime}\rm{#1}#2\,{\rm M_{\odot}\,yr^{-1}}$\fi
                                   \fi
                                \fi}
\newcommand{\SFRmod}     [2]   {\ifx&#1&%
                                   \ifx&#2&%
                                     \ifmmode{SFR^{\rm\,mod}}\else$SFR^{\rm\,mod}$\fi
                                   \else
                                     \ifmmode{#2\,{\rm M_{\odot}\,yr^{-1}}}\else$#2\,{\rm M_{\odot}\,yr^{-1}}$\fi
                                   \fi
                                \else
                                   \ifx&#2&%
                                     \ifmmode{SFR^{\rm\,mod}\rm{#1}}\else$SFR^{\rm\,mod}\rm{#1}$\fi
                                   \else
                                     \ifmmode{SFR^{\rm\,mod}\rm{#1}#2\,{\rm M_{\odot}\,yr^{-1}}}\else$SFR^{\rm\,mod}\rm{#1}#2\,{\rm M_{\odot}\,yr^{-1}}$\fi
                                   \fi
                                \fi}
\newcommand{\OLD}        [2]   {\ifx&#1&%
                                   \ifx&#2&%
                                     \ifmmode{old}\else$old$\fi
                                   \else
                                     \ifmmode{#2}\else$#2$\fi
                                   \fi
                                \else
                                   \ifx&#2&%
                                     \ifmmode{old\rm{#1}}\else$old\rm{#1}$\fi
                                   \else
                                     \ifmmode{old\rm{#1}#2}\else$old\rm{#1}#2$\fi
                                   \fi
                                \fi}
\newcommand{\OLDmod}     [2]   {\ifx&#1&%
                                   \ifx&#2&%
                                     \ifmmode{old^{\rm\,mod}}\else$old^{\rm\,mod}$\fi
                                   \else
                                     \ifmmode{#2}\else$#2$\fi
                                   \fi
                                \else
                                   \ifx&#2&%
                                     \ifmmode{old^{\rm\,mod}\rm{#1}}\else$old^{\rm\,mod}\rm{#1}$\fi
                                   \else
                                     \ifmmode{old^{\rm\,mod}\rm{#1}#2}\else$old^{\rm\,mod}\rm{#1}#2$\fi
                                   \fi
                                \fi}
\newcommand{\BD}         [2]   {\ifx&#1&%
                                   \ifx&#2&%
                                     \ifmmode{B/D}\else$B/D$\fi
                                   \else
                                     \ifmmode{#2}\else$#2$\fi
                                   \fi
                                \else
                                   \ifx&#2&%
                                     \ifmmode{B/D{\rm #1}}\else$B/D{\rm #1}$\fi
                                   \else
                                     \ifmmode{B/D{\rm #1}#2}\else$B/D{\rm #1}#2$\fi
                                   \fi
                                \fi}
\newcommand{\INCL}       [2]   {\ifx&#1&%
                                   \ifx&#2&%
                                     \ifmmode{i}\else$i$\fi
                                   \else
                                     \degrees{#2}%
                                   \fi
                                \else
                                   \ifx&#2&%
                                     \ifmmode{i{\rm #1}}\else$i{\rm #1}$\fi
                                   \else
                                     \ifmmode{i{\rm #1}\degrees{#2}}\else$i{\rm #1}\degrees{#2}$\fi
                                   \fi
                                \fi}
\newcommand{\Mcl}        [2]   {\ifx&#1&%
                                   \ifx&#2&%
                                     \ifmmode{M_{\rm cl}}\else$M_{\rm cl}$\fi
                                   \else
                                     \ifmmode{#2\,{\rm M_\odot}}\else$#2\,{\rm M_\odot}$\fi
                                   \fi
                                \else
                                   \ifx&#2&%
                                     \ifmmode{M_{\rm cl}{\rm #1}}\else$M_{\rm cl}{\rm #1}$\fi
                                   \else
                                     \ifmmode{M_{\rm cl}{\rm #1}\Msun{#2}}\else$M_{\rm cl}{\rm #1}\Msun{#2}$\fi
                                   \fi
                                \fi}
\newcommand{\hs}         [2]   {\ifx&#1&%
                                   \ifx&#2&%
                                     \ifmmode{h_{\rm s}}\else$h_{\rm s}$\fi
                                   \else
                                     \pc{#2}%
                                   \fi
                                \else
                                   \ifx&#2&%
                                     \ifmmode{h_{\rm s}{\rm #1}}\else$h_{\rm s}{\rm #1}$\fi
                                   \else
                                     \ifmmode{h_{\rm s}{\rm #1}\pc{#2}}\else$h_{\rm s}{\rm #1}\pc{#2}$\fi
                                   \fi
                                \fi}
\newcommand{\hr}         [2]   {\ifx&#1&%
                                   \ifx&#2&%
                                     \ifmmode{h_{\rm ref}}\else$h_{\rm ref}$\fi
                                   \else
                                     \pc{#2}%
                                   \fi
                                \else
                                   \ifx&#2&%
                                     \ifmmode{h_{\rm ref}{\rm #1}}\else$h_{\rm ref}{\rm #1}$\fi
                                   \else
                                     \ifmmode{h_{\rm ref}\rm{#1}\pc{#2}}\else$h_{\rm ref}\rm{#1}\pc{#2}$\fi
                                   \fi
                                \fi}
\newcommand{\hd}         [2]   {\ifx&#1&%
                                   \ifx&#2&%
                                     \ifmmode{h_{\rm d}}\else$h_{\rm d}$\fi
                                   \else
                                     \pc{#2}%
                                   \fi
                                \else
                                   \ifx&#2&%
                                     \ifmmode{h_{\rm d}{\rm #1}}\else$h_{\rm d}{\rm #1}$\fi
                                   \else
                                     \ifmmode{h_{\rm d}{\rm #1}\pc{#2}}\else$h_{\rm d}{\rm #1}\pc{#2}$\fi
                                   \fi
                                \fi}
\newcommand{\Catten}     [2]   {\ifx&#1&%
                                   \ifx&#2&%
                                     \ifmmode{\Delta m_{\lambda}}\else$\Delta m_{\lambda}$\fi
                                   \else
                                     \ifmmode{#2}\else$#2$\fi
                                   \fi
                                \else
                                   \ifx&#2&%
                                     \ifmmode{\Delta m_{\lambda}{\rm #1}}\else$\Delta m_{\lambda}{\rm #1}$\fi
                                   \else
                                     \ifmmode{\Delta m_{\lambda}{\rm #1}#2}\else$\Delta m_{\lambda}{\rm #1}#2$\fi
                                   \fi
                                \fi}
\newcommand{\Eindex}     [2]   {\ifx&#1&%
                                   \ifx&#2&%
                                     \ifmmode\beta\else$\beta$\fi
                                   \else
                                     #2%
                                   \fi
                                \else
                                   \ifx&#2&%
                                     \ifmmode\beta\rm{#1}\else$\beta\rm{#1}$\fi
                                   \else
                                     \ifmmode\beta\rm{{#1}#2}\else$\beta\rm{{#1}#2}$\fi
                                   \fi
                                \fi}
\newcommand{\chiSq}      [2]   {\ifx&#1&%
                                   \ifx&#2&%
                                     \ifmmode{\chi^2}\else$\chi^2$\fi
                                   \else
                                     #2%
                                   \fi
                                \else
                                   \ifx&#2&%
                                     \ifmmode{\chi^2\rm{#1}}\else$\chi^2\rm{#1}$\fi
                                   \else
                                     \ifmmode{\chi^2\rm{#1}#2}\else$\chi^2\rm{#1}#2$\fi
                                   \fi
                                \fi}
\newcommand {\FITS}         {\texttt{FITS}}
\newcommand {\HEALPIX}      {\texttt{HEALPix}}
\newcommand {\SMICA}        {\texttt{SMICA}}
\newcommand {\SEXTRACTOR}   {\texttt{SExtractor}}
\newcommand {\IRAF}         {\texttt{IRAF}}
\newcommand {\WREGISTER}    {\texttt{wregister}}
\newcommand {\MONTAGE}      {\texttt{Montage}}
\newcommand {\IDL}          {\texttt{IDL}}
\newcommand {\GNOMVIEW}     {\texttt{Gnomview}}
\newcommand {\HIPE}         {\texttt{HIPE}}
\newcommand {\SCANAMORPHOS} {\texttt{Scanamorphos}}
\newcommand {\BOA}          {\texttt{BoA}}
\newcommand {\MOPEX}        {\texttt{MOPEX}}
\newcommand {\COOKBOOK} [2] {\ifx&#1&%
                               \footnote{\url{#2}}%
                             \else
                               \ifx&#2&%
                                     \textit{#1}%
                                   \else
                                     \textit{#1}\footnote{\url{#2}}%
                                   \fi
                                \fi}
\newcommand {\HERMES}   {\texttt{HerM33es}}

\newcommand{\Gdust}      [2]   {\ifx&#1&%
                                   \ifx&#2&%
                                     \ifmmode{G_{\rm dust}}\else${G_{\rm dust}}$\fi
                                   \else
                                     \ifmmode{#2}\else$#2$\fi
                                   \fi
                                \else
                                   \ifx&#2&%
                                     \ifmmode{G_{\rm dust}{\rm #1}}\else${G_{\rm dust}{\rm #1}}$\fi
                                   \else
                                     \ifmmode{G_{\rm dust}{\rm #1}#2}\else${G_{\rm dust}{\rm #1}#2}$\fi
                                   \fi
                                \fi}
\newcommand{\CO}         [2]   {\ifx&#1#2&%
                                  \ifmmode{\rm CO}\else$\rm CO$\fi
                                \else
                                  \ifmmode{\rm CO(#1{\rightarrow}#2)}\else$\rm CO(#1{\rightarrow}#2)$\fi
                                \fi}
\newcommand{\Sco}        [3]   {\ifx&#1#2&%
                                   \ifx&#3&%
                                     \ifmmode{S_{\rm CO}}\else$S_{\rm CO}$\fi
                                   \else
                                     \ifmmode{#3\,\Jy{}\,\kms{}}\else$#3\,\Jy{}\,\kms{}$\fi
                                   \fi
                                \else
                                   \ifx&#3&%
                                     \ifmmode{S_{\rm CO(#1{\rightarrow}#2)}}\else$S_{\rm CO(#1{\rightarrow}#2)}$\fi
                                   \else
                                     \ifmmode{S_{\rm CO(#1{\rightarrow}#2)}{#3\,\Jy{}\,\kms{}}}\else$S_{\rm CO(#1{\rightarrow}#2)}{#3\,\Jy{}\,\kms{}}$\fi
                                   \fi
                                \fi}
\newcommand{\COtoH}      [2]   {\ifx&#1&%
                                   \ifx&#2&%
                                     \ifmmode{N{\rm (H_{2})}/I_{\rm CO(1{\rightarrow}0)}}\else$N{\rm (H_{2})}/I_{\rm CO(1{\rightarrow}0)}$\fi
                                   \else
                                     \ifmmode\rm{#2\times10^{20}\,cm^{-2}/(K\,km\,s^{-1})}\else$\rm{#2\times10^{20}\,cm^{-2}/(K\,km\,s^{-1})}$\fi
                                   \fi
                                \else
                                   \ifx&#2&%
                                     \ifmmode\rm{N(H_2)/I_{CO(1{\rightarrow}0)}#1}\else$\rm{N(H_2)/I_{CO(1{\rightarrow}0)}#1}$\fi
                                   \else
                                     \ifmmode\rm{N(H_2)/I_{CO(1{\rightarrow}0)}#1#2\times10^{20}\,cm^{-2}/(K\,km\,s^{-1})}\else$\rm{N(H_2)/I_{CO(1{\rightarrow}0)}#1#2\times10^{20}\,cm^{-2}/(K\,km\,s^{-1})}$\fi
                                   \fi
                                \fi}
\newcommand{\Xco}        [2]   {\ifx&#1&%
                                   \ifx&#2&%
                                     \ifmmode{X_{\rm CO}}\else$X_{\rm CO}$\fi
                                   \else
                                     \ifmmode{#2\times10^{20}{\rm\,cm^{-2}/(K\,km\,s^{-1})}}\else$#2\times10^{20}{\rm\,cm^{-2}/(K\,km\,s^{-1})}$\fi
                                   \fi
                                \else
                                   \ifx&#2&%
                                     \ifmmode{X_{\rm CO}{\rm #1}}\else$X_{\rm CO}{\rm #1}$\fi
                                   \else
                                     \ifmmode{X_{\rm CO}{\rm #1}#2\times10^{20}{\rm\,cm^{-2}/(K\,km\,s^{-1})}}\else$X_{\rm CO}{\rm #1}#2\times10^{20}{\rm\,cm^{-2}/(K\,km\,s^{-1})}$\fi
                                   \fi
                                \fi}
\newcommand{\HII}          {H\,{\scshape ii}}
\newcommand{\HI}           {H\,{\scshape i}}
\newcommand{\multipole}  [2]   {\ifx&#1&%
                                   \ifx&#2&%
                                     \ifmmode{\ell}\else$\ell$\fi
                                   \else
                                     \ifmmode{#2}\else$#2$\fi
                                   \fi
                                \else
                                   \ifx&#2&%
                                     \ifmmode{\ell\rm{#1}}\else$\ell\rm{#1}$\fi
                                   \else
                                     \ifmmode{\ell\rm{#1}#2}\else$\ell\rm{#1}#2$\fi
                                   \fi
                                \fi}
\newcommand{\about}      [1]   {\ifx&#1&
                                   \ifmmode\rm{\sim}\else$\rm{\sim}$\fi
                                 \else
                                   \ifmmode\rm{{\sim}#1}\else$\rm{{\sim}#1}$\fi
                                 \fi}
\newcommand{\forbline}   [3]   {[#1{\scshape\,#2}]$\,\lambda#3$}

\newcommand{\Halpha}           {\ifmmode\rm{H\alpha}\else$\rm{H\alpha}$\fi}
\def\apjl{ApJ}                              
\def\apjs{ApJS}                             
\def\ao{Appl.~Opt.}                         
\def\aap{A\&A}                              

%% file: 97_IDL/SEDsexT40/TABLES/M33_Numbers.tex
\newcommand{\Distance}{0.840}
\newcommand{\ScaleLength}{2513}
\newcommand{\Inclination}{57}
\newcommand{\HIIperc}{41.7}
\newcommand{\METave}{0.2}
\newcommand{\METmin}{0.2}
\newcommand{\METmax}{0.4}
\newcommand{\AGEave}{4.0}
\newcommand{\AGEmin}{3.5}
\newcommand{\AGEmax}{4.0}
\newcommand{\CMPave}{4.0}
\newcommand{\CMPmin}{4.0}
\newcommand{\CMPmax}{4.0}
\newcommand{\PRSave}{8}
\newcommand{\PRSmin}{8}
\newcommand{\PRSmax}{8}
\newcommand{\COVave}{0.46}
\newcommand{\COVmin}{0.36}
\newcommand{\COVmax}{0.74}
\newcommand{\CHIaveHII}{4.69}
\newcommand{\CHIminHII}{0.73}
\newcommand{\CHImaxHII}{12.15}
\newcommand{\MCLave}{1.6}
\newcommand{\MCLmin}{0.9}
\newcommand{\MCLmax}{1.9}
\newcommand{\MCDave}{2.9}
\newcommand{\MCDmin}{2.5}
\newcommand{\MCDmax}{3.5}
\newcommand{\LUMaveHII}{0.42}
\newcommand{\LUMminHII}{0.36}
\newcommand{\LUMmaxHII}{0.49}
\newcommand{\TAUave}{2.2}
\newcommand{\TAUmin}{1.6}
\newcommand{\TAUmax}{2.8}
\newcommand{\SFRave}{0.25}
\newcommand{\SFRmin}{0.20}
\newcommand{\SFRmax}{0.29}
\newcommand{\OLDave}{0.05}

\newcommand{\BDave}{0.0}

\newcommand{\CHIaveDIFF}{1.82}
\newcommand{\CHIminDIFF}{0.46}
\newcommand{\CHImaxDIFF}{4.44}
\newcommand{\MDDave}{13.8}
\newcommand{\MDDmin}{10.0}
\newcommand{\MDDmax}{17.5}
\newcommand{\LUMaveDIFF}{1.30}
\newcommand{\LUMminDIFF}{1.08}
\newcommand{\LUMmaxDIFF}{1.50}
\newcommand{\MTDave}{16.7}
\newcommand{\MTDmin}{12.5}
\newcommand{\MTDmax}{21.0}
\newcommand{\ESCaveYOUNG}{70}
\newcommand{\ESCminYOUNG}{59}
\newcommand{\ESCmaxYOUNG}{84}
\newcommand{\SFRaveTotal}{0.45}
\newcommand{\SFRminTotal}{0.31}
\newcommand{\SFRmaxTotal}{1.13}
\newcommand{\GTDave}{101}
\newcommand{\GTDmin}{80}
\newcommand{\GTDmax}{135}

%% file: 01_INTRODUCTION/Introduction.tex
%
%
\section{Introduction}
\label{sec:Introduction}
Dust grains are mixed with the gas in all the components and phases of a galaxy,
from the ionized gas and the photo-dissociation regions (PDRs)
embedded in the star forming (SF) regions,
to the diffuse interstellar medium (ISM).
Depending on their environment, dust grains are exposed to very different
radiation fields and physical conditions.
While the interstellar dust within the SF component is immersed in
the intense ultraviolet (UV) radiation coming from massive, hot stars,
dust grains in the diffuse ISM are more likely illuminated by
the radiation field of the old stellar population
\citep[][]{2003A&A...407..137H,2009A&A...493..453V}.
These differences not only change the grain heating and the subsequent emission,
but also might change the grain properties
(e.g., size distribution, chemical composition, ionization state)
and play an important role on the dust evolution
\citep[][]{2003A&A...407..137H, 2011ApJ...735....6P}.
Modeling the spectral energy distribution (SED) emitted by the dust grains
is a powerful tool to understand the grain properties as well as
the interaction between the interstellar dust and the radiation field.
One of the most simple approaches to model the dust emission is the
modified black body (MBB), where the intensity, \Snu{}{}, is expressed as:

\begin{equation}
\MBB{}{},
\end{equation}
where \BB{} is the Planck function and
\Eindex{}{} is the dust emissivity index, which in the case of idealized
spherical dielectric grains adopts the value \Eindex{=}{2}.
Although the MBB models are highly degenerated
(there is no unique solution for \Eindex{}{} and \Temp{}{},
see \citealt{2014A&A...561A..95T} for a detailed analysis),
they have been widely used to study the dust emission
\citep[e.g.,][]{2013ApJ...778...51K, 2014ApJ...797...85G}.
More sophisticated models have also been used,
from phenomenological models based on realistic dust properties
\citep[e.g.,][]{2011A&A...536A..88G},
to fully three-dimensional radiation transfer models
\citep[][]{2005MNRAS.362.1038E}.
Although the different methodologies applied make the comparison of these
studies somehow difficult, many of them have shown that it is not
possible to reproduce the SEDs of low-metallicity galaxies
due to an excess of observed emission at submillimeter (submm) wavelengths
\citep[e.g.,][]{2002A&A...382..860L,
                2003A&A...407..159G,
                2005A&A...434..867G,
                2006ApJ...645..134B,
                2009A&A...508..645G,
                2011A&A...532A..56G,
                2010A&A...519A..67I,
                2010A&A...523A..20B,
                2012ApJ...745...95D,
                2011A&A...536A..17P}.
The origin of this problem, usually known as the \textit{submm excess},
is not clear yet.
The existence of a large amount of cold dust could in principle produce the
observed shape, but the required dust mass is unreasonably high
\citep[e.g.,][]{2002A&A...382..860L}
and the gas-to-dust mass ratio too low
\citep[e.g.,][]{2011A&A...536A..88G,2011A&A...532A..56G,2014ApJ...797...85G}.
Different dust grain properties, and in particular a lower
\Eindex{}{} in the submm have also been suggested
\citep[e.g.,][]{2002A&A...382..860L,2011A&A...536A..88G,2014ApJ...797...85G}.
For example, if hydrogenated amorphous carbon grains were the most probable form
of carbonaceous grains in the ISM (instead of graphite),
this could significantly flatten the SED in the submm range
\citep[][]{2008A&A...492..127S,2013A&A...558A..62J}.
More exotic explanations have also been proposed,
from magnetic nanograins emitting at the microwave and submm wavelengths
\citep[][]{2012ApJ...757..103D}
to spinning grains
\citep[][]{1998ApJ...508..157D}
which emit however more in the millimeter (mm) range
\citep[][]{2010A&A...523A..20B}.
A step forward has been the possibility to carry out spatially resolved analysis
of the dust SED with a combination of \SPITZER{} and \HERSCHEL{} data, which
allow to investigate variations of the dust properties at small scales.
The Magellanic Clouds and \MESSIER{33} are excellent targets
to perform such studies.
In the Large Magellanic Cloud (LMC) \citet[][]{2011A&A...536A..88G}
found evidence for a submm excess in the low surface brightness areas
by fitting the dust SED locally with a detailed dust model.
\citet[][]{2014ApJ...797...85G} were able to fit the SEDs
of the LMC and the Small Magellanic Cloud (SMC)
using a MBB with a broken power-law for \Eindex{}{}.
For \MESSIER{33} \citet[][]{2014A&A...561A..95T}
carried out a MBB fit of the local SED and found
a trend of decreasing temperature T and decreasing \Eindex{}{} with radius
(this trend is the opposite of what would be
expected from the \Eindex{}{}-T degeneracy).
They concluded that a change in the dust properties is the most likely reason
for this trend.
Other studies have tried to look for variations in the dust properties
in further objects.
\citet[][]{2014MNRAS.439.2542G}
investigated the \microns{870} emission of a sample of 11 nearby galaxies
and found evidence for a submm excess in 4 of them.
For three of these objects the excess was in the outskirts of the galaxy.
\citet[][]{2013A&A...549A..70H} decomposed
the observed emission of the dwarf, low metallicity galaxy \NGC{4214}
into its SF and diffuse components.
They modeled these components separately using radiation transfer models
and found indications of submm excess in the diffuse component of \NGC{4214}.
This paper is part of a series of studies that use
radiation transfer models to fit the SED
of a sample of dwarf starburst and late-type spiral disk galaxies.
Here we apply the same methodology as in
\citet[][]{2013A&A...549A..70H} to the local-group galaxy \MESSIER{33}.
The proximity of \MESSIER{33}
\citep[\Mpc{\Distance},][]{1991ApJ...372..455F},
its moderate inclination angle
\citep[\degrees{\Inclination},][]{1987A&AS...67..509D}
and the wide data set available,
make this galaxy an excellent target to study separately
the SEDs of the SF and the diffuse component with accuracy.
Recent results of the open time key program \HERMES\
\citep[][]{2010A&A...518L..67K}
have shown that the \HERSCHEL{} observations of \MESSIER{33} can not be well
reproduced assuming standard dust properties (i.e. MBB model with \Eindex{=}{2}).
Instead, these authors had to adopt a MBB model with \Eindex{=}{1.5}
to obtain a good fit.
The value of \Eindex{=}{1.5} is extensible to the \PLANCK{} data
(see \App{app:CC} for details) and suggests that that
the dust properties or conditions might be special in \MESSIER{33}.
The aim of the present paper is to investigate if it is possible to reproduce
the SED of \MESSIER{33} with standard dust grain properties but using
more sophisticated models based on radiation transfer calculations
and a realistic geometry,
or if the anomalies found by \citet[][]{2010A&A...518L..67K} persist.
%
%
%
The paper is organized as follows.
In \Sec{sec:data} we present the data set used, from UV to radio.
\Sec{sec:Photometry} is devoted to explain the photometry measurements
and the method we have used to separate the emission coming from the SF regions
and the ISM in the galaxy.
In \Sec{sec:models} we explain briefly the models applied to fit the SED of the
two components and how we determined their input parameters.
The results of the modeling are presented in \Sec{sec:results}.
In \Sec{sec:discussion} we discuss the results
and in \Sec{sec:summary} we summarize the main conclusions of this paper.
%
%
%
%

%% file: 02_DATA/Data.tex
%
%
\section{Data \label{sec:data}}
A wide range of data from recent scientific missions
are available for \MESSIER{33} in archives.
In this section we describe the data used in our study,
which are summarized in \Tab{tab:observations}.
%
%
%
\input{./99_TABLES/M33_DataSummary}
%
%
%
\subsection{\GALEX{} \label{Data:Galex}}
To investigate the continuum UV emission
associated with the young stellar population of \MESSIER{33},
we use the data from the \textit{Galaxy Evolution Explorer}
\citep[\GALEX{},][]{2005ApJ...619L...1M}
distributed by
\citet[][]{2007ApJS..173..185G}.
The \GALEX{} far-UV (FUV \nm{\sim154}) and near-UV (NUV \nm{\sim232}) observations
and the data reduction can be found in \citet[][]{2005ApJ...619L..67T}.
The angular resolution for the \GALEX{FUV and NUV} maps
is \arcseconds{4}{2} and \arcseconds{5}{3}, respectively.
Following \citet[][]{2007ApJS..173..682M},
we conservatively adopted a calibration error of \percent{10}.
\subsection{\SLOAN{} \label{Data:Sloan}}
We used data from the Sloan Digital Sky Survey
\citep[SDSS,][]{2000AJ....120.1579Y}
to measure the stellar emission in the \textit{u}-band and to
calculate the \textit{b}-band stellar disk scale-length of \MESSIER{33}.
Due to the small FoV of the SDSS images, we needed to reconstruct a
mosaic to cover the whole extension of \MESSIER{33} in these bands.
For this purpose, we used the software \MONTAGE%
\footnote{\url{http://montage.ipac.caltech.edu}}.
\MONTAGE\ is a toolkit for creating mosaics from individual
Flexible Image Transport System (\FITS) images.
We made use of the data from the seventh data release
\citep[DR7, ][]{2009ApJS..182..543A}.
These images have a spatial coverage of
\arcminutes{13}{51}\arcminutes{\times9}{83},
an exposure time of \seconds{53.9},
and the angular resolution is \arcseconds{\sim1}{4}
(median value measured in the \SLOAN{r}).
The SDSS DR7 pipeline does not apply any sky subtraction to the raw data.
Considering that most of the frames used to reconstruct the mosaics are
fully covered by light coming from \MESSIER{33},
the usage of the DR7 data allows us to avoid possible artifacts coming
from an inaccurate sky subtraction.
Following \citet[][]{2008ApJ...674.1217P} we adopted
\percent{1} of calibration error for the \SLOAN{g}
and \percent{2} for the \SLOAN{u}.
\subsection{\texorpdfstring{\Halpha}\ \label{Data:Halpha}}
To trace the ionized gas, we use the narrow band \Halpha\ image of \MESSIER{33}
obtained with the Kitt Peak National Observatory (KPNO)
by \citet[][]{1998PhDT........16G}.
The reduction process, using standard \IRAF%
\footnote{\IRAF\ is distributed by the National Optical Astronomy Observatories,
which are operated by the Association of Universities for Research in Astronomy,
Inc., under cooperative agreement with the National Science Foundation.}
procedures to subtract the continuum emission, is described in detail in
\citet[][]{2000ApJ...541..597H}.
The total field of view (FoV) of the image is 1.75$\times$1.75~deg$^2$
(2048$\times$2048 pixels with a pixel scale of \arcseconds{2}{03})
with a \arcseconds{6}{6} resolution.
We consider typical uncertainties to be better than \percent{15}
for the \Halpha\ photometric measurements
\citep[][]{2010A&A...518L..68V}.
\subsection{\SPITZER{} \label{Data:Spitzer}}
Data of \MESSIER{33} was obtained with the \SPITZER{}
\citep[][]{2004ApJS..154....1W}
Infrared Array Camera
\citep[\CAMERAmu{IRAC}{},][]{2004ApJS..154...10F}
and the Multiband Imaging Photometer
\citep[\CAMERAmu{MIPS}{},][]{2004ApJS..154...25R}.
We use \CAMERAmu{IRAC}{3.6} and \microns{4.5} data to study the radiation field
associated with the old stellar population of \MESSIER{33}
and \CAMERAmu{MIPS}{24} to study the dust emission.
The \CAMERAmu{MIPS}{70} and \CAMERAmu{MIPS}{160} data were not used in this work
since better quality data from \CAMERAmu{PACS}{} are available at
the same wavelengths.
The emission from polycyclic aromatic hydrocarbons (PAHs),
traced by the \CAMERAmu{IRAC}{5.8} and \microns{8.0} bands,
is out of the scope of the present work.
\CAMERAmu{IRAC}{3.6} and \microns{4.5} data are described in
\citet[][]{2007A&A...476.1161V,2009A&A...493..453V,2010A&A...510A..64V}.
The \CAMERAmu{MIPS}{24} data have been processed
using the version 18.5 of the \MOPEX\ software
\citep[MOsaicker and Point source EXtractor,][]{2005PASP..117.1113M}
to combine the 16 AORs (Astronomical Observation Request) available
in the data archive.
The spatial resolution is
\arcseconds{2}{5},
\arcseconds{2}{9}, and
\arcseconds{6}{3}
for the
\CAMERAmu{IRAC}{3.6},
\CAMERAmu{IRAC}{4.5}, and
\CAMERAmu{MIPS}{24}
bands, respectively.
We adopted calibration errors of
\percent{2} and for the \CAMERAmu{IRAC}{} bands \citep[][]{2005PASP..117..978R}
and \percent{4} for \CAMERAmu{MIPS}{24} \citep[][]{2007PASP..119..994E}.
%
%
%
%
%
\subsection{\HERSCHEL{} \label{Data:Herschel}}
As part of the \HERMES\ key program
\citep[][]{2010A&A...518L..67K},
the \MESSIER{33} galaxy was mapped by \HERSCHEL{} with the
Photodetector Array Camera and Spectrometer
\citep[\CAMERAmu{PACS}{},][]{2010A&A...518L...2P}
and the Spectral and Photometric Imaging Receiver
\citep[\CAMERAmu{SPIRE}{},][]{2010A&A...518L...3G}.
\CAMERAmu{PACS}{100} and \microns{160} and \CAMERAmu{SPIRE}{250},
\microns{350}, and \microns{500} observations
were done in parallel mode in two orthogonal directions on
January 7th, 2010, covering a region of about 1.36 square degrees.
The \CAMERAmu{PACS}{70} image was obtained as a follow–up open time
cycle 2 programme, on 25 June 2012 in 2 orthogonal directions and 5 repetitions
to achieve better sensitivity
\citep[][]{2015arXiv150201347B}.
The \CAMERAmu{PACS}{} reduction has been performed using the map-making software
\SCANAMORPHOS\ \citep[][]{2013PASP..125.1126R}
as described in \citet[][]{2011AJ....142..111B}.
The \CAMERAmu{SPIRE}{} reduction has been done using the
\HERSCHEL{} Data Processing System
\citep[\HIPE,][]{2010ASPC..434..139O,2011ASPC..442..347O}
and the maps were created using a "naive" mapping projection
\citep[][]{2010A&A...518L..68V,2010A&A...518L..70B,2012A&A...543A..74X}.
The angular resolution of the \HERSCHEL{} data
range from \arcseconds{5}{6} at \microns{70} to
\arcseconds{36}{4} for the \microns{500} band.
The absolute photometric uncertainty is \percent{10} for
the \CAMERAmu{PACS}{70} and \microns{100} bands
and \percent{20} for the \CAMERAmu{PACS}{160} band
\citep[][]{2010A&A...518L...2P},
while we adopted \percent{5} as the calibration uncertainty for the three
\CAMERAmu{SPIRE}{} bands
\citep[][]{2013MNRAS.433.3062B}.
\subsection{\CAMERAmu{LABOCA}{} \label{Data:LABOCA}}
Observations of \MESSIER{33} were carried out at the
Atacama Pathfinder Experiment \citep[\APEX{},][]{2006A&A...454L..13G} using the
Large Bolometer Camera \citep[\CAMERAmu{LABOCA}{},][]{2009A&A...497..945S}
between 2010 and 2012 (Albrecht et al., in prep.).
A filter set coupled to the atmospheric window provides a bandwidth
of \microns{\sim150} (\GHz{\sim60})
around the central frequency of \microns{870} (\GHz{345}).
From observations of Uranus an almost circular beam size of
\arcseconds{19}{2}\,\arcseconds{\pm\,0}{7} was determined
\citep[][]{2009A&A...497..945S}.
\CAMERAmu{LABOCA}{} mapping was performed in on-the-fly mode parallel and orthogonal
to the  major axis of \MESSIER{33}.
Flux calibration was achieved through observations of planets as well as a set
of secondary calibrators and was found to be accurate within \percent{10}
\citep[][]{2009A&A...497..945S}.
To enable the usage of the \CAMERAmu{LABOCA}{} data in
multi-wavelength studies the flux calibration methods for \CAMERAmu{LABOCA}{}
and \CAMERAmu{SPIRE}{} were compared to ensure a correct cross-calibration.
On average \percent{2} lower flux densities from the \CAMERAmu{LABOCA}{}
calibration compared to the \CAMERAmu{SPIRE}{} method were found.
The \CAMERAmu{LABOCA}{} data were reduced using the \BOA%
\footnote{http://www.astro.uni-bonn.de/boawiki/Boa}
software package.
The applied skynoise suppression can not distinguish between correlated
atmospheric noise and astronomical emission that is uniformly extended over
areas equal to or larger than the area covered by the groups of bolometer
channels (defined by having the same amplifier box or same wiring) that are
used to derive the median noise.
This typically filters out uniform emission on scales larger than about
\arcminutes{2}{5} \citep[for details see][]{2009A&A...504..415S}.
To minimize the effect an iterative approach by subsequently improving a model
of the flux distribution was applied.
Nevertheless, given the large apparent size of \MESSIER{33},
the resulting \CAMERAmu{LABOCA}{} map does not recover the entire contribution
from the extended emission, as it is confirmed by comparing with the
corresponding \CAMERAmu{PLANCK}{850} map (see \Fig{fig:BANDSa}).
\subsection{\PLANCK{} \label{Data:Planck}}
We used \PLANCK{} data from the 2013 distribution of released products
\citep[][]{2014A&A...571A...1P}
based on the data acquired during the \PLANCK{} ``nominal'' operations period,
i.e., between 12 August 2009 and 27 November 2010.
Data can be accessed via the \PLANCK{} Legacy Archive%
\footnote{\url{http://www.sciops.esa.int/index.php?project=planck&page=Planck_Legacy_Archive}}.
In the case of the Low Frequency Instrument (\CAMERAghz{LFI}{}) \mm{10} (\GHz{30}) band,
we used the flux computed in the \PLANCK{}{} Catalog of Compact Sources
\citep[][]{2014A&A...571A..28P}.
For the other \PLANCK{} bands,
i.e., the High Frequency Instrument (\CAMERAghz{HFI}{})
\microns{350} (\GHz{857}),
\microns{550} (\GHz{545}),
\microns{850} (\GHz{353}),
\mm{1.4} (\GHz{217}),
\mm{2.1} (\GHz{143}), and
\mm{3.0} (\GHz{100}) bands
and the LFI
\mm{4.3} (\GHz{70}) and
\mm{6.8} (\GHz{44}) bands,
we extracted the fluxes from the \PLANCK{} \textit{all-sky} maps.
These maps consist of 2048 \HEALPIX\ full-sky maps
\citep[][]{2005ApJ...622..759G}
corresponding to a pixel size of \arcminutes{2}{64}
with an angular resolution that ranges from \arcminutes{4}{63} at \microns{350}
to \arcminutes{32}{3} at \mm{10} \citep[][]{2014arXiv1407.5452A}.
To generate a Gnomonic projection of the \HEALPIX\ data
centered on \MESSIER{33} we made use of the \IDL\ facility \GNOMVIEW%
\footnote{\url{http://healpix.jpl.nasa.gov/html/idlnode21.htm}}.
The cosmic microwave background (CMB) was subtracted from all the bands
(see \App{app:CMB} for details).
The contribution from other components
(e.g., cirrus, cosmic infrared background) to the background was subtracted
as the median value in a \arcminutes{5}{} width annulus enclosing \MESSIER{33}.
The resulting maps are shown in \Fig{fig:BANDSc}.
\MESSIER{33} was not detected in the \mm{4.3} and \mm{6.8} channels.
We adopted \percent{\pm10} of calibration uncertainty for the \microns{350}
and \microns{550} bands and \percent{\pm3} for all the other wavelengths
\citep[][]{2014arXiv1407.5452A}.
\subsection{\EFFELSBERG{} \label{Data:EFFELSBERG}}
\citet[][]{2007A&A...475..133T} studied the emission
at \cm{3.6} (\GHz{8.35}) of \MESSIER{33}.
Here we used their results to constrain the thermal radio emission
of the SF regions as well as to decontaminate our measurements from
synchrotron emission
(see \Sec{sec:synchrotron} for details).
They found a total flux of \mJy{761\pm63} for the inner \kpc{7.5}
and a thermal fraction of \percent{51.4\pm4.2} at this wavelength.
This fraction was obtained from the extinction corrected \Halpha\
map of the galaxy, which was derived using a map of optical depth
at \microns{160} and an extinction  law for a standard dust model
for the diffuse emission.
The authors also derived the thermal fraction using the classical method and
assuming a constant non-thermal spectral index for the whole galaxy of
$\alpha_{\rm n}=1.0\pm0.1$.
With this method they obtained a thermal fraction of \percent{63.2\pm5.3}.
As a compromise between the two values of the thermal fraction
derived by \citet[][]{2007A&A...475..133T}, we adopted the mean value as
the best estimate and we allowed it to fluctuate between
the lowest and the highest possible values.
We therefore estimated that \percent{57.3\pm11.2} (\mJy{436\pm49}) of
the total flux at \GHz{8.35} corresponds to thermal radio emission.
\ImageBands{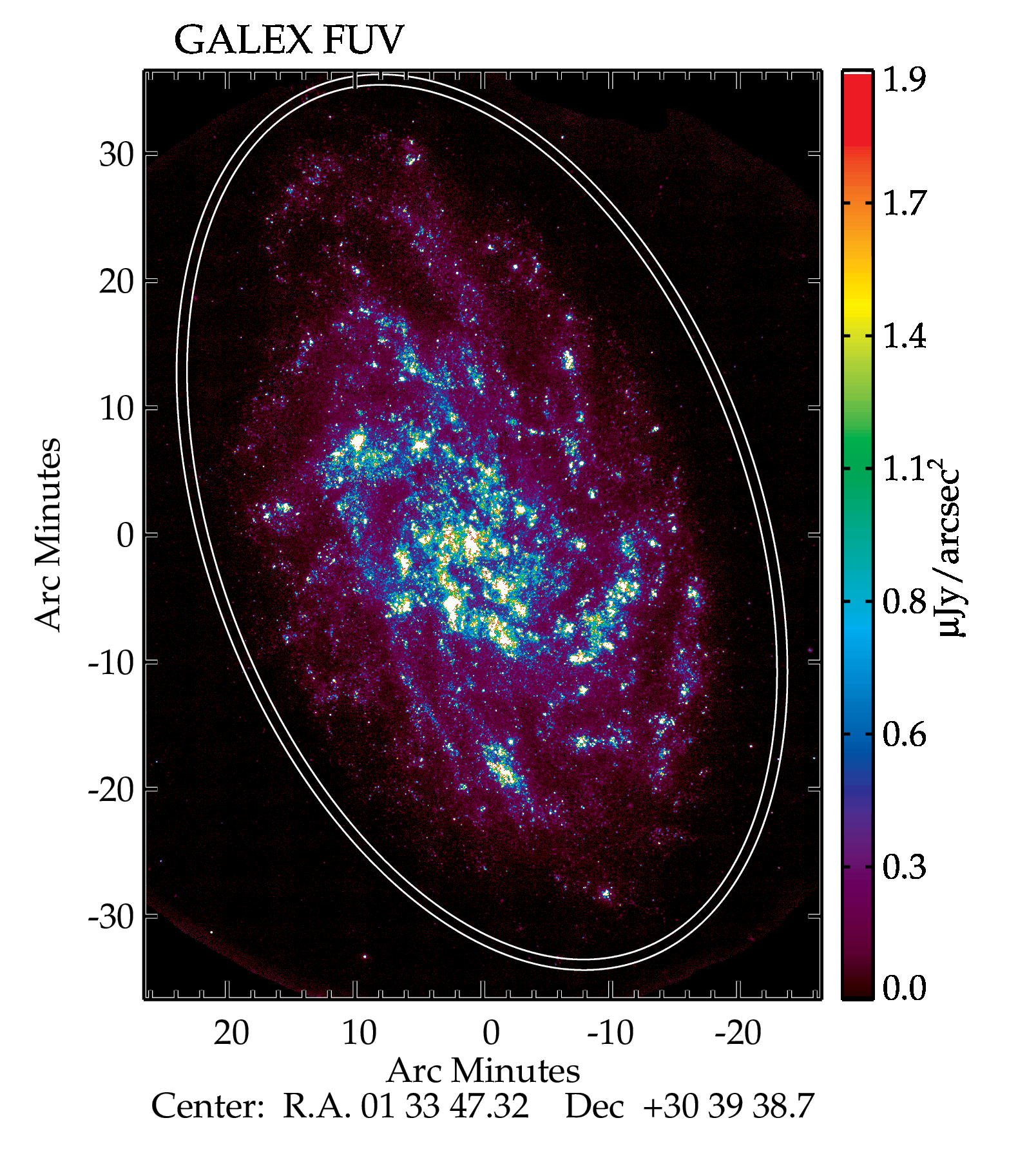}
           {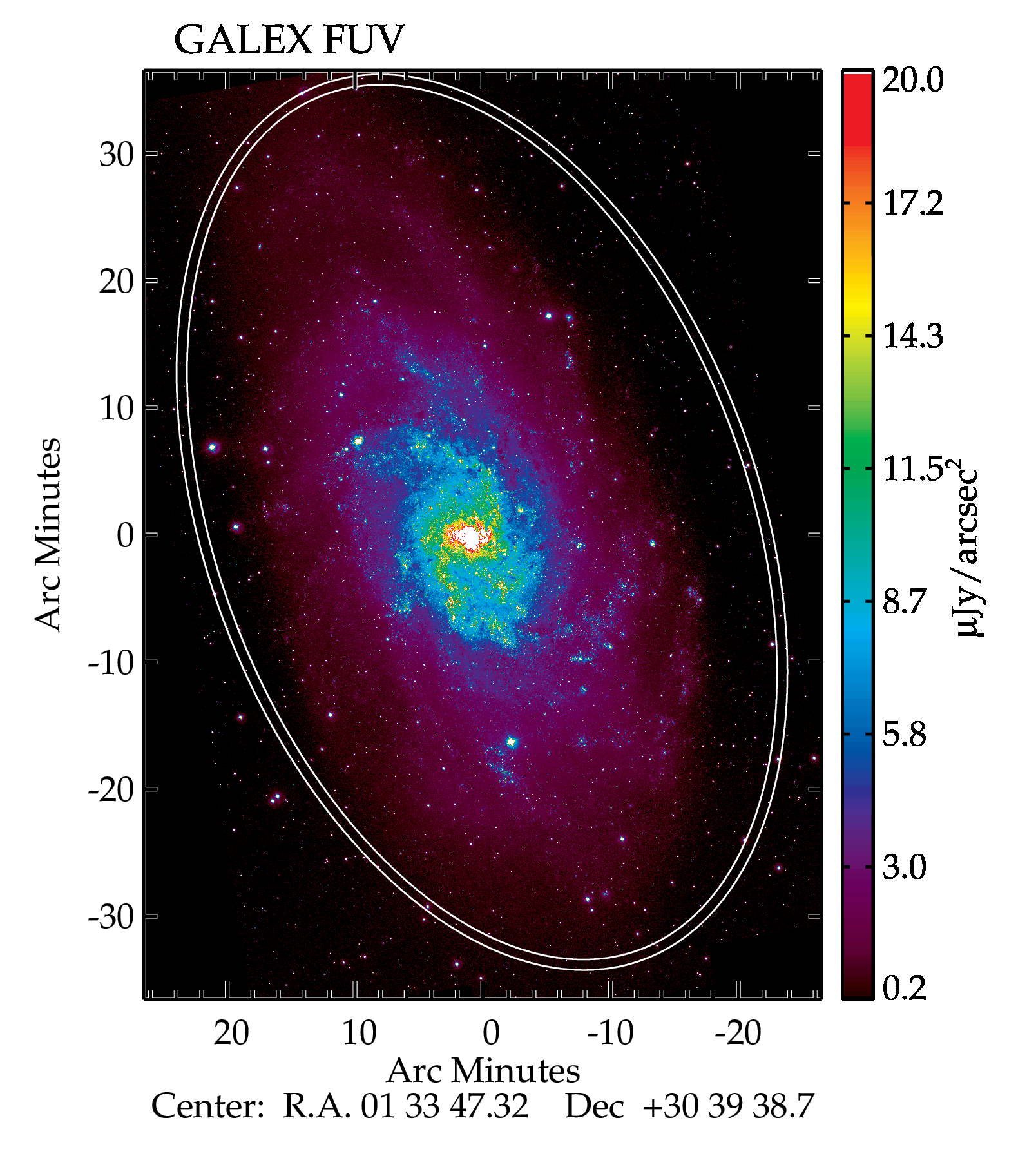}
           {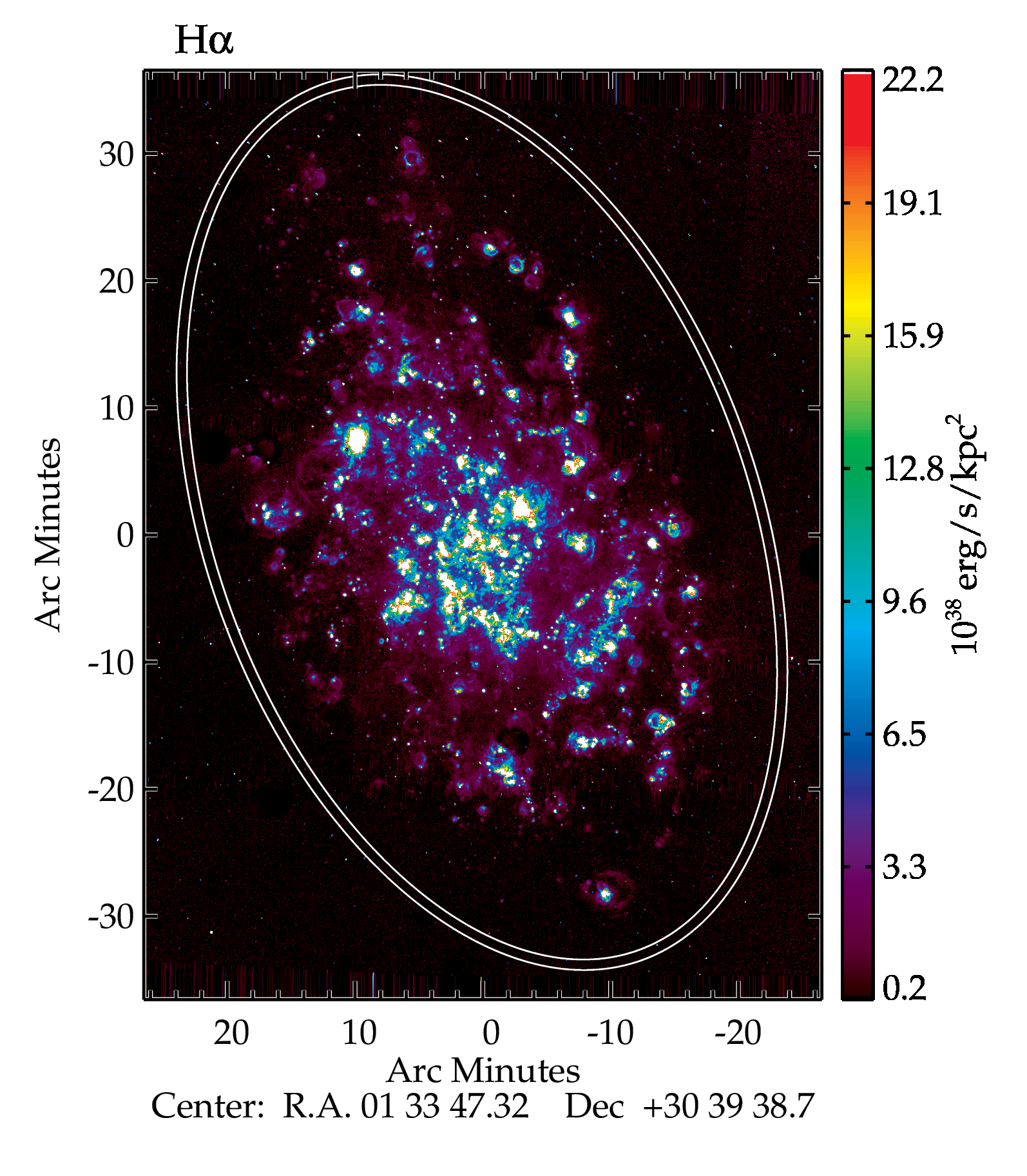}
           {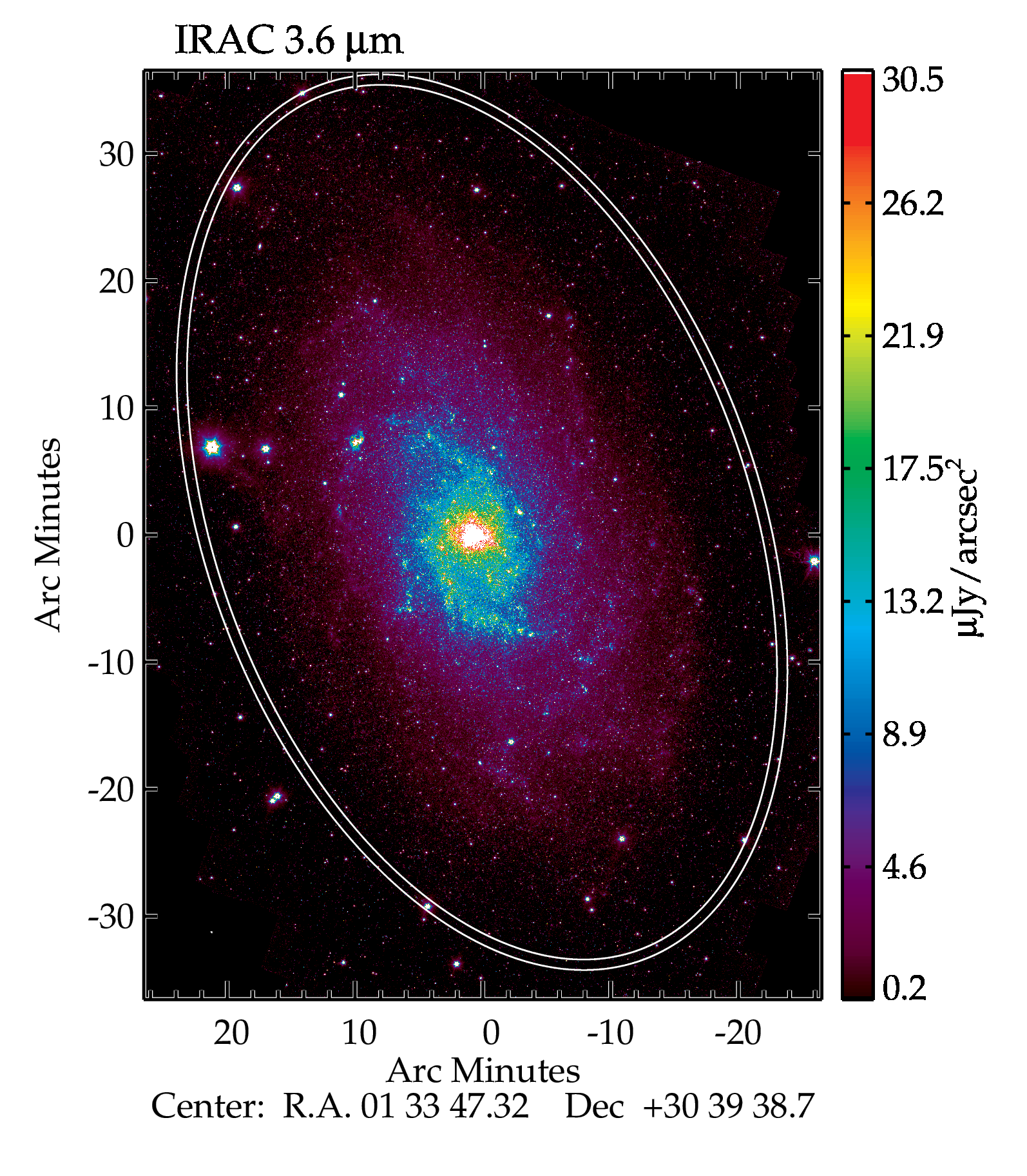}
           {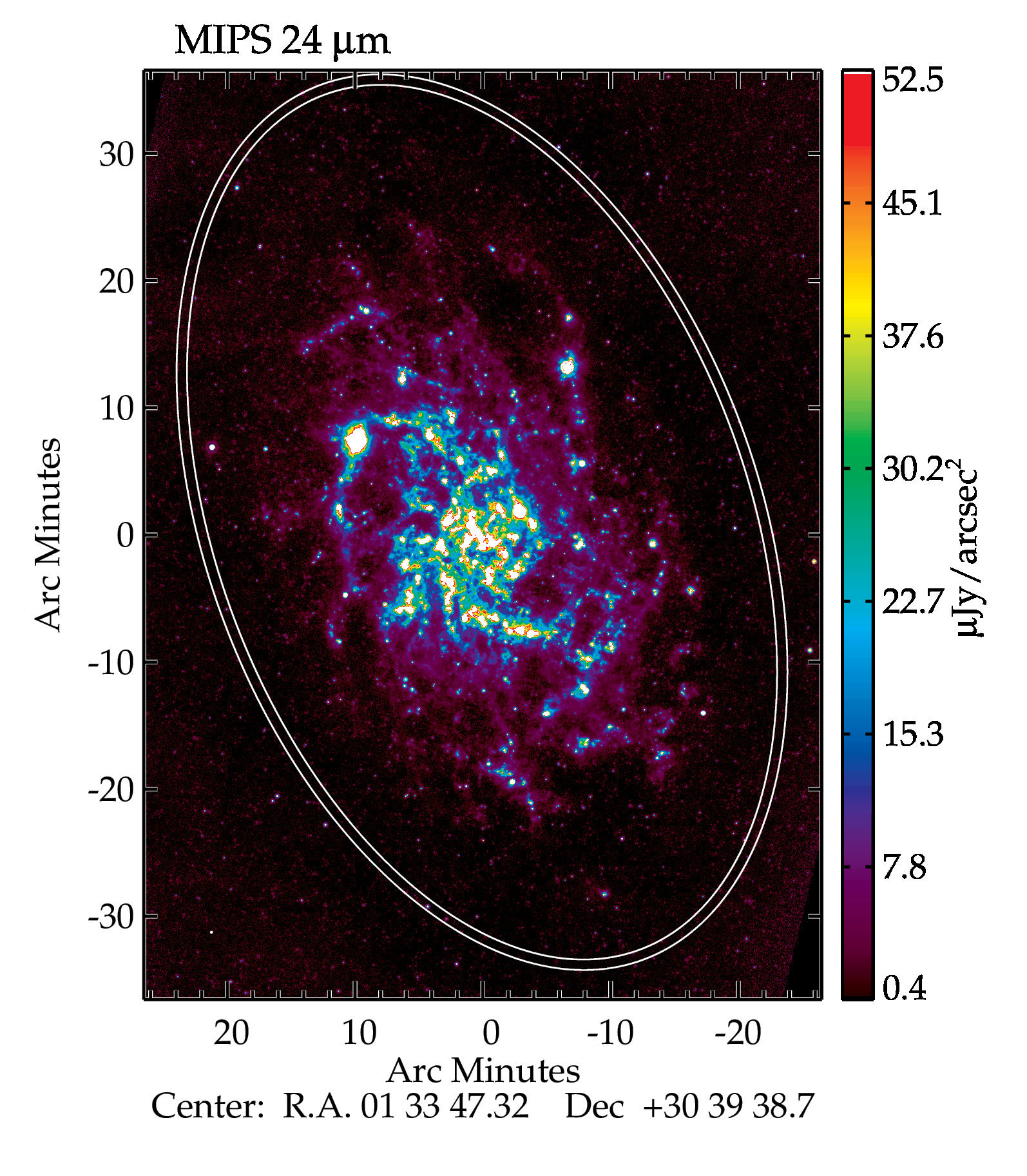}
           {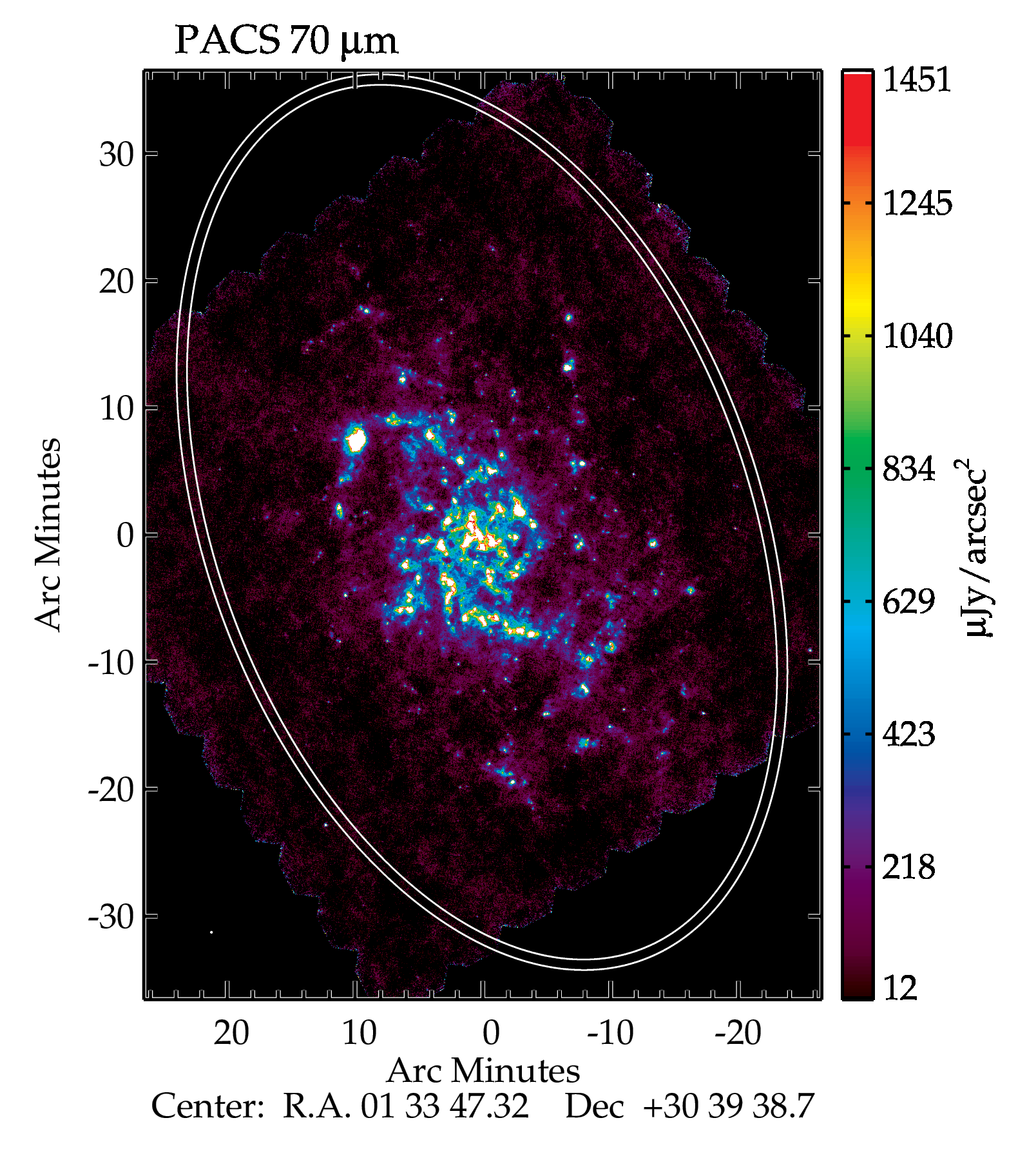}
           {fig:BANDSa}
           {Data of \MESSIER{33} at their original resolution.
            Units have been homogenized to facilitate the comparison between
            the different bands.
            At the distance of \MESSIER{33}, \arcseconds{1}{} subtends \pc{\sim4}.
            The white annulus, used to subtract the local background,
            has been included for reference.
            The white circle at the left bottom
            (too tiny to be visible for bands with \microns{\lambda\lesssim350})
            corresponds to the beam size.}
           {}
\ImageBands{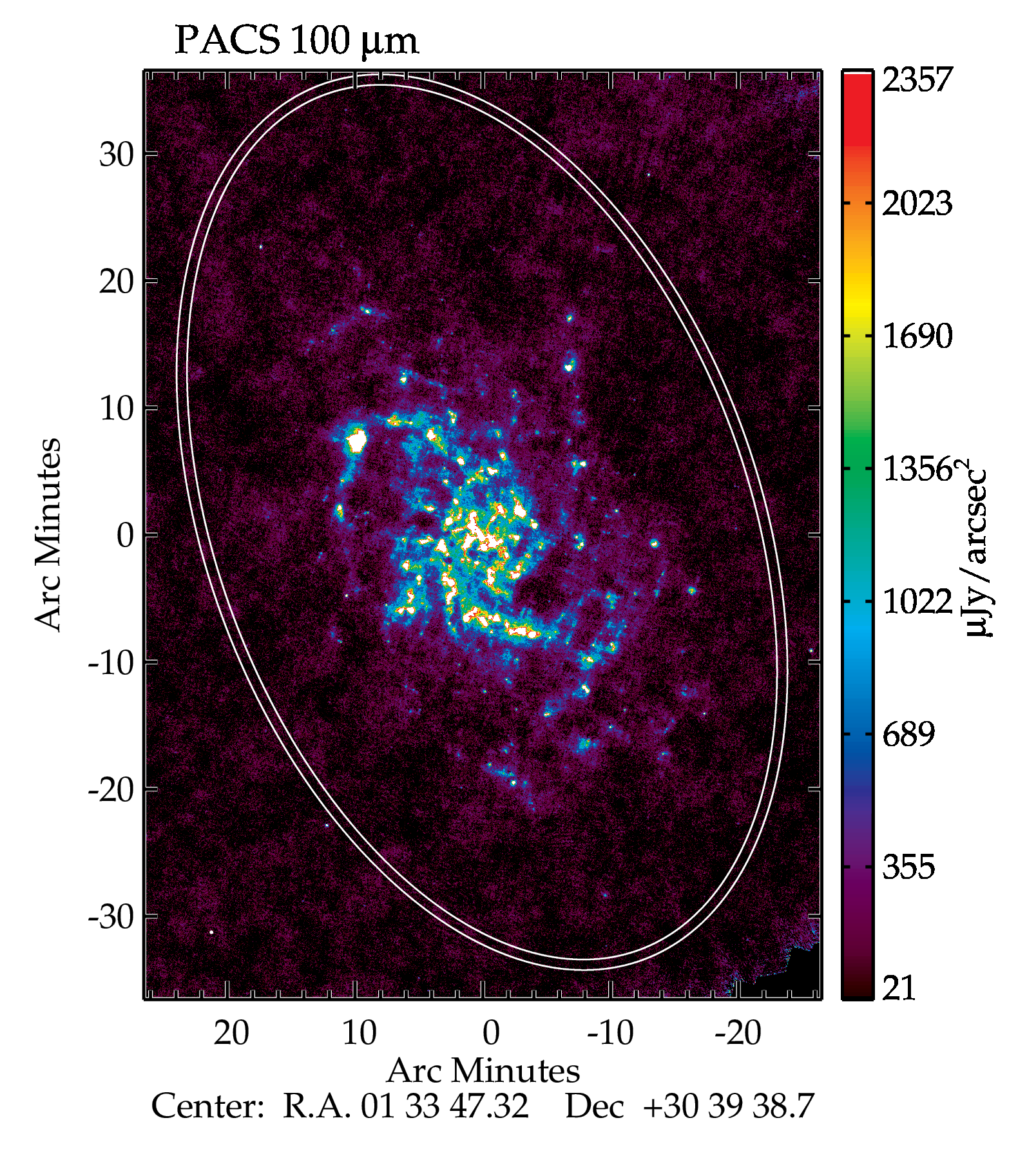}
           {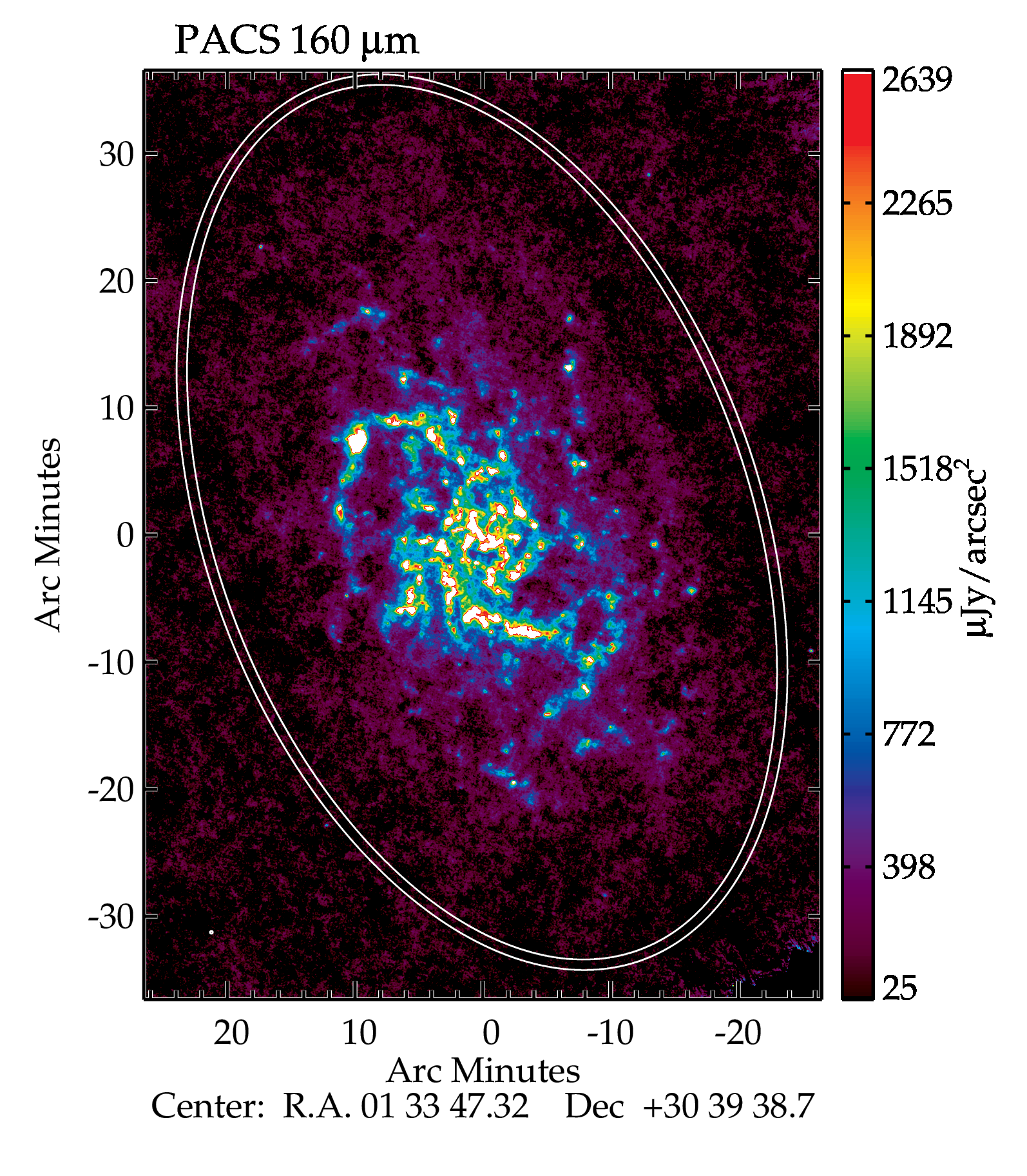}
           {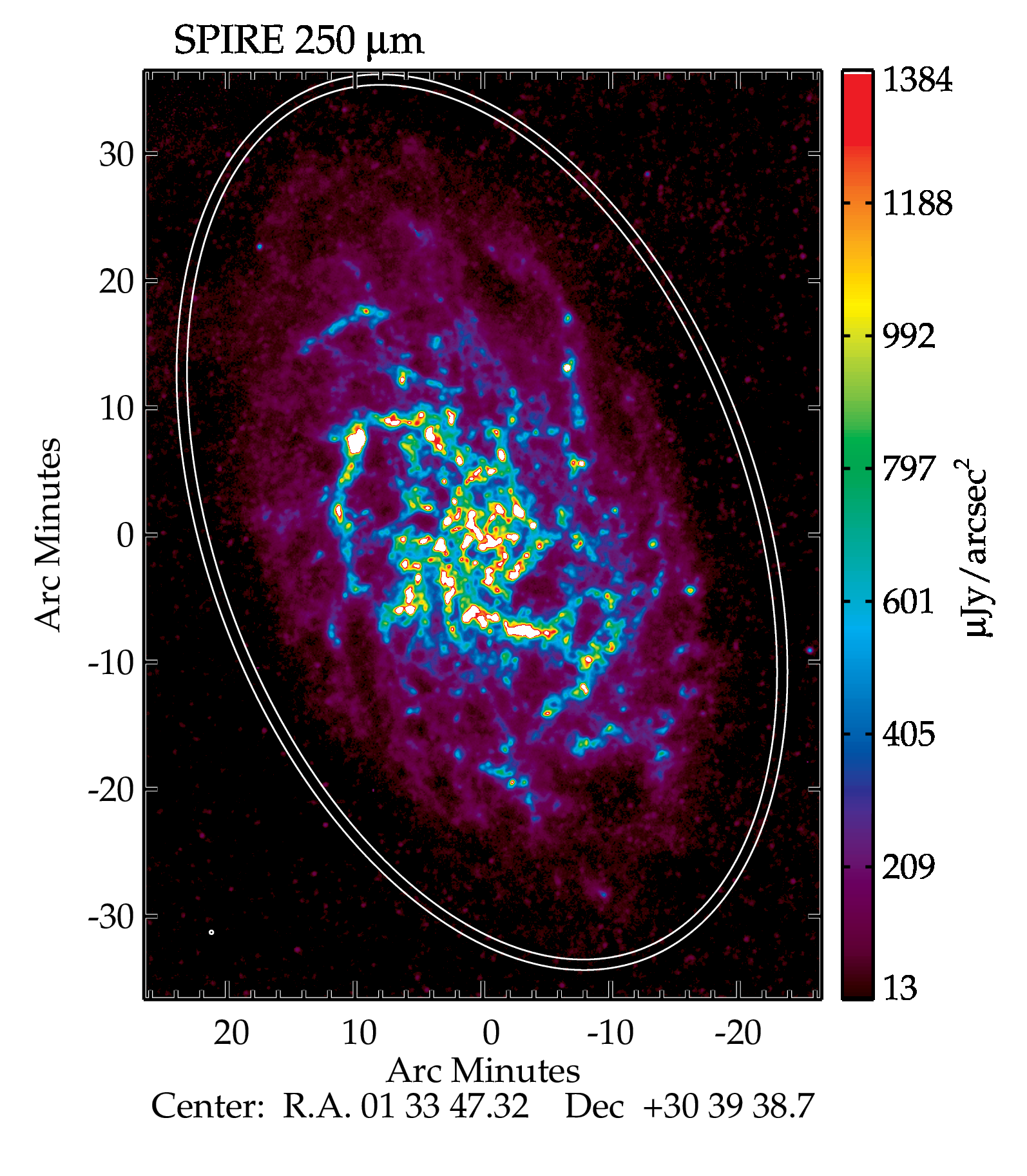}
           {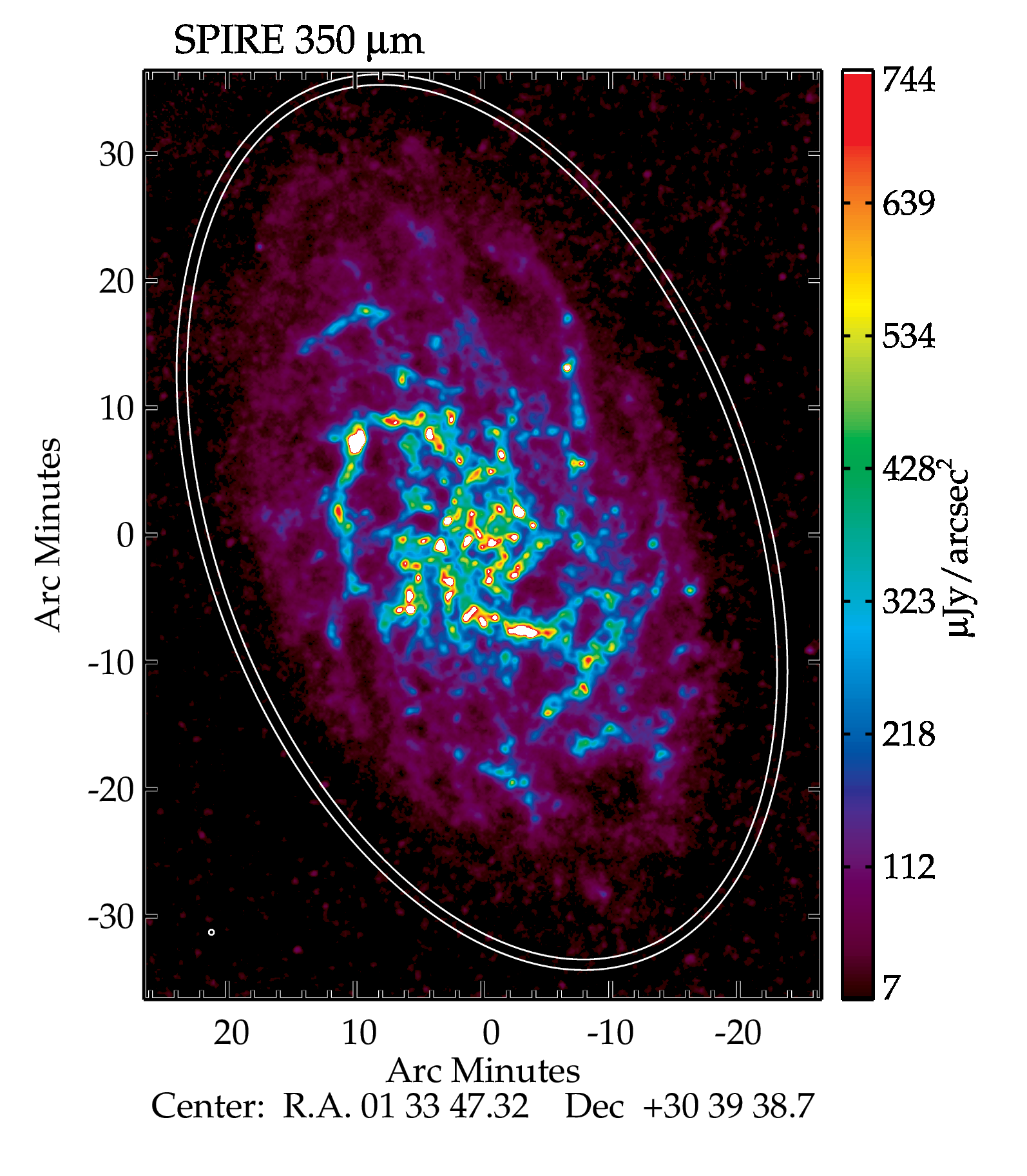}
           {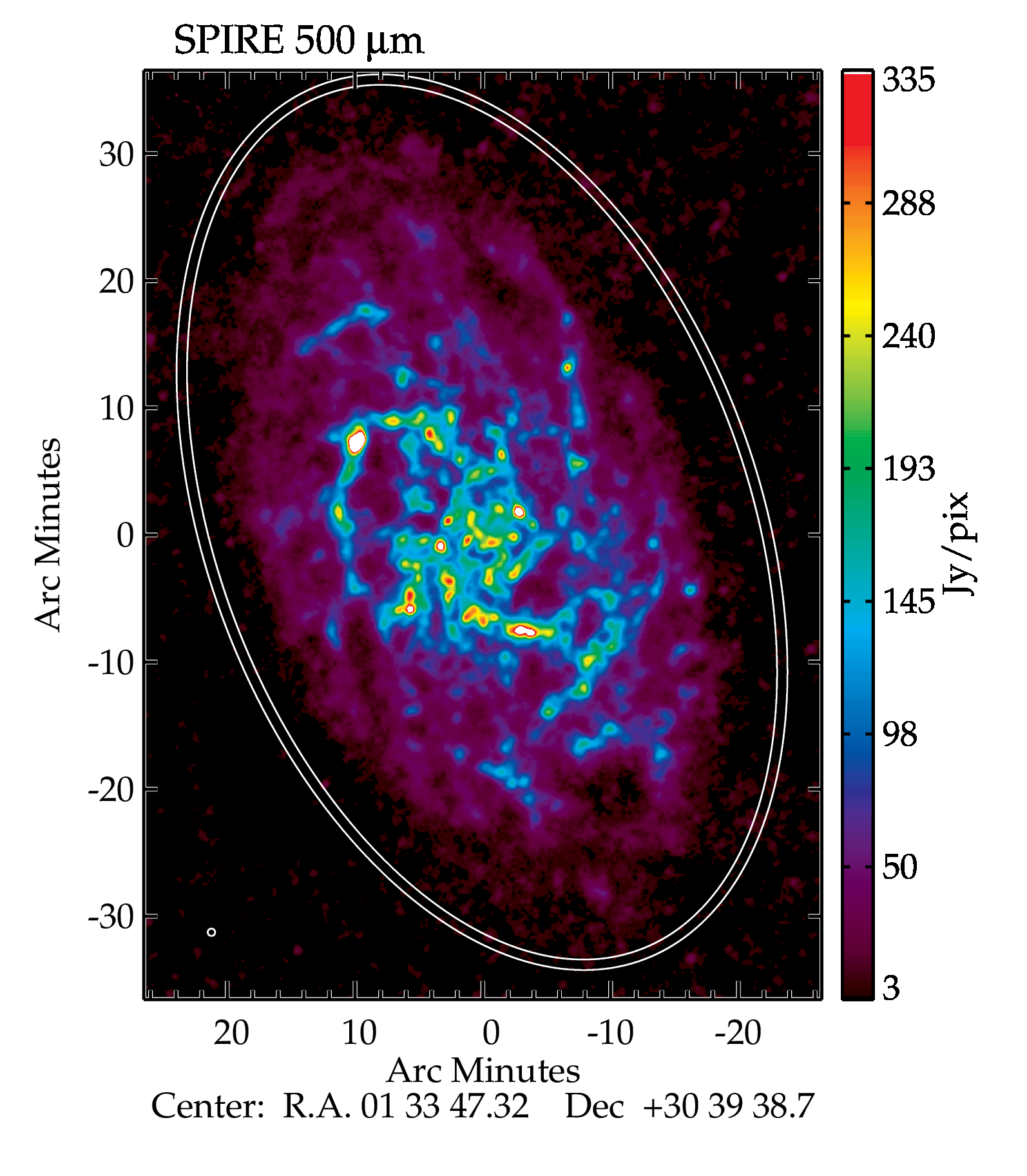}
           {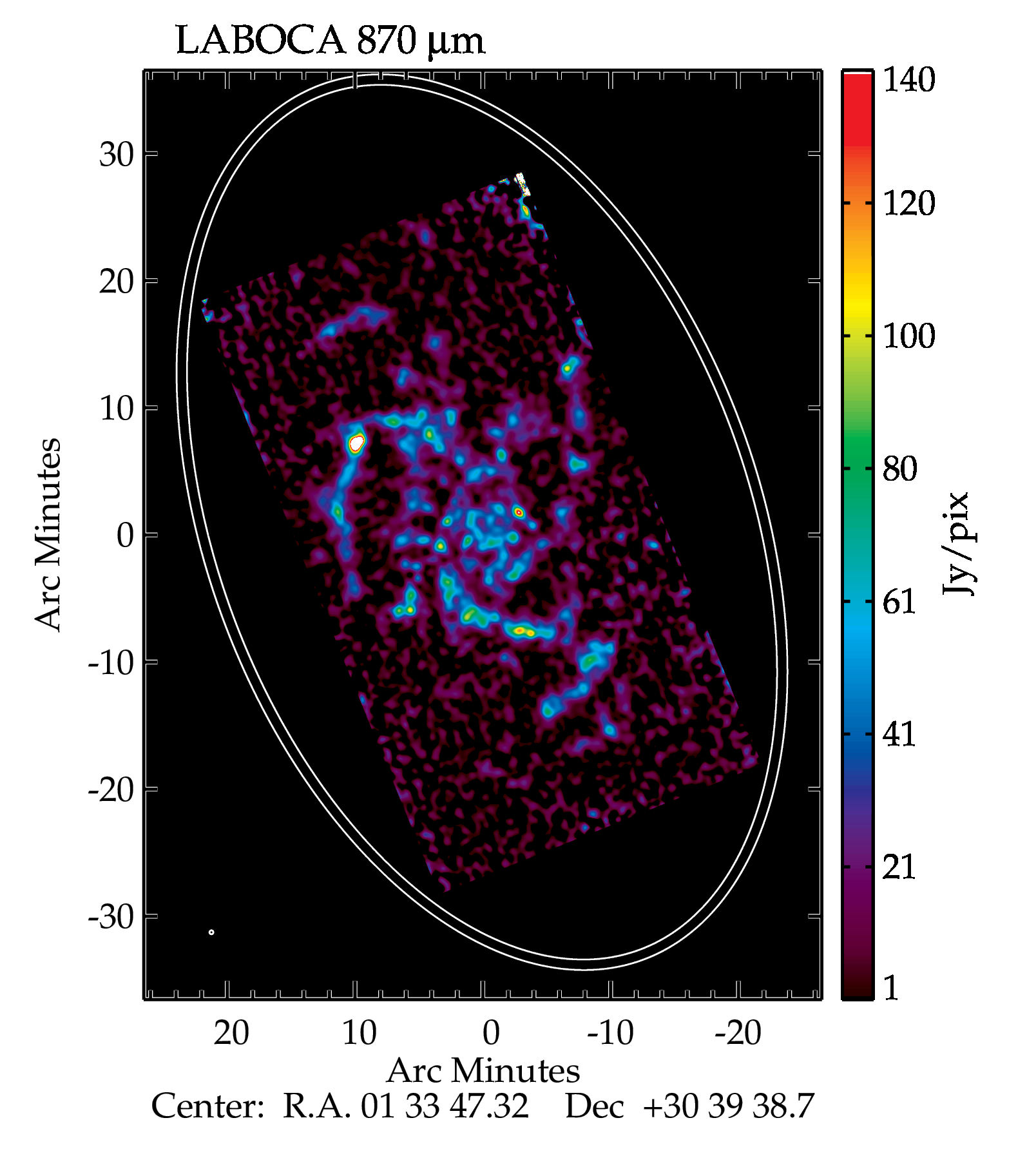}
           {fig:BANDSb}
           {Continued.}
           {\ContinuedFloat}
\ImageBands{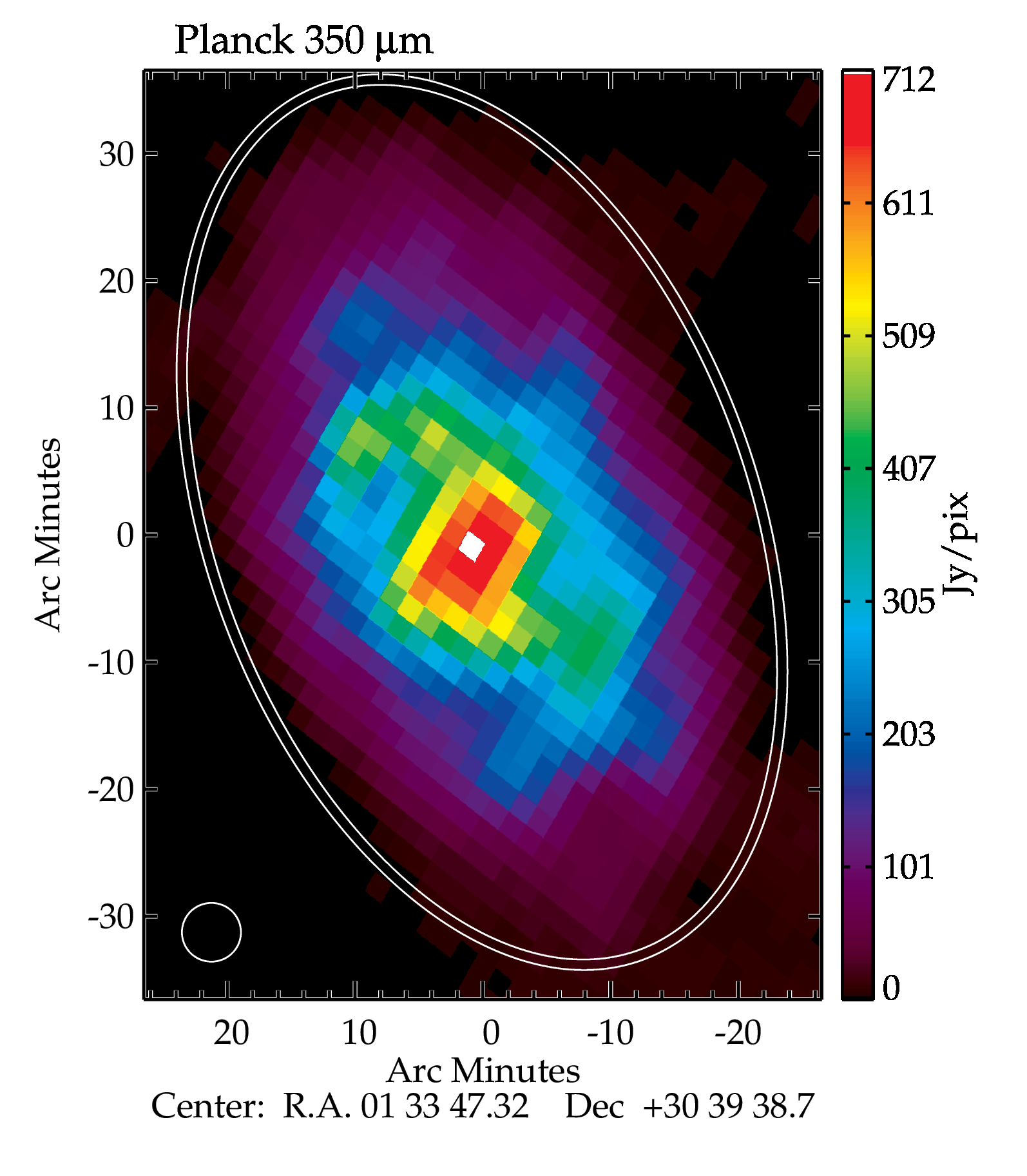}
           {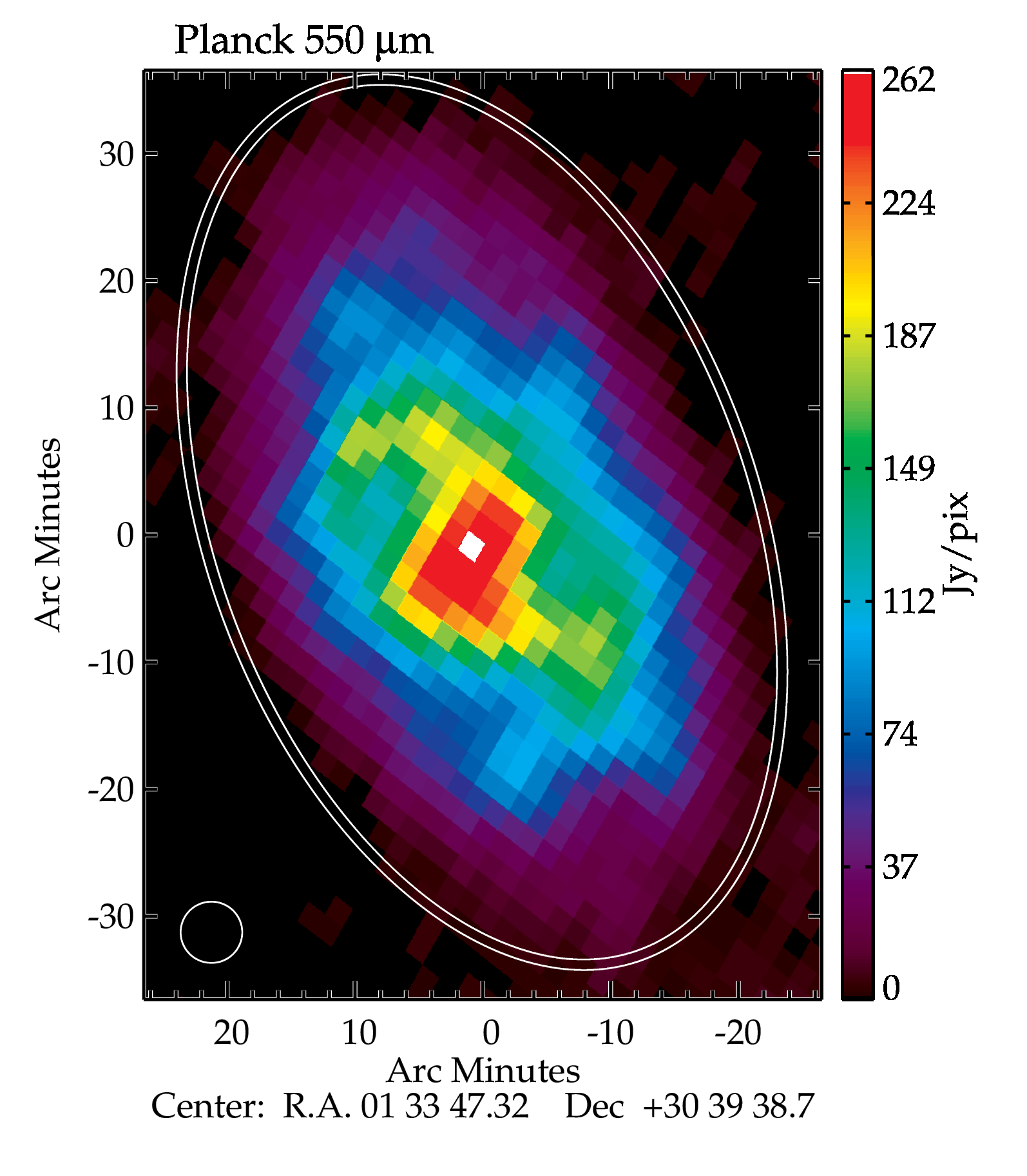}
           {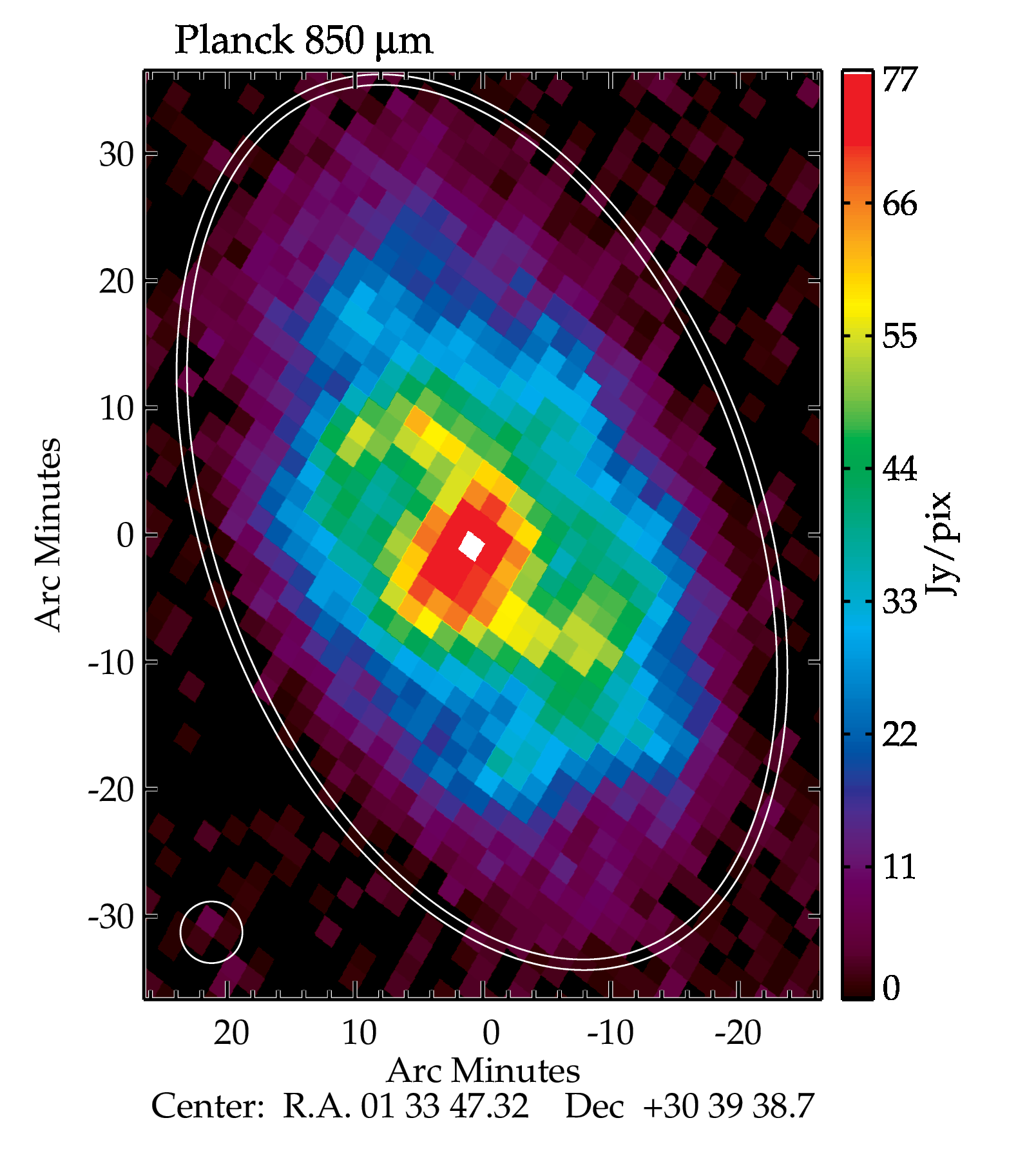}
           {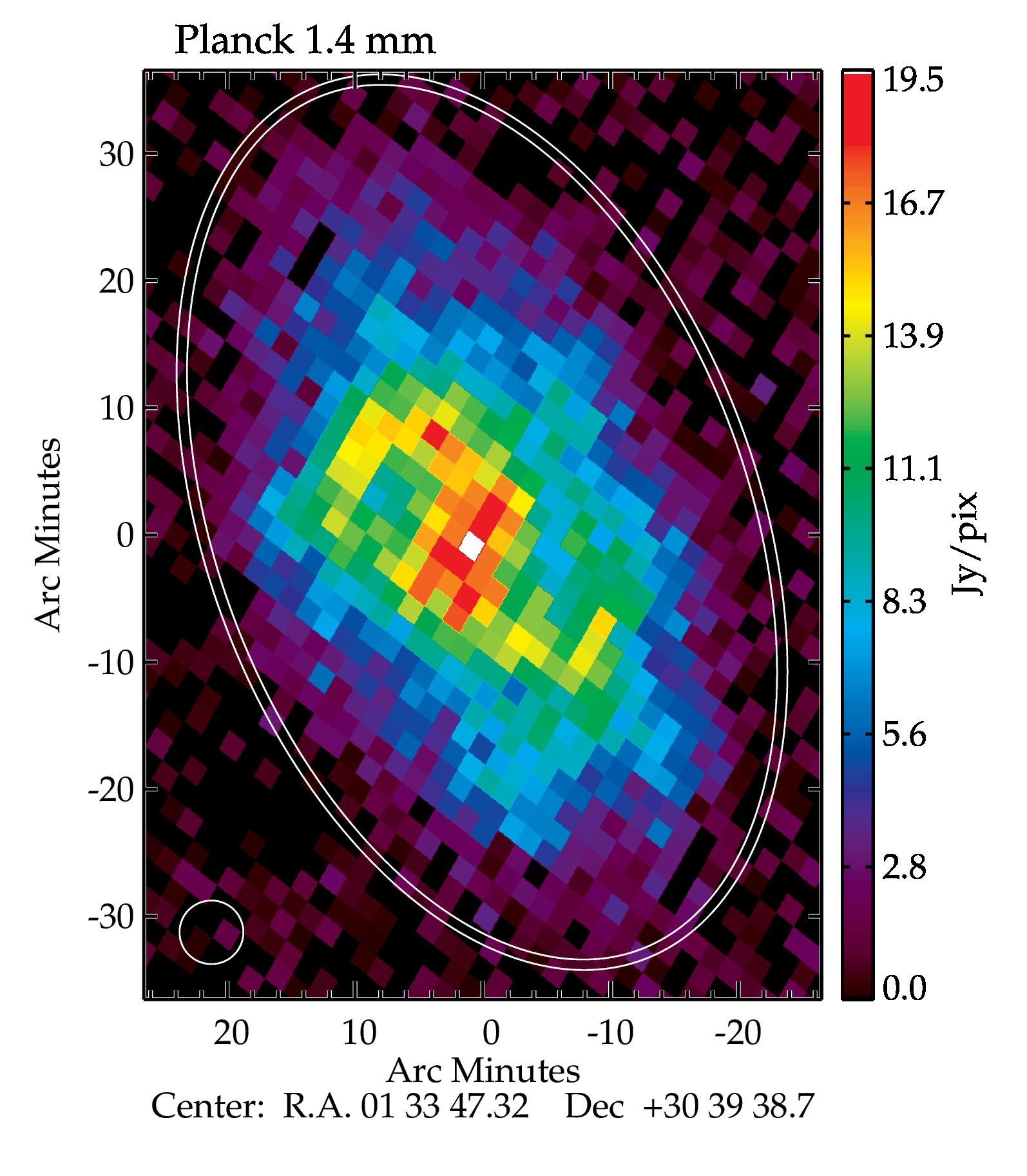}
           {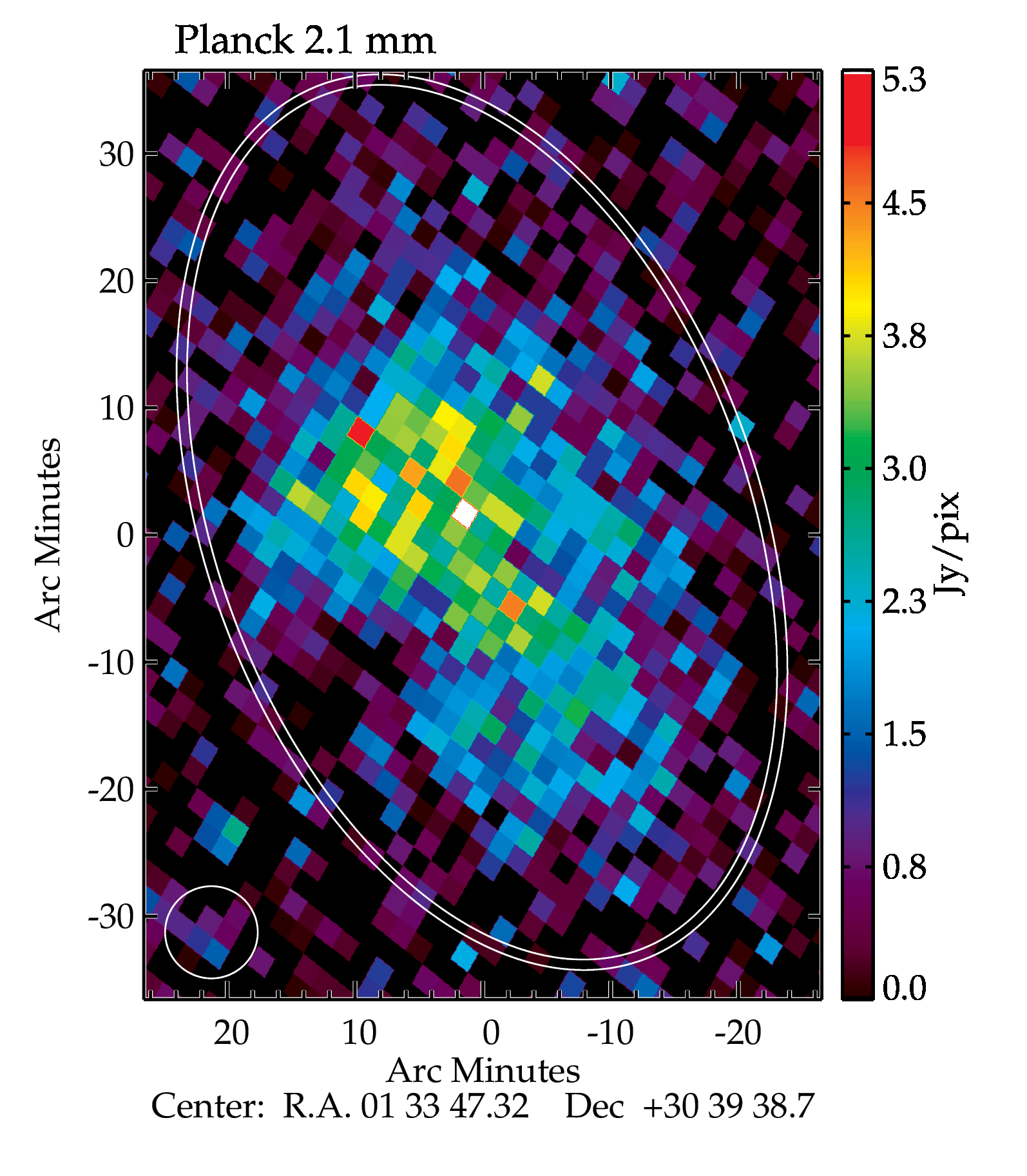}
           {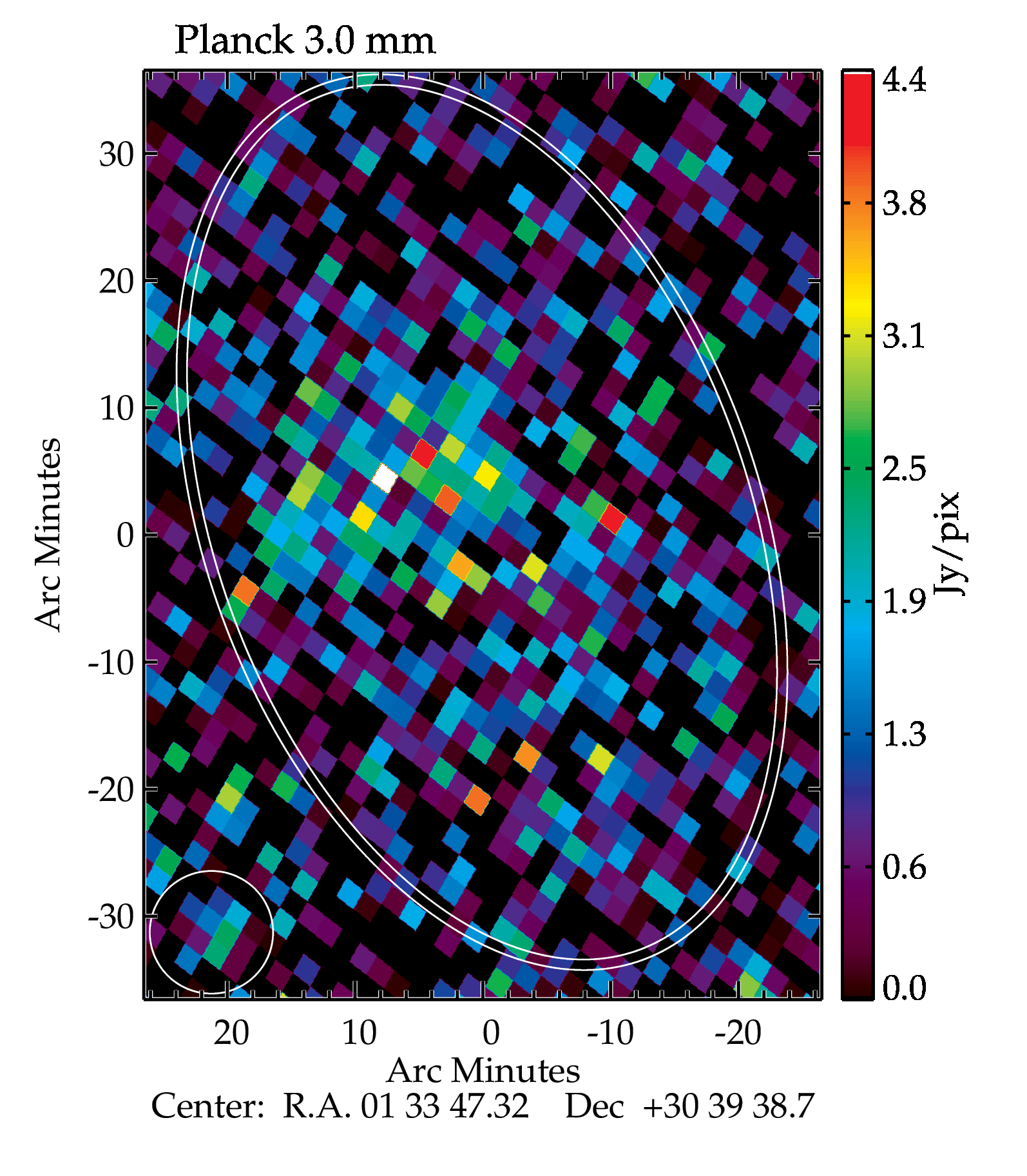}
           {fig:BANDSc}
           {Continued.}
           {\ContinuedFloat}
%
%
%

%% file: 99_TABLES/M33_DataSummary.tex
%
%
%
%

\begin{table*}[!htbp]
\centering
\vspace{26pt}
\begin{tabular}{l c c c c c l l} \hline \noalign{\medskip}
\textnormal{Telescope}               &  
\textnormal{Instrument}              &  
$\lambda_{\rm 0}$                    &  
$\nu_{\rm 0}$                        &  
\textnormal{FWHM}                    &  
$\Delta_{\rm CAL}$                   &  
\textnormal{Band name}               &  
\textnormal{References}              \\ 
\textnormal{}                        &  
\textnormal{or filter}               &  
\microns{}                           &  
\GHz{}                               &  
\textnormal{}                        &  
\%                                   &  
\textnormal{}                        &  
\textnormal{}                           
\\\noalign{\smallskip} \hline \noalign{\medskip}
%
%
\noalign{\smallskip}
\GALEX{}      & FUV                 &   0.154 & 1\,946\,704 &  \arcseconds{4}{2}    & 10 & \GALEX{FUV}                 & {\cite{2005ApJ...619L..67T}}    \\\noalign{\smallskip}
              & NUV                 &   0.232 & 1\,292\,209 &  \arcseconds{5}{3}    & 10 & \GALEX{NUV}                 & {\cite{2005ApJ...619L..67T}}    \\\noalign{\smallskip}\hline \noalign{\medskip}
\SLOAN{}      & \textit{u}          &   0.355 &    844\,486 & \arcseconds{1}{4}     &  2  & \SLOAN{u}                  & {This paper.}                   \\\noalign{\smallskip}
              & \textit{g}          &   0.469 &    639\,216 & \arcseconds{1}{4}     &  1  & \SLOAN{g}                  & {This paper.}                   \\\noalign{\smallskip}\hline \noalign{\medskip}
%
%
\textit{KPNO} & \Halpha             &   0.657 &    456\,305 & \arcseconds{6}{6}     & 15  & \Halpha                    & {\cite{1998PhDT........16G}}    \\\noalign{\smallskip}\hline \noalign{\medskip}
\SPITZER{}    & \CAMERAmu{IRAC}{}   &    3.55 &     84\,449 & \arcseconds{2}{5}     &  2  & \CAMERAmu{IRAC}{3.6}       & {\cite{2007A&A...476.1161V}}    \\\noalign{\smallskip}
              & \CAMERAmu{IRAC}{}   &    4.49 &     66\,769 & \arcseconds{2}{9}     &  2  & \CAMERAmu{IRAC}{4.5}       & {\cite{2007A&A...476.1161V}}    \\\noalign{\smallskip}
              & \CAMERAmu{MIPS}{}   &    23.7 &     12\,650 & \arcseconds{6}{3}     &  4  & \CAMERAmu{MIPS}{24}        & {This paper.}                   \\\noalign{\smallskip}\hline \noalign{\medskip}
%
%
\HERSCHEL{}   & \CAMERAmu{PACS}{}   &    70.0 &      4\,283 & \arcseconds{5}{6}     & 10  & \CAMERAmu{PACS}{70}        & {\cite{2015arXiv150201347B}}    \\\noalign{\smallskip}
              & \CAMERAmu{PACS}{}   &     100 &      2\,998 & \arcseconds{9}{8}     & 10  & \CAMERAmu{PACS}{100}       & {\cite{2011AJ....142..111B}}    \\\noalign{\smallskip}
              & \CAMERAmu{PACS}{}   &     160 &      1\,874 & \arcseconds{13}{6}    & 20  & \CAMERAmu{PACS}{160}       & {\cite{2011AJ....142..111B}}    \\\noalign{\smallskip}
              & \CAMERAmu{SPIRE}{}  &     250 &      1\,199 & \arcseconds{18}{1}    &  5  & \CAMERAmu{SPIRE}{250}      & {\cite{2012A&A...543A..74X}}    \\\noalign{\smallskip}
              & \CAMERAmu{SPIRE}{}  &     363 &         826 & \arcseconds{24}{9}    &  5  & \CAMERAmu{SPIRE}{350}      & {\cite{2012A&A...543A..74X}}    \\\noalign{\smallskip}
              & \CAMERAmu{SPIRE}{}  &     517 &         580 & \arcseconds{36}{4}    &  5  & \CAMERAmu{SPIRE}{500}      & {\cite{2012A&A...543A..74X}}    \\\noalign{\smallskip}\hline \noalign{\medskip}
\APEX{}       & \CAMERAmu{LABOCA}{} &     870 &         345 & \arcseconds{19}{2}    & 10  & \CAMERAmu{LABOCA}{870}     & Albrecht et al. (in prep.)      \\\noalign{\smallskip}\hline \noalign{\medskip}
\PLANCK{}     & \CAMERAmu{HFI}{}    &     350 &         857 & \arcminutes{4}{63}    & 10  & \CAMERAmu{PLANCK}{350}     & {This paper.}                   \\\noalign{\smallskip}
              & \CAMERAmu{HFI}{}    &     550 &         545 & \arcminutes{4}{84}    & 10  & \CAMERAmu{PLANCK}{550}     & {This paper.}                   \\\noalign{\smallskip}
              & \CAMERAmu{HFI}{}    &     850 &         353 & \arcminutes{4}{86}    &  3  & \CAMERAmu{PLANCK}{850}     & {This paper.}                   \\\noalign{\smallskip}
              & \CAMERAmm{HFI}{}    &  1\,380 &         217 & \arcminutes{5}{01}    &  3  & \CAMERAmm{PLANCK}{1.4}     & {This paper.}                   \\\noalign{\smallskip}
              & \CAMERAmm{HFI}{}    &  2\,096 &         143 & \arcminutes{7}{27}    &  3  & \CAMERAmm{PLANCK}{2.1}     & {This paper.}                   \\\noalign{\smallskip}
              & \CAMERAmm{HFI}{}    &  3\,000 &         100 & \arcminutes{9}{66}    &  3  & \CAMERAmm{PLANCK}{3.0}     & {This paper.}                   \\\noalign{\smallskip}
              & \CAMERAmm{LFI}{}    & 10\,000 &          30 & \arcminutes{32}{3}    &  3  & \CAMERAmm{PLANCK}{10}      & {This paper.}                   \\\noalign{\smallskip}\hline \noalign{\medskip}
\EFFELSBERG{} & \CAMERAmm{S}{36}    & 35\,903 &        8.35 & \arcminutes{1}{39}    &  *  & \CAMERAmm{S}{36}           & {\cite{2007A&A...472..785T}}    \\\noalign{\smallskip}\hline \noalign{\medskip}
%
%
\end{tabular}
\caption{%
Summary of the observations used in this study.
$^*$Only the total error is reported in \citet[][]{2007A&A...472..785T}.
\label{tab:observations}}
\end{table*}
%
%
%
%
%

%% file: 03_PHOTOMETRY/PhotometrySEX.tex
%
%
\section{Photometry \label{sec:Photometry}}
Prior to performing the photometry, we first homogenized all the images
to the same data unit in order to measure all fluxes in \Jy{}/pixel
(with a pixel size of \arcseconds{14}{}).
Bright foreground stars were removed from the UV and optical data.
We then convolved all the data to the resolution of \CAMERAmu{SPIRE}{500}
using the set of kernels provided by \citet[][]{2011PASP..123.1218A}%
\footnote{http://www.astro.princeton.edu/~ganiano/Kernels.html},
with the only exception of \PLANCK{} data, which have a coarser resolution
than the \CAMERAmu{SPIRE}{500} image, and therefore, they were only used
to obtain the total photometry of \MESSIER{33}.
Using the IRAF task \WREGISTER, we then regridded all the images
to the pixel size of the \CAMERAmu{SPIRE}{500} map.
Color corrections were applied to all the MIR/FIR/submm/mm fluxes
(see \App{app:CC} for details)
In this section we describe how we separated the different components,
how we corrected our measurements from undesired contributions
(e.g., \CO{}{} line emission),
and how the errors were handled.
\subsection{Component separation \label{subsec:ComponentSeparation}}
In order to decompose \MESSIER{33} into its SF and diffuse components,
we used the software \SEXTRACTOR\ \citep[][]{1996A&AS..117..393B}.
\SEXTRACTOR\ explores a given image for knots of emission by searching groups
of connected pixels that exceed some threshold above the background
in the surroundings of the region.
We run \SEXTRACTOR\ on the \Halpha\ image in order to separate
the SF from the diffuse component.
As the diffuse emission reaches a higher level in the inner part
than in the outskirts of the galaxy, the threshold should be relative
and vary to reflect these local variations.
The size of the background mesh should be large enough
to avoid a strong contamination from the SF but small enough to reproduce
the local scale variations of the diffuse emission.
We find that a mesh size of 12~pixels
(with a pixel size of \arcseconds{14}{} this corresponds to \pc{680})
yields an adequate separation between the detected SF objects
and background map accounting for the diffuse emission.
In practice, the decomposition between the SF and diffuse components
can be summarized in two main steps.
Firstly, the diffuse emission map is computed by estimating the background
(using a clipping to reject the bright SF regions)
in each cell of a grid that covers the whole image.
Secondly, SF regions are detected above a local threshold equal to
\THRESave\ times the background noise map obtained in the previous step.
The result of this separation method is shown in \Fig{fig:M33components}.%
\ImagePag{0.90}
         {./98_FIGURES/figDiffuseCompactT\THRESave.png}
         {fig:M33components}
         {%
          Example of the component separation method
          for the case of the \Halpha\ emission.
          The left panel is the original \Halpha\ image
          smoothed to the resolution of the \CAMERAmu{SPIRE}{500} band,
          the central panel shows the SF regions detected
          by \SEXTRACTOR\ using a threshold of \THRESave,
          and the right panel corresponds to the diffuse emission.
          The white contours are included to facilitate the comparison
          between panels and they correspond to the
          boundaries of the SF regions.
          The emission from the SF component represents \percent{\HIIperc}
          of the total \Halpha\ emission.
          }
We tested different local thresholds between 5 and 100 times
the background noise.
We found that for values below \THRESmin\ the SF regions detected
are significantly contaminated by diffuse emission.
On the other hand, values above \THRESmax\ produces the rejection of may of
the fainter SF regions.
A good compromise that leads to an optimal separation of the two components
was found for a threshold of \THRESave.
We adopted this intermediate value and we discuss in \App{app:THRESHOLDS}
the impact of our choice.
To replicate the same separation in the other bands, we first derive a map of
the diffuse emission in the same way as done for the \Halpha\ map.
We then obtain the difference between each original image and its diffuse emission
map to remove the background/foreground diffuse emission from the whole map.
Subsequently, we create a mask mirroring the detected SF objects in the
\Halpha\ image and applied it to the full set of multiwavelength images
to extract the exact same SF regions in all the bands,
without being contaminated by the diffuse emission.
We used the \Halpha\ image to create the mask because this is the band
that best represents the wavelength-dependent morphology of the SF regions.
Finally, for each band, the diffuse emission map is obtained by subtracting
the SF region map to the original map.
Once the SF and diffuse component are separated for each band
we determined the flux densities by integrating over the
corresponding maps (see \Tab{tab:M33_Photometry}).
We used an aperture for the
entire galaxy defined by a radius of \arcminutes{35}{} (\kpc{8.5})
and an inclination of \degrees{57} (see \Fig{fig:BANDSa}).
This aperture is slightly larger than the one used by
\citet[][]{2010A&A...518L..67K}
in order to enclose the whole emission of the \PLANCK{} bands,
which have a much poorer resolution than the data used by these authors.
In those cases where the angular resolution did not allow to extract
the emission from the SF regions
(i.e., \PLANCK{} data),
the diffuse emission was estimated by subtracting the flux of the SF component
predicted by the model of \citet[][]{2008ApJS..176..438G}
from the total flux of \MESSIER{33}.
Even though our estimates of the diffuse emission
are well constrained by this method in most wavelengths,
the lack of data points constraining the SED of the SF component
between \CAMERAmu{LABOCA}{870}
and the free-free emission at \cm{3.6} from \citet[][]{2007A&A...475..133T}
introduces some uncertainties.
As a cautionary measure, we  do not include the indirect
data points in the fitting procedure.
\begin{table*}[!htbp]
    \centering
    \input{97_IDL/SEDsexT\THRESave/TABLES/M33_Photometry}
    \caption{\label{tab:M33_Photometry}
             Flux densities of \MESSIER{33} measured for the different bands.
             The total emission ($S^{\rm obs}_{\rm TOT}$) is shown in column 2.
             Columns 3 and 4 correspond to the emission from
             the SF ($S^{\rm obs}_{\rm SF}$) and
             the diffuse ($S^{\rm obs}_{\rm TOT}$) component.
             Column 5 indicates the percentage of the total flux
             emitted by the SF component.
             }
\end{table*}
\subsection{CO contamination \label{sec:CO}}
The emission from the \CO{}{} rotational transitions
\CO{1}{0} at \mm{2.6} (\GHz{115}),
\CO{2}{1} at \mm{1.3} (\GHz{230}), and
\CO{3}{2} at \microns{870} (\GHz{345}),
fall into the bandpasses of the
\PLANCK{} \mm{3.0}, \mm{1.4}, and \microns{850} filters,
respectively \citep[][]{2014A&A...571A..13P}.
In addition, the \CAMERAmu{LABOCA}{870} filter is also affected
by the \CO{3}{2} line.
In order to correct the measured continuum fluxes for the contribution
of these major gas cooling lines,
we made use of \CO{2}{1} observations
\citep[][]{2014A&A...567A.118D}
obtained with the HEterodyne Receiver Array
\citep[HERA, ][]{2004A&A...423.1171S}
mounted on the IRAM~\meters{30} telescope on Pico Veleta (Spain).
\citet[][]{2014A&A...567A.118D} mapped the inner \kpc{\sim7} of \MESSIER{33}
with an angular resolution of \arcseconds{12}{}.
These authors reported a total \CO{2}{1} luminosity of
$2.8\times10^7\,\Kelvin{}\,\kms{}\,pc{}^2$,
which corresponds to a flux of \Sco{}{}{6.5\times10^4}.
For the \CAMERAmm{PLANCK}{1.4} filter,
which has a width of \GHz{64.5} (\kms{87\,433})
and a transmission close to \percent{100} at the frequency of the \CO{2}{1}
line \citep[see \Fig{}~1 in][]{2014A&A...571A..13P},
we measured a flux density of \Jy{17.4}.
Using these values we estimate that \percent{4.3}
of the flux density of \MESSIER{33} measured with this filter
comes from \CO{2}{1} line emission.
The \CO{1}{0} line emission was also observed with the
IRAM \meters{30} telescope in a radial cut along the major axis of \MESSIER{33}
\citep[][]{2010A&A...520A.107B}.
Comparing these data with their map, \citet[][]{2014A&A...567A.118D}
found a constant ratio $I_{\CO{1}{0}}/I_{\CO{2}{1}}$ of 1.25
in the overlapping area.
Assuming that this ratio can be extrapolated to the whole disk of \MESSIER{33},
we can estimate the flux of \CO{1}{0} as:
\begin{equation}
\label{eq:Sco10}
\Sco{1}{0}{} = 1.25 \,\times\, \left( \frac{115}{230} \right)^2 \Sco{2}{1}{} = \Sco{}{}{2.0\times10^4}.
\end{equation}
We measured a flux density of \Jy{1.56}
with the \CAMERAmm{PLANCK}{3.0} filter,
which has a width of \GHz{32.9} (\kms{97\,762})  and
a transmission of about \percent{60} at the frequency of the line.
Therefore, we estimate that the \CO{1}{0} emission is responsible for
\percent{7.9} of the flux measured with the \CAMERAmm{PLANCK}{3.0} band.
Finally, the contamination due to the emission from the \CO{3}{2} line
was estimated assuming a ratio $I_{\CO{3}{2}}/I_{\CO{2}{1}}\sim1$,
i.e.:
\begin{equation}
\label{eq:Sco32}
\Sco{3}{2}{} \sim \left( \frac{345}{230} \right)^2 \Sco{2}{1}{} = \Sco{}{}{14.6\times10^4}.
\end{equation}
With the \CAMERAmu{PLANCK}{850} filter we measured
a flux density of \Jy{74.3}.
Using a width of \GHz{101.4} (\kms{85\,032})
and a transmission \percent{\sim100},
we found that the contribution from the \CO{3}{2} line
to the \CAMERAmu{PLANCK}{850} is \percent{\sim2.3}.
In the case of the \CAMERAmu{LABOCA}{870} filter,
we measured a flux density of \Jy{37.1}.
For a width of \GHz{59.8} (\kms{52\,026})
and a transmission \percent{\sim100},
we found that the contribution from the \CO{3}{2} line
to this filter is (\percent{\sim7.6}).
\subsection{Synchrotron emission \label{sec:synchrotron}}
Unlike the free-free emission coming from the \HII\ regions,
the modeling applied in this work does not take into account the contribution
of the synchrotron emission.
For this reason, the fluxes measured with \PLANCK{} need to be decontaminated
from this component.
Following the analysis of \citet[][]{2007A&A...475..133T}
we extrapolated the synchrotron emission from \cm{3.6} to the \PLANCK{}
filters assuming a non-thermal spectral index of 0.72
(see \Sec{Data:EFFELSBERG}).
We found non-thermal contributions of \percent{30} and \percent{5}
respectively for the \CAMERAmm{PLANCK}{10} and \mm{3.0} bands,
while for the other bands this contribution was found to be negligible.
\subsection{Error of flux measurements \label{sec:error}}
In our error analysis we took two types of error into account:
i) calibration, $\Delta_{\rm cal}$, using the values presented in \Tab{tab:observations}, and
ii) measurement error due to background fluctuations, $\Delta_{\rm back}$.
The error due to the background fluctuations was calculated by assuming that
each pixel within the aperture has an error given by the standard deviation
of the background noise, $\sigma_{\rm back}$.
In addition, we have to take the error of the background into account
which was subtracted within an aperture of $N_{\rm apert}$ pixels.
This error is  $\sigma_{\rm back} N_{\rm apert} /\sqrt{N_{\rm back}}$, where
$N_{\rm back}$ is the number of pixels used to compute the level of background.
This gives \citep[see also][]{2012ApJ...745...95D} a total error
for the background subtracted flux of
\begin{equation}
  \Delta_{\rm back} =  \sigma_{\rm back}  \sqrt{N_{\rm apert} + \frac{N_{\rm apert}^2}{N_{\rm back}} }.
\end{equation}
The final error for the flux is the quadratic sum of
$\Delta_{\rm cal}$ and $\Delta_{\rm back}$.
In the case of the \PLANCK{} data, $\Delta_{\rm back}$ also includes
the error due to the CMB subtraction.
%
%
%

%% file: 97_IDL/SEDsexT40/TABLES/M33_Photometry.tex
%
%
\begin{tabular}{l c c c c c} \hline \noalign{\medskip}
\textbf{BAND} & $S^{\rm obs}_{\rm TOT}\rm\,(Jy)$ & $S^{\rm obs}_{\rm SF}\rm\,(Jy)$ & $S^{\rm obs}_{\rm DIFF}\rm\,(Jy)$ & $S^{\rm obs}_{\rm SF} / S^{\rm obs}_{\rm TOT}\rm\,(\%)$ \\\noalign{\smallskip} \hline \noalign{\medskip}
\GALEX{FUV}               &  2.40        $\pm$0.27          &                -                &               -                &       -      \\\noalign{\medskip}
\GALEX{NUV}               &  3.52        $\pm$0.39          &                -                &               -                &       -      \\\noalign{\medskip}
\SLOAN{u}                 &  5.93        $\pm$0.32          &                -                &               -                &       -      \\\noalign{\medskip}
\Halpha                   &  0.09        $\pm$0.02          &  0.04        $\pm$0.007         &                -                & 41.7         \\\noalign{\medskip}
\CAMERAmu{IRAC}{3.6}      &  23.7        $\pm$1.26          &                -                &               -                &       -      \\\noalign{\medskip}
\CAMERAmu{IRAC}{4.5}      &  15.3        $\pm$0.82          &                -                &               -                &       -      \\\noalign{\medskip}
\CAMERAmu{MIPS}{24}       &  47.6        $\pm$3.05          &  19.6        $\pm$1.25          &  29.2        $\pm$2.37          &  41.2         \\\noalign{\medskip}
\CAMERAmu{PACS}{70}       &  578         $\pm$64.7          &  223         $\pm$25.0          &  349         $\pm$42.8          &  38.7         \\\noalign{\medskip}
\CAMERAmu{PACS}{100}      &  1387        $\pm$155           &  360         $\pm$40.3          &  1026        $\pm$125           &  26.0         \\\noalign{\medskip}
\CAMERAmu{PACS}{160}      &  2162        $\pm$445           &  381         $\pm$78.6          &  1773        $\pm$376           &  17.6         \\\noalign{\medskip}
\CAMERAmu{SPIRE}{250}     &  1316        $\pm$86.8          &  197         $\pm$13.1          &  1111        $\pm$92.0          &  15.0         \\\noalign{\medskip}
\CAMERAmu{PLANCK}{350}    &  777         $\pm$89.5          &                -                & 684         $\pm$78.8          &  12.0         \\\noalign{\medskip}
\CAMERAmu{SPIRE}{350}     &  718         $\pm$47.4          &  90.5        $\pm$5.97          &  632         $\pm$52.3          &  12.6         \\\noalign{\medskip}
\CAMERAmu{SPIRE}{500}     &  319         $\pm$21.0          &  37.5        $\pm$2.47          &  279         $\pm$23.1          &  11.7         \\\noalign{\medskip}
\CAMERAmu{PLANCK}{550}    &  305         $\pm$35.1          &                -                & 279         $\pm$32.2          &  8.3          \\\noalign{\medskip}
\CAMERAmu{PLANCK}{850}    &  91.9        $\pm$6.91          &                -                & 85.7        $\pm$6.45          &  6.8          \\\noalign{\medskip}
\CAMERAmu{LABOCA}{870}    &                -                & 10.0        $\pm$1.12          &                -                &       -      \\\noalign{\medskip}
\CAMERAmm{PLANCK}{1.4}    &  21.7        $\pm$1.95          &                -                & 20.3        $\pm$1.83          &  6.5          \\\noalign{\medskip}
\CAMERAmm{PLANCK}{2.1}    &  4.27        $\pm$0.57          &                -                & 3.66        $\pm$0.49          &  14.2         \\\noalign{\medskip}
\CAMERAmm{PLANCK}{3.0}    &  1.86        $\pm$0.19          &                -                & 1.41        $\pm$0.14          &  24.2         \\\noalign{\medskip}
\CAMERAmm{PLANCK}{10}     &  0.40        $\pm$0.09          &                -                &               -                &       -      \\\noalign{\medskip}
\CAMERAmm{S}{36}          &  0.44        $\pm$0.05          &  0.44        $\pm$0.05          &                -                & 100.0        \\\noalign{\medskip}
\hline \\
\end{tabular}

%% file: 04_MODELS/Models.tex
%
%
\section{Models for the dust emission \label{sec:models}}
We analyzed the full UV to radio SED of the different emission components of the
galaxy using the radiation transfer model of \citet[][]{2011A&A...527A.109P},
which self-consistently treats the dust emission from the SF and the diffuse
components, considering the illumination of the diffuse dust both by the
distributed stellar populations and by the escaping light from the \HII\ regions.
While maintaining consistency within the framework of the
\citet[][]{2011A&A...527A.109P} model,
we used the model of \citet[][]{2008ApJS..176..438G}
to provide a detailed description of the dust emission from the SF component.
The methodology used in this work is similar to the one applied
to the low-metallicity, dwarf galaxy \NGC{4214} in \citet{2013A&A...549A..70H}.
For a detailed description of the models we refer the reader to
\citet[][]{2008ApJS..176..438G} and \citet[][]{2011A&A...527A.109P}.
To make this paper self-contained, in the following sections we present
a brief description of the physics and parameters of the models.
\subsection{The model of Popescu et al. (2011) \label{sec:DIFFmodels}}
\citet[][]{2011A&A...527A.109P} present a self-consistent model based on full
radiative transfer calculations of the propagation of starlight in disk galaxies.
These authors adopted the dust properties from \citet[][]{2001ApJ...548..296W}
and \citet[][]{2007ApJ...657..810D}, incorporating a mixture of graphite,
silicate, and PAH molecules.
To approximate the large-scale geometry of the galaxy
\citep[see \Fig{}~1 in][]{2011A&A...527A.109P},
they use two separate components:
\begin{itemize}
    \item An old component consisting of an old stellar disk,
          an old stellar bulge, and a thick disk of dust.
    \item A young component consisting of a young stellar disk
          and a thin disk of dust.
\end{itemize}
The spectral energy distribution of both old and young stellar components
are shown in \Tab{}~E.2 and in \Fig{}~8 of \citet[][]{2011A&A...527A.109P}.
These authors only considered the optical-IR radiation for the old stellar population,
and neglected any contribution in the UV.
The spectral distribution covers the wavelengths
from \angstroms{4\,430} to \angstroms{50\,000}.
The young stellar population is defined as an exponentially declining \SFR{}{}
with a time constant of \Gyr{5}, solar metallicity and Salpeter IMF
with an upper mas cut-off of \Msun{100}{}.
The spectral distribution covers from \angstroms{912} to \angstroms{50\,000}.
Apart from the diffuse component, the model includes a clumpy component,
consisting of the parent molecular clouds of young (\Myr{\leq10}) and
massive stars.
This component will be modeled here in a different way than
\citet[][]{2011A&A...527A.109P} have done (see \Sec{sec:HIImodels}),
as for \MESSIER{33} it is possible to separate the emission from the SF regions
and the diffuse dust.
The input parameters of the \citet[][]{2011A&A...527A.109P} model are:
\begin{itemize}
   \item The total central face-on B-band opacity, \TAUb{}{}.
   \item The star formation rate, \SFR{}{}.
   \item The clumpiness factor \CLUMP{}{}, which is linked to the fraction of
         photons that escape (\ESC{}{}) from the SF regions into the diffuse
         medium (\CLUMP{=}{1-\ESC{}{}}).
   \item The normalized luminosity of the old stellar disk, \OLD{}{}.
   \item The bulge-to-disk ratio, \BD{}{}.
   \item The radial scale length of the old stellar disk, \hs{}{}.
         All other spatial scales in the galaxy related to the different
         component in the model have a constant ratio with \hs{}{}
         \citep[see \Tab{}~E.1 in][]{2011A&A...527A.109P}.
   \item The inclination angle of the galaxy, \INCL{}{}.
\end{itemize}
From the primary parameters \SFR{}{} and \CLUMP{}{},
\citet[][]{2011A&A...527A.109P}
define the SFR powering the diffuse emission, \SFRp{}{},
as follows \citep[\Eq{}~45 in][]{2011A&A...527A.109P}:
\begin{equation}
 \SFRp{}{} = \SFR{}{} \times ( 1-\CLUMP{}{} ).
\end{equation}
The library of diffuse SEDs of
\citet[][]{2011A&A...527A.109P}
contains results for a four-dimensional parameter space spanned by
\TAUb{}{}, \SFRp{}{}, \OLD{}{}, and \BD{}{}.
The diffuse component is calculated as an extrinsic quantity corresponding
to a reference size (corresponding  to a reference scalelength).
To scale the intensity of the radiation field heating the diffuse dust,
the parameters \SFRp{}{} and \OLD{}{} must be scaled to the reference
size by comparing the scalelengths \citep[\Eq{}~D.3 in][]{2011A&A...527A.109P}:
\begin{equation}
 \SFRmod{}{} = \SFRp{}{} \times \left( \frac{\hr{}{}}{\hs{}{}} \right)^2,
\end{equation}
\begin{equation}
 \OLDmod{}{} = \OLD{}{} \times \left( \frac{\hr{}{}}{\hs{}{}} \right)^2,
\end{equation}
where \hr{=}{5670} is the reference B-band scalelength,
\hs{}{} is the B-band scalelength of the galaxy under study,
and \SFRmod{}{} and \OLDmod{}{} are internal parameters that allow us
to interface with the library of models.
An additional scaling is required to set the flux levels of the SEDs from
the library, \Snu{\!\!^{d,mod}}{}, to the SED that represents our galaxy,
\Snu{\!\!^d}{}:
\begin{equation}
\label{eq:SEDdiff}
\Snu{\!\!^d}{} = \left( \frac{\hs{}{}}{\hr{}{}} \right)^2 \times \Snu{\!\!^{d,mod}}{},
\end{equation}
where \Snu{\!\!^{d,mod}}{} is determined by the parameters
\TAUb{}{}, \SFRmod{}{}, \OLDmod{}{}, and \BD{}{}
\citep[][\Eq{}~D.2]{2011A&A...527A.109P}.
We would like to stress that in these models the absolute flux level of
the predicted dust SED is fixed by the input parameters.
In the model of \citet[][]{2011A&A...527A.109P},
the stellar radiation field leaving the galaxy, \Snu{^{\!\!s}}{},
can be calculated as:
\begin{equation}
   \label{eq:SEDstars}
   \Snu{^{\!\!s}}{} = \Snu{^{\!\!s,int}}{}\,10^{-\Catten{}{}/2.5},
\end{equation}
where \Snu{^{\!\!s,int}}{} is the intrinsic stellar radiation field,
and \Catten{}{} is the composite attenuation.
In the case of the UV and optical emission coming from
the young stellar population, \Catten{}{} can be written
as:
\begin{equation}
   \label{eq:Catten}
   \Catten{}{} \simeq -2.5 \log ( 1-\CLUMP{}{}\,f_{\lambda} ) + \Catten{^{\!\!tdisk}}{}.
\end{equation}
The first part of \Eq{eq:Catten} takes into account the attenuation
in the SF component, while \Catten{^{\!\!tdisk}}{}
is responsible for the attenuation in the diffuse component.
The wavelength dependence of the escape fraction, $f_{\lambda}$,
is tabulated in \citet[][\Tab{}~A.1]{2004A&A...419..821T}.
Therefore, by combining \Eq{eq:SEDstars} and \Eq{eq:Catten},
it is possible to calculate the radiation field associated with the
young stellar population that escapes the galaxy (\Snu{^{\!\!s,young}}{}).
The emission of the old stars that leaves the galaxy
(\Snu{^{\!\!s,old}}{}) is calculated in a similar way,
but in this case the composite attenuation
is \Catten{=}{}\Catten{^{\!\!tdisk}}{},
since this radiation field is only attenuated by the diffuse ISM.
It is noteworthy that in the \citet[][]{2011A&A...527A.109P} model
the old stellar population emits exclusively in the optical-IR wavelength range
(see their \Tab{}~E.2),
i.e., the emission corresponds to the radiation field emitted
by a remarkably old stellar population.
Thus, what \citet[][]{2011A&A...527A.109P} calls 'old' stellar population
does not correspond to the definition of 'old' stellar population
usually found in the literature
(see the \Fig{}~1 panel Scd of \citealt[]{2008MNRAS.386..697R}
for an example of a moderate old stellar population with
a SED extending into the optical-UV range).
As a consequence, in the \citet[][]{2011A&A...527A.109P} model
the old stellar population has little relevance for the dust heating.
\subsection{The model of Groves et al. (2008) \label{sec:HIImodels}}
In \citet[][]{2011A&A...527A.109P} the clumpy component corresponding
to the SF regions is modeled with an average template representing
the emission of all SF regions.
Although this is a good approach for unresolved galaxies,
the high spatial resolution of our data allowed us to apply
a detailed modeling of the SF component.
The \citet[][]{2008ApJS..176..438G} model describes the luminosity
evolution of a star cluster of mass \Mcl{}{}, and incorporates the
expansion of the \HII\ region and PDR due to the mechanical energy
input of stars and supernovae.
The dust emission from the \HII\ region and the surrounding PDR is calculated
from radiation transfer calculations.
The hydrogen column density of the PDR is fixed to the value
\ColDensity{=}{10^{22}}.
The \citet[][]{2008ApJS..176..438G} model assumes a standard mixture of dust
consisting of graphite, silicate, and PAHs, with standard grain properties from
\citet[][]{1993ApJ...402..441L},
\citet[][]{2001ApJ...554..778L}, and
\citet[][]{2001ApJ...548..296W}.
The input parameters of the model are:
\begin{itemize}
    \item The metallicity of the star cluster, \MET{}{},
          in units of the solar metallicity \Zsun{}{}
          \citep[][]{2005ASPC..336...25A}.
    \item The age of the star cluster, \AGE{}{}, in \Myr{}.
    \item The ambient pressure, expressed as
          \begin{equation}
             \PRS{}{}=\log\left(\frac{n_{\rm0}\,T_{\rm0}}{\cm{}^{-3}\,\Kelvin{}}\right),
          \end{equation}
          where $n_{\rm0}$ and $T_{\rm0}$ are respectively the number density
          and the temperature of the surrounding ISM.
    \item The compactness parameter, \CMP{}{}, which parametrizes the heating
          capacity of the star cluster and depends on \Mcl{}{} and \PRS{}{} as:
          \begin{equation}
             \label{eq:CMP}
             \CMP{}{}=\frac{3}{5}\log\left(\frac{\Mcl{}{}}{\Msun{}{}}\right) + \frac{2}{5}\,\PRS{}{}.
          \end{equation}
          The \CMP{}{} parameter determines the dust grain temperature distribution.
    \item The covering factor, \COV{}{}, which represents the fraction of the
          surface of the \HII\ region covered by the PDR.
          Note that \COV{}{} is the same parameter as the \CLUMP{}{} factor
          in the model of \citet[][]{2011A&A...527A.109P}.
          Hereafter the abbreviation \COV{}{} will be used to refer to both parameters.
\end{itemize}
\subsection{Observational constraints on the input parameters \label{sec:parameters} }
The large amount of ancillary data for \MESSIER{33},
as well as the results of previous studies from the literature,
allowed us to constrain part of the input parameters for both models.
For the model of \citet[][]{2011A&A...527A.109P}
we were able to place the following constraints:
\begin{itemize}
    \item \textit{Bulge-to-disk ratio:}
          Using deep H-band (\microns{1.65}) observations of \MESSIER{33},
          \citet[][]{1993ApJ...410L..79M} found an excess of emission in the
          innermost \arcminutes{2}{} that they identified with a small bulge.
          \citet[][]{1994ApJ...434..536R} claimed that if a compact bulge is
          present, it does not make a large contribution to the spheroidal light.
          More recently, \citet[][]{2007ApJ...669..315C} analyzed gas and stellar
          radial velocities in the innermost \kpc{0.5} of \MESSIER{33}.
          Their study showed that the central part of the galaxy does not
          exhibit kinematic signature of a significant bulge.
          Since there is no convincing evidence for a bulge in \MESSIER{33},
          here we opted for a bulge-to-disk ratio (\BD{}{}) equal to 0.
    \item \textit{Inclination angle:}
          We assumed an inclination angle of \INCL{}{57} based on
          \cm{21} line observations of \MESSIER{33} with the
          Westerbork Synthesis Radio Telescope
          \citep[][]{1987A&AS...67..509D}.
    \item \textit{B-band scalelength:}
          We obtained the radial stellar scale length in the B-band, \hs{}{},
          using the \SLOAN{g} image.
          The fit of an exponential function to the surface brightness profile
          gives a value of \hs{=}{2513\pm80} (see \Fig{fig:SCALElength}).
          This value agrees with the scalelength for young star formation
          tracers such as FUV, NUV and \Halpha\ \citep[][]{2009A&A...493..453V}.
    \item \textit{Dust scalelength:}
          In the model of \citet[][]{2011A&A...527A.109P} the scale of the different
          components are normalised to the B-band scalelength of the disk, \hs{}{}.
          In particular, the \citet[][]{2011A&A...527A.109P} model assumes that the scalelength
          of the thick disk of dust (\hd{}{}) is a factor 1.4 larger than \hs{}{}.
          To compare the actual geometry of \MESSIER{33} with the model assumptions
          we calculated the scalelength traced by the \CAMERAmu{SPIRE}{500} emission.
          Shorter wavelengths are more affected by the dust temperature distribution,
          while \PLANCK{} data at longer wavelengths lack the angular resolution
          necessary to extract isophotal fluxes.
          For \CAMERAmu{SPIRE}{500} we found a scalelength of \pc{3200\pm100}
          (see \Fig{fig:SCALElengthDUST}).
          If we approximate \hd{}{} with this value, we obtain $\hd{}{}/\hs{}{}=1.27\pm0.08$,
          close to the value 1.4 assumed in the model.
    \item \textit{old:}
          We derived this parameter as \OLD{}{0.05} by assuming that the
          \CAMERAmu{IRAC}{3.6} and \microns{4.5} bands mainly trace the
          emission from the old stellar population
          \citep[][]{2015MNRAS.448..135B}.
          Then we integrated the luminosity of \MESSIER{33} under these two
          bands and derived \OLD{}{} as the ratio of this luminosity and the
          integrated luminosity of the model galaxy for the old stellar component
          \citep[][\Tab{}~E2]{2011A&A...527A.109P}.
          \ImageCol{0.46}
                   {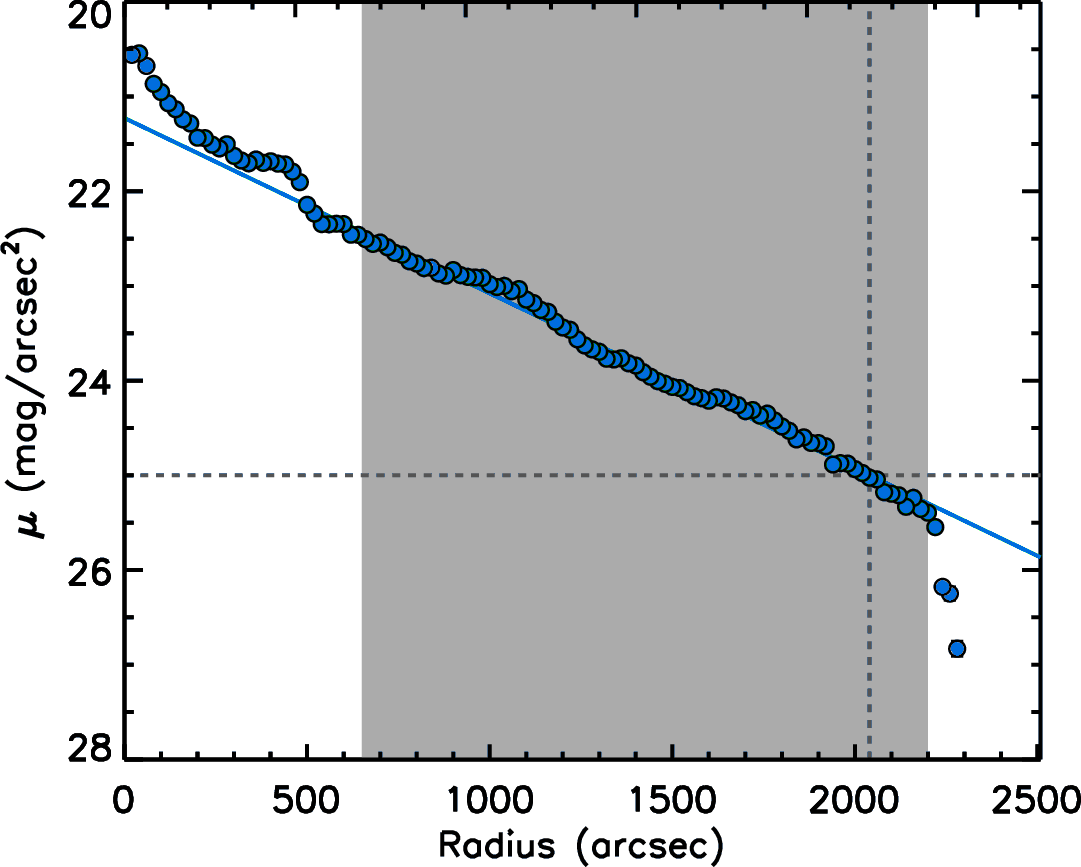}
                   {fig:SCALElength}
                   {B-band surface brightness of the stellar disk of \MESSIER{33}
                    plotted against deprojected radius.
                    The background subtracted \SLOAN{g} was used to obtain the
                    average value of circular apertures in the outer parts of \MESSIER{33} disk
                    where previously foreground stars and \HII\ regions were masked.
                    The fit (solid line) was achieved using the isophotes in the radial
                    range marked by the grey-shaded area.
                    We obtain a final scalelength of \pc{2513\pm80}.
                    The horizontal and vertical dashed gray lines indicate the 25
                    $\rm mag/arcsecond^2$ isophote.
                    The error bars are smaller than the data points.}
          \ImageCol{0.46}
                   {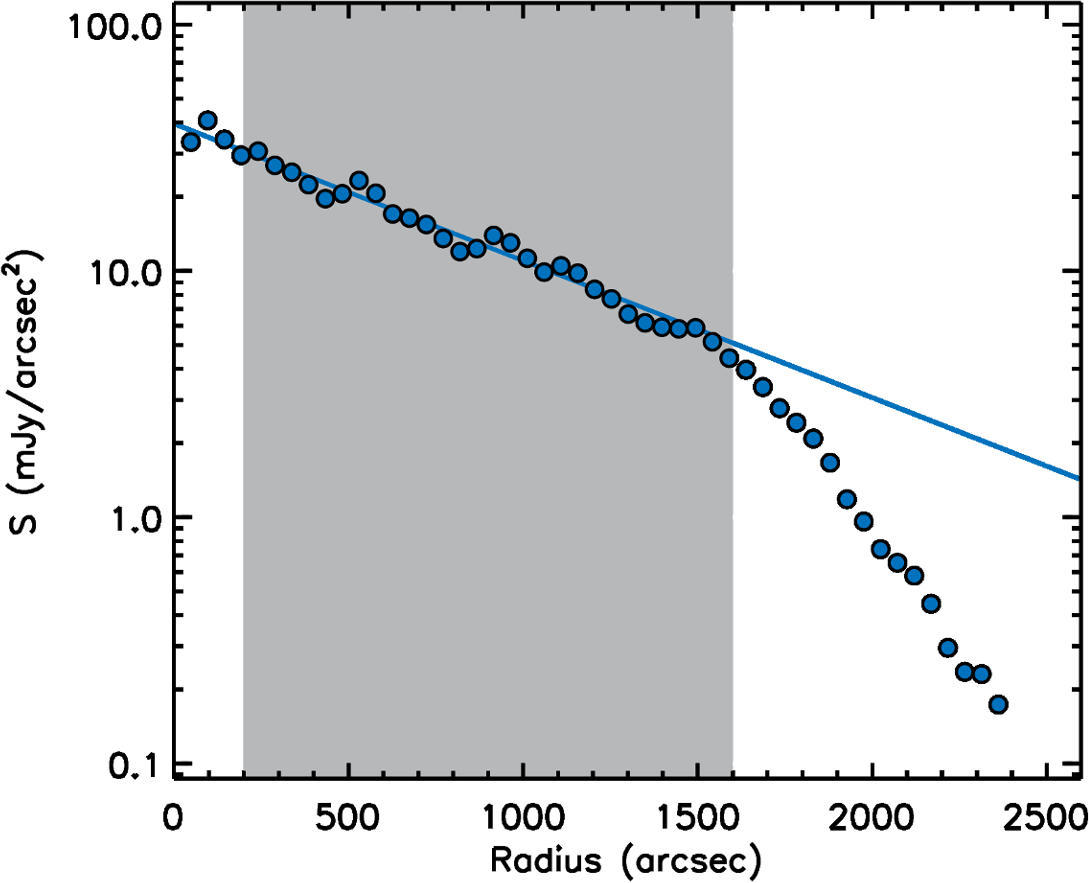}
                   {fig:SCALElengthDUST}
                   {\CAMERAmu{SPIRE}{500} surface brightness of \MESSIER{33}
                    plotted against deprojected radius.
                    The fit (solid line) was achieved using the isophotes in the
                    radial range marked by the grey-shaded area.
                    The error bars are smaller than the data points.}
\end{itemize}
In the case of the model of  \citet[][]{2008ApJS..176..438G},
we established the following constraints:
\begin{itemize}
   \item \textit{Metallicity:}
         \citet{2007A&A...470..865M} studied the radial oxygen abundance
         gradient in \MESSIER{33} using a list of 83 \HII\ regions with
         \forbline{O}{iii}{4363} detections.
         These authors found that 12+log[O/H] ranges from \about{8.5~to~8.1}
         (see their \Fig{}~3).
         For the solar metallicity presented in \citet[][]{2005ASPC..336...25A}
         these limits corresponds to values of \Zsun{} from \about{0.7~to~0.3}.
         To cover this range we constrained the metallicity to
         the discrete values \Zsun{1.0,~0.4,~and~0.2} allowed by the
         \citet[][]{2008ApJS..176..438G} model.
    \item \textit{Age:}
          A small sample of young stellar clusters was studied by
          \citet[][]{2010A&A...521A..41G} and
          \citet[][]{2014ApJS..211...22F},
          finding that most of the star clusters have ages of \Myr{\sim3}.
          A larger sample was studied by \citet[][]{2011A&A...534A..96S},
          who found representative values about \Myr{5} in the inner
          part of disk (galactocentric radii \kpc{\lesssim4})
          and about \Myr{7} for star clusters within
          the outer disk (from \kpc{4~to~7}).
          Therefore, we restricted the age to the range from \Myr{3~to~7}.
    \item \textit{Compactness and ambient pressure:}
          Unfortunately we could not constrain either \CMP{}{} or \PRS{}{},
          so we left both parameters free in the fitting procedure and checked
          later if they are within the expected range of values derived from
          \Eq{eq:CMP} and the mass range derived by
          \citet[][]{2011A&A...534A..96S}
          for the star clusters in the inner part of \MESSIER{33}
          (from \Msun{}{2} to \Msun{}{5}).
\end{itemize}
The remaining parameters were left free in the fitting procedure.
\subsection{Fitting procedure \label{sec:fitting} }
In order to find the best-fitting model and its uncertainty for both
the SF and the diffuse components we used the following procedure:
for each component we generated a total of 100
slightly different data sets by allowing the observed
fluxes to randomly vary following a Gaussian distribution
within the error bars.
For each one of these sets we found the best-fitting model
using a \chiSq{}{} minimization technique.
To build the best-fitting solution
(i.e. most probable SED and its range of uncertainty)
we extracted the median, the minimum, and the maximum SEDs
from the set of 100 best-fitting models.
To express the range of values obtained through the iteration process
we will use the nomenclature \Vlim{Xmin}{Xmed}{Xmax}
to report median (Xmed), maximum (Xmax), and minimum (Xmin) values.
%
%
%

%% file: 05_RESULTS/Results.tex
%
%
\section{Results \label{sec:results}}
\subsection{Best fit for the SF component \label{sec:fitHII}}
The best-fitting solution for the SF component was obtained by exploring
the parameter space of the model of \citet[][]{2008ApJS..176..438G}
with the constraints presented in \Tab{tab:M33_BestParameters}
(see \Sec{sec:parameters} for details).
For each one of the 100 iterations, a total of 6885 different parameter
combinations were tested to find the minimum \chiSq{}{}.
Note that \PLANCK{} data could not be used to fit the SF component
due to their lack of angular resolution.
The best fitting model is shown in \Fig{fig:fitHII},
and the range of values found for the parameters is presented
in \Tab{tab:M33_BestParameters}.
The reduced \chiSq{}{} is \Vlim{\CHIminHII}{\CHIaveHII}{\CHImaxHII}.
We found that the metallicity of the best-fitting solution is
\Vlim{\METmin}{\METave}{\METmax},
at the low end of the values reported by \citet{2007A&A...470..865M}.
\ImagePag{0.70}
         {97_IDL/SEDsexT\THRESave/PLOTS/HII+PDR.png}
         {fig:fitHII}
         {%
          \textit{Top:}
          Results of the fitting procedure for the SF component of \MESSIER{33}.
          The green filled circles are the observed fluxes for the SF component
          presented in \Tab{tab:M33_Photometry}.
          The best-fitting model is represented by the white dashed line while
          its uncertainty is defined by the gray shaded area.
          For clarity errors are only shown in the bottom panel.
          \textit{Bottom:}
          Residuals of the fitting procedure.
          The green filled circles corresponds to the percentage difference of the observed fluxes
          and the best-fitting model (white dashed line).
          The gray shaded area gives the uncertainty of the best-fitting model in terms of percentage.
         }
The best-fitting solution gives a total mass of the stellar clusters of
\Msun{\Vlim{\MCLmin}{\MCLave}{\MCLmax}}{6}.
Taking into account the number of SF regions detected we can estimate
an average value of about \Msun{5}{3} for the individual stellar clusters,
well centered within the limits estimated by \citet[][]{2011A&A...534A..96S}.
The dust luminosity associated to the SF regions is
\Lsun{\Vlim{\LUMminHII}{\LUMaveHII}{\LUMmaxHII}}{9}.
As \Fig{fig:fitHII} shows,
the model of \citet[][]{2008ApJS..176..438G}
is not able to reproduce the \CAMERAmu{LABOCA}{870} observations
(see \Tab{tab:M33_RatioObsModel} for details).
A complete discussion of this excess is given in \Sec{sec:discussion}.
\subsection{Best fit for the diffuse component \label{sec:fitDIFF}}
We searched the library of the diffuse dust SEDs of
\citet[][]{2011A&A...527A.109P}
for the best-fitting solution to the data in the MIR/submm range,
leaving \TAUb{}{} and \SFRp{}{} as free parameters and
keeping fixed \OLD{~to~}{\OLDave} and \BD{~to~}{\BDave}
(see \Tab{tab:M33_BestParameters} for details).
For each one of the 100 iterations we tested a total of 784 different
combinations of \TAUb{}{} and \SFRp{}{}.
The flux level was scaled using \Eq{eq:SEDdiff} and \hs{=}{\ScaleLength}.
As we mentioned in \Sec{sec:Photometry},
\PLANCK{} data were not used in the fitting procedure
since their diffuse fluxes were not directly measured but inferred,
while \CAMERAmu{LABOCA}{870} band was not used due to the observational
limitations discussed in \Sec{Data:LABOCA}.
The \CAMERAmm{S}{36} data point was neither used since
the thermal radio continuum is not related to the diffuse ISM
but to the ionized gas within the SF component.
The results are shown in \Fig{fig:fitDIFF} and \Tab{tab:M33_RatioObsModel}.
The reduced \chiSq{}{} is
\Vlim{\CHIminDIFF}{\CHIaveDIFF}{\CHImaxDIFF}
and the dust luminosity associated to the diffuse component is
\Lsun{\Vlim{\LUMminDIFF}{\LUMaveDIFF}{\LUMmaxDIFF}}{9}.
The model of \citet[][]{2011A&A...527A.109P}
systematically underpredicts the submm and mm estimated fluxes from \PLANCK{}
(see \Tab{tab:M33_RatioObsModel} for details).
This excess of emission is discussed in \Sec{sec:discussion}.
\ImagePag{0.70}
         {97_IDL/SEDsexT\THRESave/PLOTS/DIFFUSE.png}
         {fig:fitDIFF}
         {%
          \textit{Top:}
          Results of the fitting procedure for the diffuse component of \MESSIER{33}.
          The red filled circles are the observed fluxes for the diffuse component
          presented in \Tab{tab:M33_Photometry}.
          The best-fitting model is represented by the white dashed line while
          its uncertainty is defined by the gray shaded area.
          For clarity errors are only shown in the bottom panel.
          \textit{Bottom:}
          Residuals of the fitting procedure.
          The red filled circles corresponds to the percentage difference of the observed fluxes
          and the best-fitting model (white dashed line).
          The gray shaded area gives the uncertainty of the best-fitting model in terms of percentage.
         }
\input{./99_TABLES/M33_BestParameters}
\subsection{Total emission \label{sec:fitTOTAL}}
Once we separately fitted the emission from the SF and the diffuse component,
we combined both SEDs to match the global MIR to radio emission of the galaxy,
which is shown in \Fig{fig:fitDUST}.
At MIR/FIR wavelengths the observations are reasonably well fitted,
while at submm and mm wavelengths the excess of emission found in the
individual components translates into a significant excess in the global
SED of \MESSIER{33} (\Fig{fig:fitDUST}), showing up at submm wavelengths
and reaching compelling values at \mm{} wavelengths
(see \Tab{tab:M33_RatioObsModel}).
It is important to note that no traces of the excess are visible
at the \CAMERAmm{PLANCK}{10} point extracted from the
\PLANCK{} Catalog of Compact Sources (see \Sec{Data:Planck}).
From the values that we obtained for the radiation field leaving the
SF regions and powering the diffuse dust emission, \SFRp{}{},
and for the average fraction of the SF component covered by PDRs, \COV{}{},
we estimated the intrinsic \SFR{}{} of \MESSIER{33} as:
\begin{equation}
   \SFR{}{}=\frac{\SFRp{}{}}{1-\COV{}{}} =
   \SFR{}{\Vlim{\SFRminTotal}{\SFRaveTotal}{\SFRmaxTotal}},
\end{equation}
in good agreement with the value \SFR{}{0.45\pm0.10}
found by \citet[][]{2009A&A...493..453V} using
extinction corrected UV and \Halpha\ measurements.
\ImagePag{0.70}
         {97_IDL/SEDsexT\THRESave/PLOTS/TOTAL.png}
         {fig:fitDUST}
         {%
          \textit{Top:}
          Combination of the best-fitting models for the SF component
          (green dashed line, see \Fig{fig:fitHII} for details) and
          the diffuse component
          (red dashed-dotted line, see \Fig{fig:fitDIFF} for details).
          The blue filled circles are the observed fluxes for the entire galaxy
          presented in \Tab{tab:M33_Photometry}.
          The white dashed line and the gray shaded area correspond
          to the direct sum of the best-fitting models
          and their uncertainties obtained
          for the SF and the diffuse components.
          For clarity errors are only shown in the bottom panel.
          \textit{Bottom:}
          Residuals of the fitting procedure.
          The blue filled circles corresponds to the percentage difference
          of the observed fluxes and the best-fitting model (white dashed line).
          The gray shaded area gives the uncertainty
          of the best-fitting model in terms of percentage.
         }
\subsection{Energetic balance \label{sec:BALANCE}}
The comparison between the predicted radiation field escaping \MESSIER{33}
(see \Sec{sec:DIFFmodels} for details)
and the observations is shown in \Fig{fig:fitSTELLAR}.
The attenuation and its uncertainty were calculated using
\TAUb{=}{\Vlim{\TAUmin}{\TAUave}{\TAUmax}} and
\COV{=}{\Vlim{\COVmin}{\COVave}{\COVmax}}
as explained in \Sec{sec:DIFFmodels}.
The good agreement in the infrared part comes from the fact that the parameter
\OLD{}{} was scaled by hand to match \Snu{^{\!\!s,old}}{} to the observations
(see \Sec{sec:parameters}).
On the other hand,
in the UV-optical part we found that the integrated luminosity calculated from the
observations is \percent{\Vlim{\ESCminYOUNG}{\ESCaveYOUNG}{\ESCmaxYOUNG}}
higher than the expected luminosity.
This discrepancy is discussed in \Sec{sec:discussion}.
\ImageCol{0.40}
         {97_IDL/SEDsexT\THRESave/PLOTS/STELLAR.png}
         {fig:fitSTELLAR}
         {%
          \textit{Top:}
          Results of the fitting procedure for the stellar component of \MESSIER{33}.
          The purple and orange filled circles are respectively the observed fluxes of
          the young and the old stellar populations presented in \Tab{tab:M33_Photometry}.
          The white dashed line and the gray shaded area correspond to the combination
          of the attenuated emission and uncertainty of the young and old stellar populations
          (\Snu{^{\!\!s,young}}{} and \Snu{^{\!\!s,old}}{}, respectively).
          For clarity errors are only shown in the bottom panel.
          \textit{Bottom:}
          Residuals of the fitting procedure.
          The filled circles corresponds to the percentage difference of the observed fluxes
          and the best-fitting model (white dashed line).
          The gray shaded area gives the uncertainty of the best-fitting model in terms of percentage.
         }
\subsection{Gas-to-dust ratio \label{sec:GasToDust}}
In the model of \citet[][]{2008ApJS..176..438G}, the mass of dust contained
in the SF regions depends on the combination of parameters and it scales
with \Mcl{}{}.
For the SF component of \MESSIER{33} we obtained:
\begin{equation}
    M^{\rm SF}_{\rm dust} = \Msun{\Vlim{\MCDmin}{\MCDave}{\MCDmax}}{6}.
\end{equation}
The total mass of diffuse dust of \MESSIER{33} can be calculated from
the \Eq{}~44 of  \citet[][]{2011A&A...527A.109P}:
\begin{equation}
    M^{\rm diff}_{\rm dust} =
    0.992 \times \TAUb{}{} \times \hs{}{}^{2} \pc{}^{-2}\,\Msun{}{}=
    \Msun{\Vlim{\MDDmin}{\MDDave}{\MDDmax}}{6}.
\end{equation}
\citet[][]{2014A&A...567A.118D} estimated that the total molecular mass of
\MESSIER{33} is \Msun{0.3}{9} using a CO-to-H$_2$ conversion factor
\Xco{=\COtoH{}{}=}{4},
including helium.
\citet[][]{2010A&A...522A...3G} used Very Large Array (\VLA{})
observations of the atomic gas and derived a total mass of \Msun{1.4}{9}
within a \kpc{8.5} radius.
From all these values we can derive a gas-to-dust ratio of:
\begin{equation}
   \label{eq:Gdust}
   \Gdust{}{} =
   \frac{ M_{\rm atom} + M_{\rm mol} } { M^{\rm SF}_{\rm dust} + M^{\rm diff}_{\rm dust} } =
   \frac{\Msun{1.7}{9}}{\Msun{\Vlim{\MTDmin}{\MTDave}{\MTDmax}}{6}} =
   \Vlim{\GTDmin}{\GTDave}{\GTDmax},
\end{equation}
at the low end of the range determined by \citet[][]{2010A&A...518L..67K},
who found, using MBB models, that \Gdust{}{} ranges from 200 to 120 depending
on the value of \Eindex{}{}.
Given the solar value \Gdust{=}{137} \citep[][\Tab{}~2]{2007ApJ...663..866D},
and assuming that the gas-to-dust mass ratio scales linearly with metallicity
(i.e., the fraction of metals incorporated in the dust is constant),
due to the sub-solar metallicity of \MESSIER{33}
we would expect a higher value of \Gdust{}{} between 200 and 450
(see \Sec{sec:parameters} for details).
Possible origins for this discrepancy are discussed in \Sec{sec:discussion}.
\begin{table*}
    \centering
    \input{97_IDL/SEDsexT\THRESave/TABLES/M33_RatioObsModel}
    \caption{\label{tab:M33_RatioObsModel}
             Comparison of the observations and the models
             for the SF component (columns 2 and 3),
             the diffuse component (columns 4 and 5), and
             the total emission (columns 6 and 7) for \MESSIER{33}.
             In the three cases $\Delta$ gives the deviation
             of the observed data points from the model expectations,
             while $\delta^{\rm obs}$ and $\delta^{\rm mod}$ corresponds to the
             relative error of the observations and the models, respectively.
             }
\end{table*}'
%
%
%

%% file: 99_TABLES/M33_BestParameters.tex
\begin{table}[!htbp]
\centering
\begin{tabular}{l c c c c} \hline \noalign{\medskip}
Parameter              &  Abbr.        & Units        & Range     & Value                                \\\noalign{\smallskip}\hline\noalign{\medskip}
Metallicity            &  {\MET{}{}}   & {\Zsun{}}     & 0.2-1.0   & \Vlim{\METmin}{\METave}{\METmax}    \\\noalign{\medskip}
Age                    &  {\AGE{}{}}   & {\Myr{}}      & 3.0-7.0   & \Vlim{\AGEmin}{\AGEave}{\AGEmax}    \\\noalign{\medskip}
Compactness            &  {\CMP{}{}}   & {-}           & Free      & \Vlim{\CMPmin}{\CMPave}{\CMPmax}    \\\noalign{\medskip}
Ambient pressure       &  {\PRS{}{}}   & {-}           & Free      & \Vlim{\PRSmin}{\PRSave}{\PRSmax}    \\\noalign{\medskip}
Covering factor        &  {\COV{}{}}   & {\%}          & Free      & \Vlim{\COVmin}{\COVave}{\COVmax}    \\\noalign{\medskip}
Opacity                &  {\TAUb{}{}}  & {-}           & Free      & \Vlim{\TAUmin}{\TAUave}{\TAUmax}    \\\noalign{\medskip}
Star formation rate    &  {\SFRp{}{}}  & {\SFRp{}{~}}  & Free      & \Vlim{\SFRmin}{\SFRave}{\SFRmax}    \\\noalign{\medskip}
Old stellar luminosity &  {\OLD{}{}}   & {-}           & Fixed     & $\OLDave$                           \\\noalign{\medskip}
Bulge-to-disk ratio    &  {\BD{}{}}    & {-}           & Fixed     & $\BDave$                            \\\noalign{\medskip}
Inclination angle      &  {\INCL{}{}}  & {deg}         & Fixed     & $\Inclination$                      \\\noalign{\medskip}
B-band scalelength     &  {\hs{}{}}    & {\pc{}}       & Fixed     & $\ScaleLength$                      \\\noalign{\medskip}
\hline \\
\end{tabular}
\caption{%
         List of the parameters (column 1) with their
         abbreviations (column 2),
         units (column 3), and
         observational constraints (column 4)
         used to obtain the best-fitting solutions.
         The median, minimum, and maximum values derived from the fitting
         procedure are shown in column 5.
         \label{tab:M33_BestParameters}
         }
\end{table}

%% file: 97_IDL/SEDsexT40/TABLES/M33_RatioObsModel.tex
%
%
\begin{tabular}{l c c c c c c} \hline \noalign{\medskip}
\textbf{BAND} & $ \frac{ S^{\rm obs}_{\rm SF} } { S^{\rm mod}_{\rm SF} } $ & $ \frac{ \Delta_{\rm SF} } { \sqrt{ \left( \delta^{\rm obs}_{\rm SF} \right)^2 + \left( \delta^{\rm mod}_{\rm SF} \right)^2} } $ & $ \frac{ S^{\rm obs}_{\rm DIFF} } { S^{\rm mod}_{\rm DIFF} } $ & $ \frac{ \Delta_{\rm DIFF} } { \sqrt{ \left( \delta^{\rm obs}_{\rm DIFF} \right)^2 + \left( \delta^{\rm mod}_{\rm DIFF} \right)^2} } $ & $ \frac{ S^{\rm obs}_{\rm TOT} } { S^{\rm mod}_{\rm TOT} } $ & $ \frac{ \Delta_{\rm TOT} } { \sqrt{ \left( \delta^{\rm obs}_{\rm TOT} \right)^2 + \left( \delta^{\rm mod}_{\rm TOT} \right)^2} } $ \\ \noalign{\smallskip} \hline \noalign{\medskip}
\GALEX{FUV}               &             -           &   -   &             -           &   -   & \Vlim{2.64}{4.28}{197.88} & 5.22  \\\noalign{\medskip}
\GALEX{NUV}               &             -           &   -   &             -           &   -   & \Vlim{2.91}{3.98}{8.18} & 7.76  \\\noalign{\medskip}
\SLOAN{u}                 &             -           &   -   &             -           &   -   & \Vlim{1.35}{1.68}{2.35} & 2.70  \\\noalign{\medskip}
\CAMERAmu{MIPS}{24}       & \Vlim{0.89}{1.11}{1.35} & 0.44  & \Vlim{0.87}{1.02}{1.27} & 0.12  & \Vlim{0.86}{1.03}{1.27} & 0.17  \\\noalign{\medskip}
\CAMERAmu{PACS}{70}       & \Vlim{0.81}{0.98}{1.09} & -0.13 & \Vlim{0.88}{1.09}{1.47} & 0.33  & \Vlim{0.86}{1.05}{1.30} & 0.24  \\\noalign{\medskip}
\CAMERAmu{PACS}{100}      & \Vlim{0.81}{0.99}{1.13} & -0.03 & \Vlim{0.84}{0.98}{1.22} & -0.07 & \Vlim{0.83}{0.99}{1.20} & -0.07 \\\noalign{\medskip}
\CAMERAmu{PACS}{160}      & \Vlim{0.86}{1.01}{1.16} & 0.04  & \Vlim{0.88}{0.97}{1.10} & -0.12 & \Vlim{0.88}{0.98}{1.11} & -0.08 \\\noalign{\medskip}
\CAMERAmu{SPIRE}{250}     & \Vlim{0.84}{0.99}{1.17} & -0.04 & \Vlim{0.80}{0.88}{1.01} & -0.78 & \Vlim{0.81}{0.90}{1.04} & -0.68 \\\noalign{\medskip}
\CAMERAmu{PLANCK}{350}    &             -           &   -   & \Vlim{0.87}{1.01}{1.19} & 0.03  & \Vlim{0.87}{1.01}{1.19} & 0.03  \\\noalign{\medskip}
\CAMERAmu{SPIRE}{350}     & \Vlim{0.91}{1.08}{1.32} & 0.42  & \Vlim{0.88}{1.02}{1.21} & 0.11  & \Vlim{0.88}{1.02}{1.21} & 0.12  \\\noalign{\medskip}
\CAMERAmu{SPIRE}{500}     & \Vlim{1.05}{1.25}{1.57} & 1.26  & \Vlim{0.95}{1.13}{1.37} & 0.63  & \Vlim{0.97}{1.15}{1.40} & 0.74  \\\noalign{\medskip}
\CAMERAmu{PLANCK}{550}    &             -           &   -   & \Vlim{1.11}{1.32}{1.61} & 1.41  & \Vlim{1.08}{1.29}{1.57} & 1.27  \\\noalign{\medskip}
\CAMERAmu{PLANCK}{850}    &             -           &   -   & \Vlim{1.29}{1.58}{1.95} & 2.45  & \Vlim{1.24}{1.52}{1.88} & 2.23  \\\noalign{\medskip}
\CAMERAmu{LABOCA}{870}    & \Vlim{1.47}{1.75}{2.23} & 3.43  &             -           &   -   &             -           &   -   \\\noalign{\medskip}
\CAMERAmm{PLANCK}{1.4}    &             -           &   -   & \Vlim{1.65}{2.04}{2.55} & 4.07  & \Vlim{1.55}{1.91}{2.39} & 3.66  \\\noalign{\medskip}
\CAMERAmm{PLANCK}{2.1}    &             -           &   -   & \Vlim{1.55}{1.93}{2.43} & 3.29  & \Vlim{1.41}{1.73}{2.15} & 2.74  \\\noalign{\medskip}
\CAMERAmm{PLANCK}{3.0}    &             -           &   -   & \Vlim{2.27}{2.84}{3.58} & 6.78  & \Vlim{1.64}{1.98}{2.44} & 4.24  \\\noalign{\medskip}
\CAMERAmm{PLANCK}{10}     &             -           &   -   &             -           &   -   & \Vlim{0.74}{0.86}{1.06} & -0.48 \\\noalign{\medskip}
\CAMERAmm{S}{36}          & \Vlim{0.70}{0.81}{0.99} & -0.86 &             -           &   -   & \Vlim{0.70}{0.81}{0.99} & -0.86 \\\noalign{\medskip}
\hline \\
\end{tabular}

%% file: 06_DISCUSSION/Discussion.tex
%
%
\section{Discussion \label{sec:discussion}}
Our study has revealed three compelling discrepancies between the models and
the observations:
excess of emission at submm and mm wavelengths (\Fig{fig:fitDUST}),
deficit of absorption of UV photons (\Fig{fig:fitSTELLAR}), and
abnormally low value of \Gdust{}{} (\Sec{sec:GasToDust}).
In the following, we discuss in detail different hypothesis that could explain
these discrepancies.
\subsection{Cold dust component}
A population of very cold dust (VCD) with temperatures below (\Kelvin{10})
has been proposed to explain the excess of emission at submm wavelengths
\citep[e.g.][]{2003A&A...407..159G,
               2005A&A...434..867G,
               2009A&A...508..645G,
               2010A&A...518L..55G}.
Although in general the VCD component is able to fit well the submm excess,
this hypotheses is considered unlikely because
the large mass of dust required and
the strong shielding from the ambient radiation field
necessary to keep the grains at such low temperatures
\citep[][]{2002A&A...382..860L}.
In this line, \citet[][]{2011A&A...536A..88G} found that the submm excess
in the LMC can not be produced by a cold dust component.
To explore the VCD hypothesis in the case of \MESSIER{33} we added
to its total SED a MBB component with \Eindex{=}{2}
(see top panel of \Fig{fig:EXCESS}).
We found that the submm and mm excess can be well fitted using
a temperature of \Kelvin{\sim6}.
The mass of dust required to produce such emission is:
\begin{equation}
    M_{\rm VCD} \sim \Msun{40}{6},
\end{equation}
which is about three times higher than the  combined mass of dust
found in the SF and the diffuse components.
Adding this extra mass of dust would imply an unreasonably low value
of \Gdust{\lesssim}{50}.
For reference, note that \citet[][]{2014MNRAS.444L..90B} found a value
\Gdust{<}{14.5} for the extreme case of \NGC{5485},
an early type galaxy with no atomic or molecular gas detected
but with a prominent dust lane perpendicular to the photometric major axis.
\ImageCol{0.47}
         {97_IDL/SEDsexT\THRESave/PLOTS/EXCESS.png}
         {fig:EXCESS}
         {%
          Total SED of \MESSIER{33}
          (see \Fig{fig:fitDUST} for a detailed description)
          combined with an extra component (purple line-dashed)
          to account for the emission from
          very cold dust (top panel),
          spinning dust grains (mid panel),
          and magnetic nanoparticles (bottom panel).
         }
\subsection{Spinning grains}
\citet[][]{1998ApJ...508..157D} proposed that electric dipole radiation
from spinning grains are responsible for the anomalous emission of the
diffuse Galactic background peaking near \mm{10}.
In the \citet[][]{1998ApJ...508..157D} model the very small grains are
disk-like shaped while large grains are spherical,
and both rotate around their axis of major inertia,
which produces that the frequency of photons emitted is
identical to the angular frequency.
According to these authors, the rotational excitation of the spinning
grains is dominated by collisions with ions and plasma drag
and the rotation rate depends on several factors such as the
intensity of the radiation field and the physical parameters of the gas phase.
For example, while the peak of the emission from spinning grains  ranges from
\mm{{\sim}7~to~15} in the cold neutral, warm neutral and warm ionized medium,
it can reach values below \mm{2} in PDRs
(see \citealt[][]{2010ApJ...715.1462H} for a refined version of the model of
\citealt[][]{1998ApJ...508..157D}).%
To study the possibility of spinning dust grains as responsible for
the excess of emission detected in \MESSIER{33}
we included their contribution in the total SED.
As the mid panel of \Fig{fig:EXCESS} shows,
the excess of emission at wavelengths \mm{\lesssim2} can not be reproduced
even if we assume spinning grains rotating at the highest rate
allowed by \citet[][]{2010ApJ...715.1462H}.
Moreover, the contribution from spinning grains degrade the good agreement
previously found between models and observations at wavelengths around \mm{10}.
\subsection{Magnetic nanoparticles}
\citet[][]{2013ApJ...765..159D} claimed that if part of the interstellar Fe
forms ferro- or ferri-magnetic grains, then the magnetic dipole radiation
from these grains might contribute significantly to the total dust SED.
These authors considered three materials as potential candidates:
metallic Fe,
magnetite (Fe$_3$O$_4$), and
maghemite ($\gamma$-Fe$_2$O$_3$).
\citet[][]{2012ApJ...757..103D} show how magnetic grains are able
to explain the strong excess of submm and mm emission found in the SMC.
In order to test if magnetic nanoparticles can explain
the excess of emission found in \MESSIER{33}
we fitted the SED expected for \nm{10} metallic Fe grains at \Kelvin{18}
(see right panel in \Fig{}~8 of \citealt[][]{2013ApJ...765..159D})
by adjusting the intensity until the best fit to the FIR-mm
data points was achieved,
as it is shown in the bottom panel of \Fig{fig:EXCESS}.
The contribution from magnetic grains can alleviate the
mm and submm excess but is not able to satisfactory fit
the data in this wavelength range.
In addition, the fit at the \CAMERAmm{PLANCK}{10}.
We also tested the possibility of \nm{100} metallic Fe grains at \Kelvin{18}.
The fit is worse than for the \nm{10} grains
in the submm/mm/\CAMERAmm{PLANCK}{10} range.
In addition, they have a strong impact on the peak of the SED
thus degrading the fit at wavelengths \microns{\sim150}.
Thus, for \MESSIER{33}, in contrast to the SMC,
magnetic grain emission is not an entirely satisfactory explanation
for the mm and submm excess.
\subsection{\texorpdfstring{CO-to-H$_2$}\ ~conversion factor}
The low value of \Gdust{}{} obtained here depends in part on the
mass of molecular gas obtained by \citet[][]{2014A&A...567A.118D},
who determined a conversion factor \Xco{=}{4} for \MESSIER{33},
about twice higher than the Galactic value which is reasonable
for the $\sim2$ times lower metallicity.
A value of \Xco{\gtrsim}{20} would provide enough molecular gas
to achieve a \Gdust{}{} in the expected range of \Gdust{\gtrsim}{200}.
However, this is a unreasonably high value for the case of \MESSIER{33},
where combining CO, \HI, and IR measurements,
\citet[][]{2011ApJ...737...12L} did not find evidence
of such deviations in the value of \Xco{}{}.
\subsection{Porosity}
A porous ISM would allow a higher fraction of the UV radiation to
escape the galaxy without interacting with the dust
thus alleviating the UV discrepancy.
Several theoretical studies try to explain how the internal
properties of galaxies affect the escape fraction.
Among the properties that can influence the escape fraction are:
the covering factor of the clumps in the ISM,
as well as the density of the clumped and inter clump medium
\citep[][]{2011ApJ...731...20F},
blow-out of shells created by supernova (SN) remnants
that provide the photons with a clear path to escape the galaxy
\citep[][]{1994ApJ...430..222D},
and the porosity of the ISM given as the fraction of the ISM that is
devoid of \HI\ owing to the expansion of SN-driven bubbles.
Therefore, the UV discrepancy derived from our modeling might be due
to large shell structures across the whole galactic disk
\citep[][]{2013A&A...552A.140R}.
\subsection{Dust heated by evolved stars}
Could an unrealistic SED of the stellar populations that are heating the dust
be responsible for the mm and submm excess and/or the UV discrepancy?
\citet[][]{2011A&A...527A.109P} discussed in detail the influence
of the spectral shape of the stellar radiation on the dust SED
(see their \Sec{}~4.3).
They concluded that the stellar SED does not affect significantly
the shape of the dust SED and that the most relevant parameters
shaping the dust SED are the overall stellar luminosities
as well as the geometry and the properties of the dust.
Therefore, the exact shape of the stellar SED is unlikely to be
a very important source of error in our study.
In order to quantify in \MESSIER{33} the contribution to the
dust heating by the evolved stellar population,
we set \SFR{}{} to 0 and checked the amplitude of the diffuse SED keeping the
remaining values the same as in our best-fitting solution.
We found that this contribution is indeed negligible with the exception of the
wavelengths around the peak, where it contributes slightly to the dust emission.
It is noteworthy that this result is at first sight in contradiction
with the results of \citet[][]{2011AJ....142..111B},
who found a tight correlation between the
\microns{250{/}350} ratio and the \microns{3.6} luminosity ($L_\microns{3.6}$)
at scales of \arcseconds{42}{} (\pc{\sim170})
that they interpret as an indication of the old stars
being the main heating source of the dust emitting in the
\microns{250{-}350} wavelength range.
This apparent discrepancy between our study and
\citet[][]{2011AJ....142..111B} is not real but
it comes from different definitions of what is considered
``old'' stellar population
(see our \Sec{sec:DIFFmodels} for details).
\subsection{Geometry}
The discrepancies revealed by our study might be explained by differences
between the actual geometry of \MESSIER{33} and the geometry assumed by
the model of \citet[][\Tab{}~E.1]{2011A&A...527A.109P}.
As we mentioned in \Sec{sec:parameters}, there is no evidence
of any discrepancy regarding the radial dimensions of the disk of dust.
Regarding the vertical dimensions, for a B-band scalelength of \hs{}{2513},
the model of \citet[][]{2011A&A...527A.109P} predicts scaleheights of
\pc{186} for the old stellar disk, \pc{121} for the thick disk of dust,
and \pc{40} for both the young stellar disk and the thin disk of dust.
Using the break of the power-spectrum in different emission bands,
\citet[][]{2012A&A...539A..67C} were able to give an estimation of the
thickness of the stellar and dust disks of \MESSIER{33}:
cold dust lies in a \pc{350} disk, warm dust lies in a \pc{100} disk,
and the thickness of the stellar disk in the FUV and NUV bands
was found to be \pc{40} and \pc{56}, respectively.
Unfortunately, the conversion of the breaks found by
\citet[][]{2012A&A...539A..67C} to scaleheights is not straightforward
since it depends on the conditions in the disk as the simulations
of \citet[][\Tab{}~4]{2012A&A...539A..67C} have shown.
Therefore, we can not exclude that the discrepancies found in our study
are due to differences between the vertical dimensions of \MESSIER{33}
and the model assumptions.
\subsection{Different dust properties}
Differences between the actual dust properties of
the low metallicity galaxy \MESSIER{33}
and those adopted in the models
\citep[][]{1993ApJ...402..441L,
           2001ApJ...554..778L,
           2001ApJ...548..296W,
           2007ApJ...657..810D}
might explain simultaneously the three discrepancies found.
Recently \citet[][]{2013A&A...558A..62J} proposed that interstellar dust
in the diffuse ISM might be characterized by a
lower emissivity index \Eindex{}{} at long wavelengths.
If this is the case, these dust grains would provide flatter emission at
submm and mm wavelengths, which would result in a better agreement between
our modeling and the observations.
Additionally, a lower emissivity index resulting in a
higher extinction coefficient
would allow a lower dust surface density
to account for the dust emission,
which would result in a decrement of the UV absorption,
and thus a lower UV excess,
and a lower dust mass,
thus increasing the value of \Gdust{}{}.
%
%
%

%% file: 07_SUMMARY/Summary.tex
%
%
\section{Summary and conclusions \label{sec:summary}}
We have modeled separately the emission coming from the SF regions and
from the diffuse ISM in the spiral galaxy \MESSIER{33},
using radiation transfer models based on dust grains with standard properties.
Thanks to the wide set of data available for this galaxy,
as well as the numerous previous studies from other authors,
we could constrain an important part of the parameters used by the models.
For the SF component, we found that the model of \citet[][]{2008ApJS..176..438G}
is able to reproduce the MIR-FIR observations and the thermal radio continuum.
However, we found that the model underestimates the \CAMERAmu{LABOCA}{} data
and it is barely compatible with the \CAMERAmu{SPIRE}{500} data.
Regarding the diffuse component, we found that the model of
\citet[][]{2011A&A...527A.109P} fits well the MIR/FIR data points
but underpredicts severely the observations longwards \microns{800}.
We also found that our modeling underpredicts by
\percent{\Vlim{\ESCminYOUNG}{\ESCaveYOUNG}{\ESCmaxYOUNG}}
the UV radiation that escapes the galaxy without interacting with the dust.
We derived a gas-to-dust ratio
\Gdust{=}{\Vlim{\GTDmin}{\GTDave}{\GTDmax}},
significantly lower than the expected value based
on the sub-solar metallicity of \MESSIER{33}.
We discussed various processes that could explain these discrepancies.
A cold dust component emitting in the submm and mm is unlikely
due to the large mass of dust that would be needed.
Spinning grains and the magnetic nanoparticles can not reproduce well
the submm and mm excess and they seriously degrade
the fit at other wavelengths.
Previous studies of the molecular gas in \MESSIER{33} discards the possibility
of anomalies in the \Xco{}{} conversion factor that could explain the
low value of \Gdust{}{}.
Since in the model of \citet[][]{2011A&A...527A.109P} the exact shape of the
stellar radiation field has little influence in the dust emission,
dust heating from evolved stars is unlikely to be a source of error.
Differences in the geometry assumed by the model and the actual geometry
of \MESSIER{33} could explain the discrepancies but there
is no observational evidence for such differences.
A  porous ISM could explain to the excess escape of UV radiation.
As a final option, we discussed different physical properties
of the dust grains.
A lower \Eindex{}{} would flatten the dust emission at submm and mm wavelengths,
decrease the UV attenuation,
and bring the gas-to-dust ratio closer to the expected value,
thus explaining simultaneously the three discrepancies.
%
%
%

%% file: 08_CC/CC.tex
%
%
\section{MBB fit and color corrections \label{app:CC}}
As pointed out in \Sec{sec:Introduction},
MBB models are one of the simplest and widely used techniques
to fit the dust SED.
\citet[][]{2010A&A...518L..67K}
found that a MBB with \Eindex{=}{1.5} can reproduce the dust SED of \MESSIER{33}
up to of \microns{500},
the longest wavelength available in their study.
Here we confirm that the value \Eindex{=}{1.5} found by
these authors is still valid up to wavelengths of \mm{3} as
revealed by \PLANCK{} data (see \Fig{fig:CC}).
\ImageCol{0.48}
         {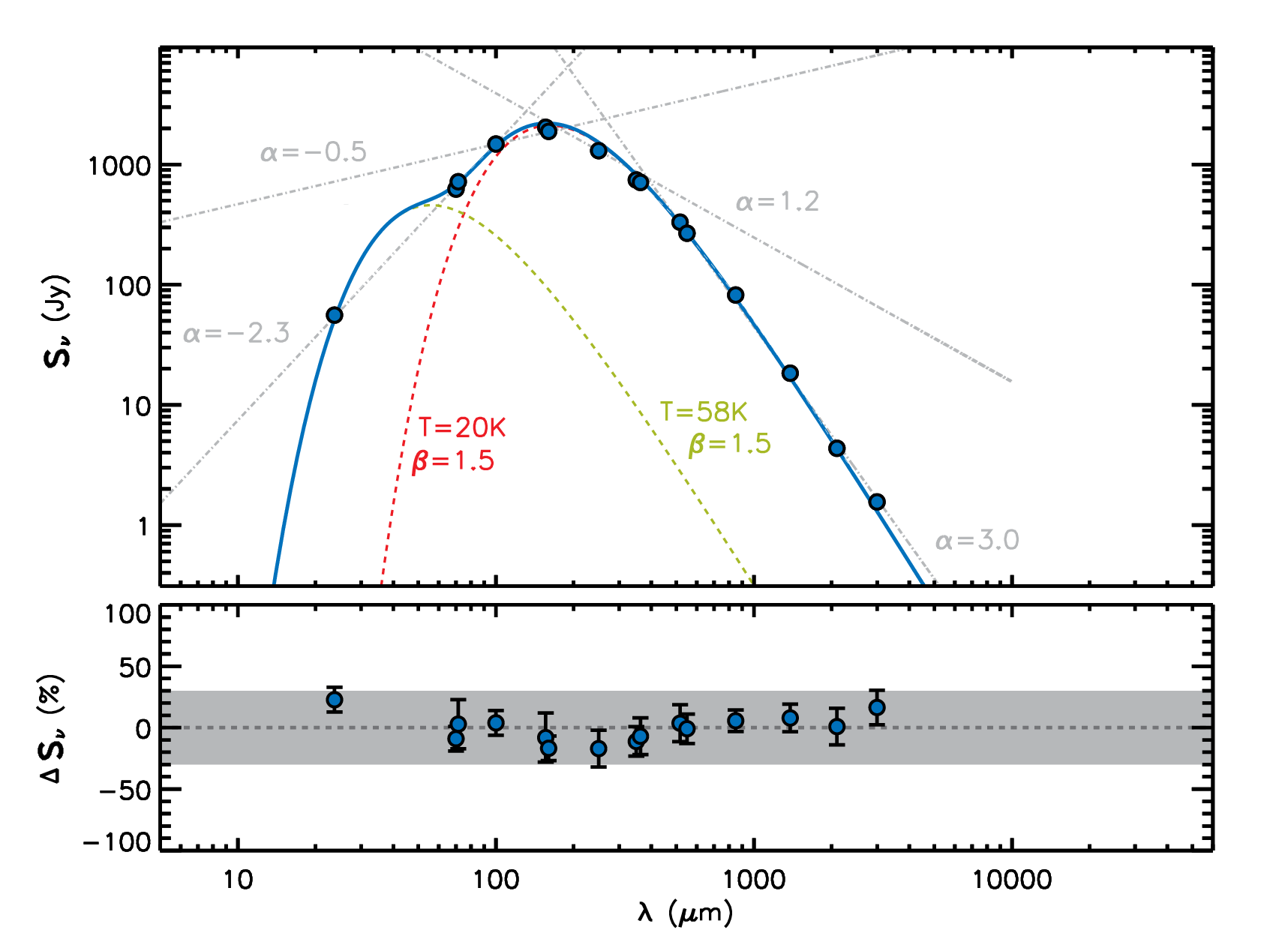}
         {fig:CC}
         {Characterization of the SED of \MESSIER{33} using MBB and power-law spectra.
          The residuals shown in the bottom panel correspond to the MBB fit.
          The gray shaded area of \percent{\pm30} is included to facilitate the reading.
         }
As we mentioned in \Sec{sec:Photometry},
our photometric measurements were color corrected.
Color corrections are necessary to compare monochromatic fluxes with
the models of \citet[][]{2011A&A...527A.109P} and \citet[][]{2008ApJS..176..438G}.
In order to select appropriate SED profiles in the immediacy of each data point,
we mimicked the observations of \MESSIER{33} using MBB models and
power-laws, \PLAW{}.
We found that the observations of \MESSIER{33} can be well reproduced with both a
two-component modified black body of \Eindex{=}{1.5} and
\Temp{=}{20} and \Temp{=}{57} or a combination of four power-laws with indices
ranging from -2.3 to 3.0 (see \Fig{fig:CC} for details).
Then we searched in the tables for
\CAMERAmu{MIPS}{}\COOKBOOK{}{http://irsa.ipac.caltech.edu/data/SPITZER/docs/mips/mipsinstrumenthandbook/},
\CAMERAmu{PACS}{}\COOKBOOK{}{http://herschel.esac.esa.int/Docs/PACS/html/pacs_om.html},
\CAMERAmu{SPIRE}{}\COOKBOOK{}{http://herschel.esac.esa.int/Docs/SPIRE/html/spire_om.html}, and
\PLANCK{} \citep[][]{2014A&A...571A...9P}
for the listed color corrections that best correspond to the profiles found.
For example, for \CAMERAmu{MIPS}{24} we used the value 0.960 tabulated for a
power-law of index -2.0 (no values are listed for -2.3).
For the remaining \CAMERAmu{PACS}{}, and \CAMERAmu{SPIRE}{}
bands we found lower corrections.
%
In the case of \PLANCK{} data, color corrections are higher, with
values ranging from 0.963 at \GHz{143} to 0.854 at \GHz{545}.
%
%
%

%% file: 09_CMB/CMB.tex
%
%
\section{CMB subtraction \label{app:CMB}}
The angular size of the fluctuations of the cosmic microwave background (CMB)
in the line of sight of \MESSIER{33} is comparable to the size of the galaxy
itself (see \Fig{fig:CMB}).
For this reason, the contribution of the CMB
to the flux of \MESSIER{33} must be removed in a pixel-by-pixel basis
instead of of simply assuming an homogeneous background.
Four different CMB maps obtained with different component separation techniques
are available in the \PLANCK{} Legacy Archive
\citep[see \Tab{}~1 in][]{2014A&A...571A..12P}.
The Spectral Matching Independent Component Analysis method (\SMICA),
reconstructs the CMB map as a linear combination of all the
nine \PLANCK{} frequency channels harmonically transformed up to
a multipole \multipole{=}{4000}
\citep[see \App{}~D in][for details]{2014A&A...571A..12P}.
\citet[][]{2014arXiv1407.5452A} used the \SMICA\ map to subtract the CMB
from \MESSIER{31}, as from a visual inspection this appears to be the cleanest
map of the CMB in that region from the four methods.
Following these authors, we also made use of the \SMICA\ map to remove the CMB
in the line of sight of \MESSIER{33}.
The Gnomonic projection of the \SMICA\ map was generated using the same method
explained in \Sec{Data:Planck}.
\ImageCol{0.48}
         {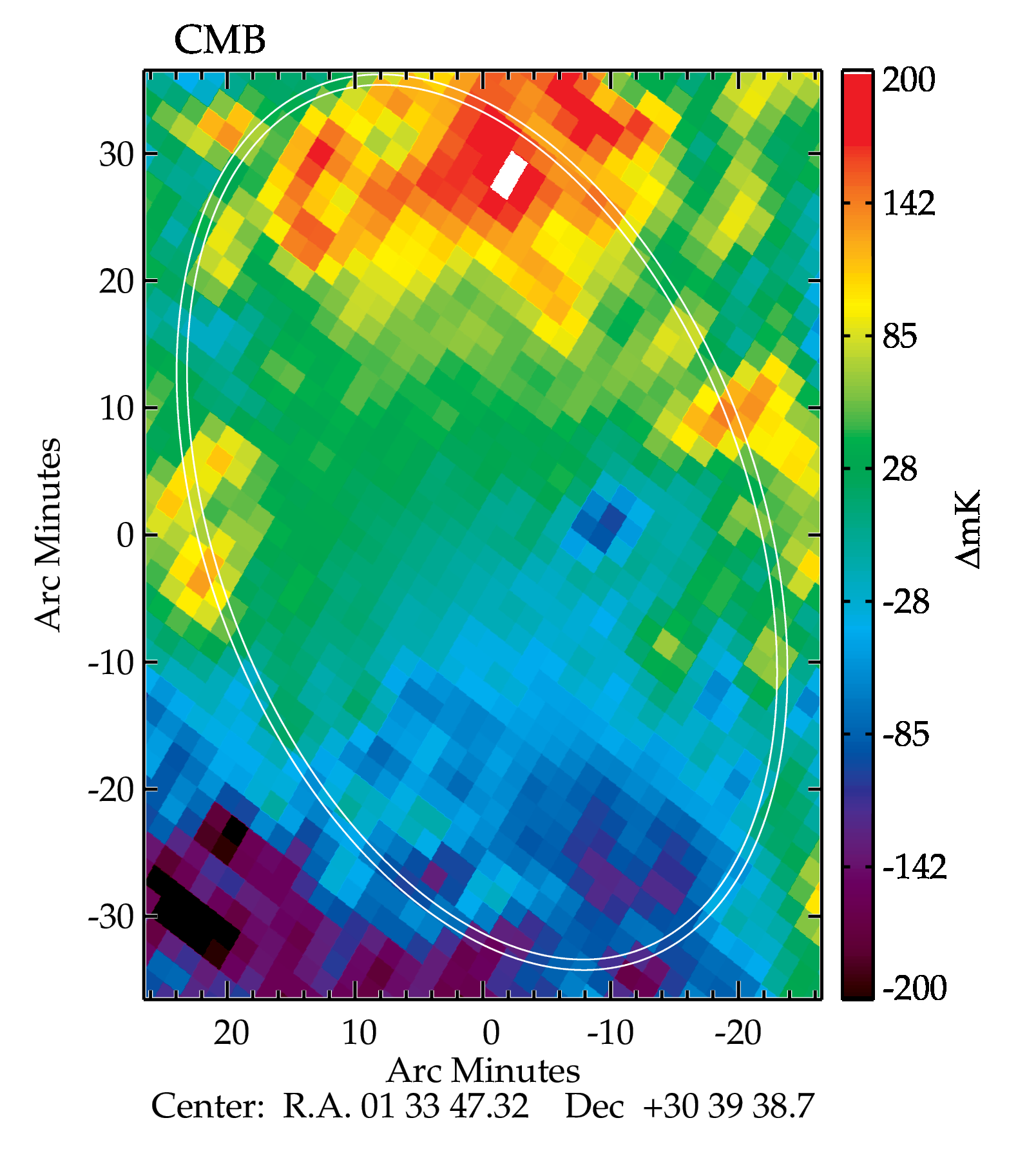}
         {fig:CMB}
         {Gnomonic projection of the CMB map of \PLANCK{} in the line of sight
          of \MESSIER{33} generated with the \SMICA\ method.
          The white annulus used to subtract the local background in
          \Fig{fig:BANDSc} was included here for reference.
          The pixel size is \arcminutes{2}{6}.}
%

%

%% file: 10_THRESHOLD/THRESHOLD.tex
%
%
\section{Alternative thresholds \label{app:THRESHOLDS}}
In addition to the threshold of \THRESave\ times the background noise
used in this paper to separate the SF and the diffuse components
(see \Sec{subsec:ComponentSeparation} for details),
we also examined the results we obtain when adopting thresholds of
\THRESmin\ (see \Fig{fig:M33componentsTmin}) and
\THRESmax\ (see \Fig{fig:M33componentsTmax}).
As \Tab{tab:M33_Photometry_THRESmin} and \Tab{tab:M33_Photometry_THRESmax} show,
the results derived from our modeling does not depend significantly
on the value of the threshold used,
i.e., independently of the threshold we found
i) similar deviations between the models and the observations,
ii) similar dust masses, and
iii) similar fractions of UV radiation escaping \MESSIER{33}.
The main reason is that the emission from the SF component constitutes
only a small fraction of the total SED for most wavelengths
($\microns{100}\lesssim\lambda\lesssim\mm{1}$).
\ImagePag{0.90}
         {./98_FIGURES/figDiffuseCompactT\THRESmin.png}
         {fig:M33componentsTmin}
         {Same as \Fig{fig:M33components} for a
          threshold of \THRESmin\ times the background noise.}
\begin{table*}[!htbp]
    \centering
    \input{97_IDL/SEDsexT\THRESmin/TABLES/M33_RatioObsModel}
    \caption{\label{tab:M33_Photometry_THRESmin}
             Same as \Tab{tab:M33_Photometry} for a
             threshold of \THRESmin\ times the background noise.
             }
\end{table*}
\ImagePag{0.90}
         {./98_FIGURES/figDiffuseCompactT\THRESmax.png}
         {fig:M33componentsTmax}
         {Same as \Fig{fig:M33components} for a
          threshold of \THRESmax\ times the background noise.}
\begin{table*}[!htbp]
    \centering
    \input{97_IDL/SEDsexT\THRESmax/TABLES/M33_RatioObsModel}
    \caption{\label{tab:M33_Photometry_THRESmax}
             Same as \Tab{tab:M33_Photometry} for a
             threshold of \THRESmax\ times the background noise.
             }
\end{table*}
%
%
%
%
%
%

%% file: 97_IDL/SEDsexT20/TABLES/M33_RatioObsModel.tex
%
%
\begin{tabular}{l c c c c c c} \hline \noalign{\medskip}
\textbf{BAND} & $ \frac{ S^{\rm obs}_{\rm SF} } { S^{\rm mod}_{\rm SF} } $ & $ \frac{ \Delta_{\rm SF} } { \sqrt{ \left( \delta^{\rm obs}_{\rm SF} \right)^2 + \left( \delta^{\rm mod}_{\rm SF} \right)^2} } $ & $ \frac{ S^{\rm obs}_{\rm DIFF} } { S^{\rm mod}_{\rm DIFF} } $ & $ \frac{ \Delta_{\rm DIFF} } { \sqrt{ \left( \delta^{\rm obs}_{\rm DIFF} \right)^2 + \left( \delta^{\rm mod}_{\rm DIFF} \right)^2} } $ & $ \frac{ S^{\rm obs}_{\rm TOT} } { S^{\rm mod}_{\rm TOT} } $ & $ \frac{ \Delta_{\rm TOT} } { \sqrt{ \left( \delta^{\rm obs}_{\rm TOT} \right)^2 + \left( \delta^{\rm mod}_{\rm TOT} \right)^2} } $ \\ \noalign{\smallskip} \hline \noalign{\medskip}
\GALEX{FUV}               &             -           &   -   &             -           &   -   & \Vlim{3.01}{5.70}{190.39} & 5.22  \\\noalign{\medskip}
\GALEX{NUV}               &             -           &   -   &             -           &   -   & \Vlim{3.02}{4.30}{7.85} & 7.49  \\\noalign{\medskip}
\SLOAN{u}                 &             -           &   -   &             -           &   -   & \Vlim{1.35}{1.70}{2.31} & 2.68  \\\noalign{\medskip}
\CAMERAmu{MIPS}{24}       & \Vlim{0.95}{1.15}{1.33} & 0.69  & \Vlim{0.83}{1.01}{1.21} & 0.02  & \Vlim{0.86}{1.04}{1.23} & 0.25  \\\noalign{\medskip}
\CAMERAmu{PACS}{70}       & \Vlim{0.87}{0.95}{1.05} & -0.32 & \Vlim{0.83}{1.06}{1.38} & 0.20  & \Vlim{0.85}{1.02}{1.23} & 0.11  \\\noalign{\medskip}
\CAMERAmu{PACS}{100}      & \Vlim{0.84}{0.95}{1.10} & -0.26 & \Vlim{0.80}{0.98}{1.17} & -0.11 & \Vlim{0.81}{0.97}{1.15} & -0.15 \\\noalign{\medskip}
\CAMERAmu{PACS}{160}      & \Vlim{0.86}{0.98}{1.15} & -0.06 & \Vlim{0.85}{0.98}{1.06} & -0.11 & \Vlim{0.86}{0.98}{1.08} & -0.09 \\\noalign{\medskip}
\CAMERAmu{SPIRE}{250}     & \Vlim{0.87}{1.00}{1.17} & -0.01 & \Vlim{0.81}{0.89}{0.97} & -0.93 & \Vlim{0.82}{0.91}{1.01} & -0.76 \\\noalign{\medskip}
\CAMERAmu{PLANCK}{350}    &             -           &   -   & \Vlim{0.90}{1.02}{1.16} & 0.11  & \Vlim{0.89}{1.02}{1.16} & 0.10  \\\noalign{\medskip}
\CAMERAmu{SPIRE}{350}     & \Vlim{0.96}{1.10}{1.30} & 0.63  & \Vlim{0.90}{1.03}{1.17} & 0.17  & \Vlim{0.91}{1.03}{1.18} & 0.23  \\\noalign{\medskip}
\CAMERAmu{SPIRE}{500}     & \Vlim{1.12}{1.29}{1.56} & 1.76  & \Vlim{0.98}{1.14}{1.33} & 0.75  & \Vlim{1.00}{1.17}{1.37} & 1.07  \\\noalign{\medskip}
\CAMERAmu{PLANCK}{550}    &             -           &   -   & \Vlim{1.15}{1.36}{1.58} & 1.71  & \Vlim{1.12}{1.31}{1.53} & 1.50  \\\noalign{\medskip}
\CAMERAmu{PLANCK}{850}    &             -           &   -   & \Vlim{1.36}{1.62}{1.92} & 2.95  & \Vlim{1.30}{1.54}{1.84} & 2.66  \\\noalign{\medskip}
\CAMERAmu{LABOCA}{870}    & \Vlim{1.58}{1.80}{2.21} & 4.39  &             -           &   -   &             -           &   -   \\\noalign{\medskip}
\CAMERAmm{PLANCK}{1.4}    &             -           &   -   & \Vlim{1.74}{2.11}{2.52} & 4.83  & \Vlim{1.63}{1.95}{2.33} & 4.38  \\\noalign{\medskip}
\CAMERAmm{PLANCK}{2.1}    &             -           &   -   & \Vlim{1.63}{1.99}{2.39} & 3.83  & \Vlim{1.47}{1.75}{2.08} & 3.21  \\\noalign{\medskip}
\CAMERAmm{PLANCK}{3.0}    &             -           &   -   & \Vlim{2.40}{2.93}{3.53} & 7.88  & \Vlim{1.68}{1.99}{2.36} & 4.73  \\\noalign{\medskip}
\CAMERAmm{PLANCK}{10}     &             -           &   -   &             -           &   -   & \Vlim{0.73}{0.85}{1.00} & -0.57 \\\noalign{\medskip}
\CAMERAmm{S}{36}          & \Vlim{0.69}{0.79}{0.94} & -1.05 &             -           &   -   & \Vlim{0.69}{0.79}{0.94} & -1.05 \\\noalign{\medskip}
\hline \\
\end{tabular}

%% file: 97_IDL/SEDsexT60/TABLES/M33_RatioObsModel.tex
%
%
\begin{tabular}{l c c c c c c} \hline \noalign{\medskip}
\textbf{BAND} & $ \frac{ S^{\rm obs}_{\rm SF} } { S^{\rm mod}_{\rm SF} } $ & $ \frac{ \Delta_{\rm SF} } { \sqrt{ \left( \delta^{\rm obs}_{\rm SF} \right)^2 + \left( \delta^{\rm mod}_{\rm SF} \right)^2} } $ & $ \frac{ S^{\rm obs}_{\rm DIFF} } { S^{\rm mod}_{\rm DIFF} } $ & $ \frac{ \Delta_{\rm DIFF} } { \sqrt{ \left( \delta^{\rm obs}_{\rm DIFF} \right)^2 + \left( \delta^{\rm mod}_{\rm DIFF} \right)^2} } $ & $ \frac{ S^{\rm obs}_{\rm TOT} } { S^{\rm mod}_{\rm TOT} } $ & $ \frac{ \Delta_{\rm TOT} } { \sqrt{ \left( \delta^{\rm obs}_{\rm TOT} \right)^2 + \left( \delta^{\rm mod}_{\rm TOT} \right)^2} } $ \\ \noalign{\smallskip} \hline \noalign{\medskip}
\GALEX{FUV}               &             -           &   -   &             -           &   -   & \Vlim{2.40}{3.54}{197.88} & 5.19  \\\noalign{\medskip}
\GALEX{NUV}               &             -           &   -   &             -           &   -   & \Vlim{2.78}{3.67}{8.18} & 7.81  \\\noalign{\medskip}
\SLOAN{u}                 &             -           &   -   &             -           &   -   & \Vlim{1.33}{1.64}{2.35} & 2.66  \\\noalign{\medskip}
\CAMERAmu{MIPS}{24}       & \Vlim{0.90}{1.08}{1.25} & 0.37  & \Vlim{0.89}{1.02}{1.29} & 0.13  & \Vlim{0.87}{1.02}{1.24} & 0.09  \\\noalign{\medskip}
\CAMERAmu{PACS}{70}       & \Vlim{0.88}{0.99}{1.12} & -0.03 & \Vlim{0.91}{1.08}{1.52} & 0.36  & \Vlim{0.91}{1.06}{1.36} & 0.23  \\\noalign{\medskip}
\CAMERAmu{PACS}{100}      & \Vlim{0.88}{1.02}{1.17} & 0.09  & \Vlim{0.84}{0.97}{1.22} & -0.13 & \Vlim{0.85}{0.98}{1.21} & -0.09 \\\noalign{\medskip}
\CAMERAmu{PACS}{160}      & \Vlim{0.90}{1.03}{1.18} & 0.11  & \Vlim{0.86}{0.96}{1.08} & -0.15 & \Vlim{0.87}{0.98}{1.09} & -0.10 \\\noalign{\medskip}
\CAMERAmu{SPIRE}{250}     & \Vlim{0.85}{0.98}{1.14} & -0.10 & \Vlim{0.81}{0.88}{1.04} & -0.69 & \Vlim{0.82}{0.90}{1.05} & -0.62 \\\noalign{\medskip}
\CAMERAmu{PLANCK}{350}    &             -           &   -   & \Vlim{0.93}{1.01}{1.21} & 0.05  & \Vlim{0.92}{1.01}{1.21} & 0.03  \\\noalign{\medskip}
\CAMERAmu{SPIRE}{350}     & \Vlim{0.91}{1.06}{1.28} & 0.36  & \Vlim{0.95}{1.02}{1.24} & 0.20  & \Vlim{0.94}{1.02}{1.23} & 0.11  \\\noalign{\medskip}
\CAMERAmu{SPIRE}{500}     & \Vlim{1.04}{1.21}{1.49} & 1.19  & \Vlim{1.03}{1.14}{1.40} & 1.05  & \Vlim{1.04}{1.15}{1.42} & 1.20  \\\noalign{\medskip}
\CAMERAmu{PLANCK}{550}    &             -           &   -   & \Vlim{1.19}{1.32}{1.63} & 2.03  & \Vlim{1.16}{1.29}{1.59} & 1.80  \\\noalign{\medskip}
\CAMERAmu{PLANCK}{850}    &             -           &   -   & \Vlim{1.39}{1.57}{1.97} & 3.77  & \Vlim{1.34}{1.52}{1.90} & 3.39  \\\noalign{\medskip}
\CAMERAmu{LABOCA}{870}    & \Vlim{1.48}{1.72}{2.14} & 3.62  &             -           &   -   &             -           &   -   \\\noalign{\medskip}
\CAMERAmm{PLANCK}{1.4}    &             -           &   -   & \Vlim{1.78}{2.04}{2.57} & 6.02  & \Vlim{1.68}{1.92}{2.42} & 5.38  \\\noalign{\medskip}
\CAMERAmm{PLANCK}{2.1}    &             -           &   -   & \Vlim{1.68}{1.93}{2.46} & 4.56  & \Vlim{1.51}{1.75}{2.18} & 3.57  \\\noalign{\medskip}
\CAMERAmm{PLANCK}{3.0}    &             -           &   -   & \Vlim{2.47}{2.86}{3.64} & 9.90  & \Vlim{1.71}{2.02}{2.51} & 5.01  \\\noalign{\medskip}
\CAMERAmm{PLANCK}{10}     &             -           &   -   &             -           &   -   & \Vlim{0.74}{0.89}{1.09} & -0.38 \\\noalign{\medskip}
\CAMERAmm{S}{36}          & \Vlim{0.69}{0.84}{1.02} & -0.75 &             -           &   -   & \Vlim{0.69}{0.84}{1.02} & -0.75 \\\noalign{\medskip}
\hline \\
\end{tabular}